\documentclass[aps,physrev,groupedaddress]{revtex4-2}
\usepackage{graphicx}
\usepackage{amsmath, amsfonts, amssymb}
\usepackage{stmaryrd,mathrsfs}
\usepackage[normalem]{ulem}
\usepackage[utf8]{inputenc}

\usepackage[colorlinks = true,
linkcolor = blue,
urlcolor  = blue,
citecolor = blue,
anchorcolor = blue]{hyperref}

\newcommand{\const}{\text{const}}
\def\T{\mathcal{T}}
\def\dd{\text{d}}
\def\TH{T_\text{H}}
\def\L{\mathcal{L}}
\def\C{\mathcal{C}}
\def\H{\mathcal{H}}
\def\P{\mathcal{P}}
\def\R{\mathcal{R}}

\def\ren{\text{ren}}

\def\Re{\text{Re}}
\def\spacetime{spacetime}
\def\worldsheet{world sheet}
\def\antiquark{antiquark}

\newcommand{\deriv}[2]{\frac{\dd #1}{\dd #2}}
\newcommand{\pderiv}[2]{\frac{\partial #1}{\partial #2}}

\begin{document}

\title{The Static Heavy Quark-Antiquark Potential within String Theory in Arbitrary Stationary Backgrounds}

\author{Nikita Tsegelnik}
\email[]{tsegelnik@theor.jinr.ru}
\affiliation{Joint Institute for Nuclear Research, Dubna, 141980 Russia}

\date{\today}

\begin{abstract}
	We consider a static open string in arbitrary stationary spacetime, which can represent a heavy quark-\antiquark{} pair within the holographic framework or effective theory.
	Generally, the string profile is not symmetric with respect to the turning point, and the symmetry restores for a simple string configuration in backgrounds with certain constraints.
	We identify a wide family of metrics for which the symmetry is preserved, enabling a direct isolation of the linear-in-distance term in the static potential for simple symmetric string configurations, even in non-diagonal backgrounds.
    As a first example, we apply our formulas to the black brane dual to the $\mathcal{N}=4$ SYM plasma at finite temperature.
    We find that the separation distance between quarks, $L$, is given in terms of a hypergeometric function, while the potential, $V$, consists of two distinct contributions: a term linear in the separation and a term that involves its derivative by temperature.
    Analysis of the leading terms in the series expansion reveals that the temperature corrections of the separation distance lead to the "swallowtail" behavior.
    Further, applying our formulas to the Rindler-AdS spacetime dual to an accelerated $\mathcal{N}=4$ SYM plasma, we obtain analytic expressions for the distance and potential in terms of the elliptic integrals, which in the large Hawking temperature (large acceleration or small curvature) limit come to the conformal results for pure AdS.
	Then, we show that the distance between quarks decreases, the static potential between them increases, and the characteristic temperatures increase with an acceleration, $a_c$.
    However, we observe that an acceleration-scaled potential, $a_c V$, as a function of the acceleration-scaled distance, $a_c L$, does not depend on the certain value of the acceleration.
\end{abstract}


\maketitle

\setcounter{equation}{0}
\section{Introduction}

Since the beginning of the 21st century, there has been active research into the Quantum Field Theory (QFT) in a presence of external fields or under extreme conditions.
Such conditions --- extreme temperatures, densities, rotation, acceleration, background electromagnetic fields or condensates --- can fundamentally alter the theory dynamics.
They may affect phase transitions, stabilize or destabilize various topological structures, or give rise to entirely new novel phenomena.
These studies are crucial not only from a theoretical standpoint, but also for interpreting modern high-energy experiments, where such extreme states are routinely created.

A prominent example is the physics of heavy-ion collisions (HIC), where generated electromagnetic fields are among the strongest in the known universe.
It has been predicted that the magnetic fields reaching up to $10^{18}$ Gauss may arise in non-central collisions~\cite{Voskre-mag, SIT-2009, Bzdak:2011yy, Deng:2012pc, Bloczynski:2012en, Gursoy:2014aka, Inghirami:2016iru, Toneev:2016tgj, Gursoy:2018yai}.
Such fields can significantly modify the Quantum Chromodynamics (QCD) behavior, catalyzing the deconfinement and chiral transitions via the inverse magnetic catalysis~\cite{Galilo:2011nh, Bali:2012zg, Ayala:2014iba, Mueller:2015fka, Mao:2016fha, Farias:2016gmy, Braguta:2019yci}, inducing anomalous transport within the Chiral Magnetic Effect~\cite{Fukushima:2008xe, Sadofyev:2010is, Toneev:2012zx, Kharzeev:2013ffa}, and generating the spin polarization~\cite{PhysRevC.95.054902, Guo:2019joy, Liu:2021nyg, Xu:2022hql, Buzzegoli:2022qrr, Liu:2024hii}.
Recently, a possibility of relatively long-lived strong electric fields has also been explored~\cite{Taya:2024wrm}, offering a direct probe of non-perturbative vacuum decay and pair production via the Schwinger mechanism~\cite{Schwinger:1951nm, Baur:2007zz, Gelis:2015kya}.

Another rapidly growing direction involves non-inertial effects in QFT, which have gained considerable community interest following the observation of global spin polarization in HIC~\cite{STAR:2007ccu, STAR:2017ckg, STAR:2020xbm, HADES:2022enx}.
Contrary to early naive expectations of zero averaged polarization~\cite{PhysRevC.33.1999} and first controversial experimental evidences~\cite{PhysRevLett.47.229, Anikina:1984}, these observations have spurred a comprehensive re-examination of the spin theory~\cite{PhysRevLett.109.232301, Becattini:2013fla, PhysRevC.95.054902, Sorin:2016smp, Teryaev:2017wlm, Csernai:2018yok, BECATTINI2021136519, Liu:2021uhn, Yi:2021ryh, PhysRevD.104.054043, Lin:2022tma}, establishing the thermal vorticity~\cite{Becattini:2013fla, Becattini:2013vja, Becattini:2015ska}, consisting of rotation, acceleration, and temperature gradients, as the dominant source of the observed polarization signal in HIC.

Large angular momentum of the initial system, huge pressure and temperature gradients lead to the generation of the vorticity~\cite{PhysRevC.77.024906, Gao:2007bc, Jiang:2016woz, Sass:2022ucj, Tsegelnik:2022eoz}.
Detailed modelling of the velocity and vorticity fields~\cite{Becattini:2013fla, Becattini:2013vja, Becattini:2015ska, Jiang:2016woz, Teryaev:2015gxa, PhysRevC.93.031902, PhysRevC.93.064907, Ivanov:2017dff, Xia:2018tes, Kolomeitsev:2018svb, PhysRevC.101.064908, Tsegelnik:2022eoz} indicate that the medium created in heavy-ion collisions exhibits extreme local rotation $\omega \sim 10^{21}-10^{23} \text{sec}^{-1}$ and undergoes rapid collective expansion, whose early-time Hubble parameter exceeds the cosmological value by roughly 40 orders of magnitude.
Notable, these predictions of the extreme values of the vorticity lead to the statement that QGP is a fastest-rotating fluid in nature~\cite{STAR:2017ckg, Petersen2017}.
Recently, it was also shown that acceleration of the nuclear medium may reach extremely high values $a\sim 0.1\mbox{--}1\,$GeV~\cite{Prokhorov:2025vak}.

All of these extreme kinematic conditions motivate wide theoretical studies of QFT in rotating~\cite{Unruh:1974bw, Vilenkin:1979ui, Letaw:1979wy, Iyer:1982ah, Davies:1996ks, Duffy:2002ss, Fischer:2003zz, Yamamoto:2013zwa, Ambrus:2014uqa, Ebihara:2016fwa, Jiang:2016wvv, Chernodub:2016kxh, Wang:2018sur, Zhang:2018ome, Chernodub:2020qah, Fujimoto:2021xix, Chen:2022smf, Chen:2023cjt, Sun:2023kuu, Ambrus:2023bid, Voskresensky:2023znr, Sun:2024anu, Chernodub:2025jwy, Kiefer:2025xdp} and accelerated~\cite{Unruh:1983ac, Ohsaku:2004rv, Korsbakken:2004bv, Ebert:2006bh, Castorina:2012yg, Takeuchi:2015nga, Benic:2015qha, Becattini:2017ljh, Prokhorov:2018bql, Prokhorov:2018qhq, Scardigli:2018jlm, Prokhorov:2019cik, Palermo:2021hlf, Akhmedov:2021agm, Chernodub:2024wis, Chernodub:2025ovo, Bordag:2025zbt, Prokhorov:2026swu} medium.
It is also essential to note combined researches including rotation and electromagnetic fields~\cite{Chen:2015hfc, Liu:2017spl, Liu:2017zhl, Fukushima:2018grm, Sadooghi:2021upd, Voskresensky:2024vfx}.
Notably, recent lattice QCD calculations reveal striking results: the deconfinement (or chiral restoration) critical temperature exhibits opposite responses to rotation in the gluonic and fermionic sectors~\cite{Braguta:2020biu, Braguta:2021jgn, Braguta:2021ucr, Braguta:2022str, Braguta:2023aio, Yang:2023vsw}, negative moment of inertia and rotational instability of QGP~\cite{Chernodub:2022veq, Braguta:2023kwl, Braguta:2023yjn, Braguta:2023qex, Braguta:2023tqz, Braguta:2023iyx, Braguta:2025yud}.

These extreme conditions --- strong external fields, rotation, acceleration, and high temperature/density --- often correspond to strongly coupled regimes where traditional perturbative QCD methods fail.
This strongly motivates the use of non-traditional methods, such as gauge/gravity duality, particularly the AdS/CFT correspondence~\cite{Maldacena:1997re}, as a powerful non-perturbative tool~\cite{DeWolfe:2013cua, Casalderrey-Solana:2011dxg, Arefeva:2014kyw}.
By mapping a gauge theory in $D-1$ dimensions to a classical gravitational theory in $D$-dimensional anti-de Sitter (AdS) space, holography provides geometric insight into a strongly coupled regime of the gauge theory.

In particular, the holography framework is successfully used in application to the calculations of the potential energy of the static quark-\antiquark{} pair~\cite{PhysRevLett.80.4859, REY1998171, BRANDHUBER199836, Brandhuber:1999jr, KINAR2000103, Rey:1998ik, Boschi-Filho:2006hfm, Liu:2006he, White:2007tu, Giataganas2012, Giataganas:2011nz}, studies of the confinement and deconfinement phases~\cite{Witten:1998zw, SUNDBORG2000349, SPRADLIN2005199, Polchinski:2000uf, Cho:2002hq, Aharony:2003sx, SPRADLIN2005199, Karch:2006pv, BallonBayona:2007vp, Gursoy:2008za, Marolf:2013ioa, Brodsky:2014yha, PhysRevLett.120.071605}, evaluations of the radiative energy losses~\cite{Sin:2004yx, Herzog:2006se, Herzog:2006gh, Gubser:2006bz, Liu:2006ug, Chernicoff:2012gu}, and other problems~\cite{Akhmedov:1998vf, Son:2002sd, Polchinski:2001tt, Boschi-Filho:2002xih, Boschi-Filho:2002wdj, Kovtun:2004de, Casalderrey-Solana:2006fio, Liu:2006nn, Ryu:2006bv, Shuryak:2005ia, Grumiller:2008va, Bobev:2005cz, Gubser:2009sx, Akhmedov:2010mz, Mateos:2011ix, Strickland:2013uga, Giataganas:2017koz, Chu:2019uoh, Dimov:2019fxp, Giataganas:2022wqj, Avramov:2023eif, Arkhipova:2024iem, Astrakhantsev:2025ujt}.
There are also a lot of researches in presence of the electromagnetic fields (finite chemical potential and baryon density)~\cite{Hawking:1999dp, Caceres:2006dj, DHoker:2009ixq, Ballon-Bayona:2013cta, Dudal:2014jfa, Evans:2016jzo, Li:2016gfn, Arefeva:2018hyo, Braga:2018zlu, Dudal:2018rki, Cheng:2025inr}, rotation~\cite{Hawking:1998kw, Hawking:1999dp, Gibbons:2004ai, Bhattacharyya:2007vs, NataAtmaja:2010hd, Mcinnes:2018wzw, Arefeva:2020jvo, Golubtsova:2021agl, Braga:2022yfe, Chen:2022mhf, McInnes:2022xxt, Yadav:2022qcl, Braga:2023qee, Braga:2023fac, PhysRevD.107.106017, Chen:2022obe, Chen:2023yug, Braga:2023qej, Cai:2023cjl, Chen:2024edy, Ferreira:2025iqe, Zhu:2025bom, Braga:2025eiz}, both of them~\cite{McInnes:2014haa, McInnes:2016dwk, Chen:2020ath, Hou:2021own, Mamani:2022qnf, Zhao:2022uxc, Wang:2024szr, Chen:2024jet, Ahmed:2025bwi}, and uniform acceleration alone or in combinations with aforementioned fields~\cite{Deser:1997ri, Hamilton:2006az, Podolsky:2006px, Chernicoff:2008sa, Russo:2008gb, Ghoroku:2010sp, Hirayama:2010xi, Chernicoff:2010yv, Czech:2012be, Parikh:2011aa, Fareghbal:2014oba, Almheiri:2014lwa, Astorino:2016xiy, Appels:2016uha, Parikh2018, Anabalon:2018qfv, Anabalon:2018ydc, Gregory:2019dtq, Ferrero:2020twa, Cassani:2021dwa, Sugishita:2022ldv, Barrientos:2022bzm, Wu:2023meo, Ju:2023bjl, Arenas-Henriquez:2023hur, Ju:2023dzo}.

A powerful probe for understanding how these extreme conditions affect strongly coupled nuclear matter is the static heavy quark-\antiquark{} potential.
Within the string theory, applicable both in holographic approach and in 4-dimensional string-based effective models of QCD, this potential is computed from the \worldsheet{} action of a (classical) string, the endpoints of which represent the (infinitely) heavy static quark and \antiquark{}.

Specifically, in a gauge-theory context, the expectation value of a rectangular temporal Wilson loop, $\langle W(C) \rangle$, defined over a time interval $T \to \infty$ and a spatial separation $L$, is governed by the dynamics of the underlying flux tube.
In the limit where the flux tube is well-described by a thin, classical string, the dominant contribution comes from the minimal-area \worldsheet{} spanning the contour $C$.
This yields the relation
\begin{align}
\langle W(C) \rangle \sim e^{-T V(L)} = e^{-S_{\textrm{NG}}[\Sigma]},
\end{align}
where $S_{\textrm{NG}}[\Sigma]$ is the (renormalized) Nambu--Goto action evaluated on the corresponding surface $\Sigma$.
The static quark-antiquark potential is therefore extracted as
\begin{align}\label{eq:vqq-def}
V(L) = \lim_{T \to \infty} \frac{S_{\textrm{NG}}[\Sigma]}{T}.
\end{align}

The key advantage of this approach is its generality: the background metric in the string action can encode not only the geometry of a holographic bulk but also the effective spacetime felt by a flux tube in a thermal, rotating, electromagnetically polarized and other medium.
In this paper, we will focus on arbitrary stationary spacetimes for several compelling reasons.
First, a vast and physically relevant class of backgrounds is stationary: pure AdS, AdS black branes and black holes, Rindler-AdS, geometries with instantons or scalar condensates, and many solutions describing equilibrium states with rotation or external fields.
Second, since we are interested in the static quark-\antiquark{} potential ($T \to \infty$), the relevant string \worldsheet{} probes a time-independent configuration.
Third, while stationary metrics retain a high degree of physical generality, their time-independence substantially simplifies the equations of motion (EoM), allowing us to derive closed analytic expressions even in the presence of non-diagonal metric components.
By working with a general stationary metric, we therefore obtain universal, tractable formulas for $V(L)$ that are applicable across a wide spectrum of holographic and effective string-based models of strongly coupled matter.

The paper is organized in the following way.
In Sec.~\ref{sec:general}, starting from arbitrary stationary spacetime, we calculate the separation distance and static potential for the quark-\antiquark{} pair and analyze several important simplifications of the background metrics.
In Sec.~\ref{sec:brane} we obtain a precise analytic expression for the static quark-\antiquark{} potential within the finite-temperature AdS/CFT framework formulated within the black brane; the details of the analytic calculation are provided in App.~\ref{sec:app:brane}.
In Sec.~\ref{sec:rindler-ads} we review the Rindler-AdS background dual to the accelerated $\mathcal{N}=4$ super Yang-Mills plasma and find precise analytic expressions for the separation distance and static potential between quarks, the details of the analytical calculations of which we give in the App.~\ref{sec:app:rindler-ads}.
In Sec.~\ref{sec:summary} we conclude and discuss obtained results and prospects.

\section{The static quark-antiquark potential in arbitrary stationary background}\label{sec:general}

\subsection*{The background and string configuration}

We consider a $D$-dimensional stationary \spacetime{} ($D \geq 3$) with the non-diagonal metric allowing for off-diagonal components that mix different spatial directions:
\begin{align}\label{eq:metric}
	ds^2 = G_{MN} dx^M dx^N,
\end{align}
where the metric components $G_{MN}$ are functions of the spatial coordinates, $x_i$, but independent of the time coordinate, $x_0$.
The string propagation in such background is governed by the Nambu--Goto action
\begin{align}\label{eq:action:general}
	S_{\textrm{NG}} = \frac{1}{2\pi \alpha'} \int  \dd \tau\, \dd \sigma\, \sqrt{-g}.
\end{align}
Here, $1/2\pi \alpha'$ is the string tension coefficient, $\alpha'=\ell_s^2$ is the Regge slope parameter and the fundamental string length scale $\ell_s$, and $g$ is the determinant of the induced metric on the \worldsheet{}.
Parametrizing the string \worldsheet{} by the coordinate pair $(\tau, \sigma)$, one then needs to find the induced metric tensor $g_{\alpha\beta}$, which is defined as
\begin{align}\label{eq:induced-metric:general}
	g_{\alpha\beta} = G_{MN} \partial_\alpha X^M \partial_\beta X^N
\end{align}
with the embedded coordinates $X^M$ and components of the \spacetime{} metric $G_{MN}$.

We employ a static gauge choice for the string \worldsheet{}, parametrizing it in a such way that the time coordinate on the \worldsheet{}, $\tau$, is identified with the background time, $x_0$.
The spatial coordinate on the string \worldsheet{}, $\sigma$, is identified with a specific spatial direction in the global \spacetime{}, $x_p$.
The remaining coordinates, $x_i$ with $i \neq 0, p$, are the dynamical fields $x_i(\sigma)$, describing how the string bends away from the straight line along $\sigma$.
This configuration looks as follows:
\begin{align}\label{eq:string-config}
	x_0 = \tau, \qquad x_p  = \sigma, \qquad x_i = x_i(\sigma), \qquad x_i(\sigma =-L/2) = x_i (\sigma = +L/2) = x_{i_{\max}},
\end{align}
where the string endpoints are fixed at $\sigma=\pm L/2$ and $x_i = x_{i_{\max}}$.
Hence, the distance between endpoints in terms of $\sigma$ has a simple form:
\begin{align}\label{eq:L}
	L = \int_{-L/2}^{L/2} \dd \sigma.
\end{align}

\begin{figure}[!htb]\centering
	\begin{minipage}{0.25\linewidth}
        \includegraphics[width=1\linewidth]{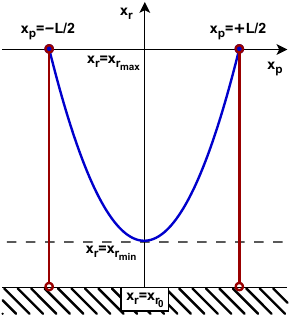}
    \end{minipage}\hspace*{0.2\linewidth}
	\begin{minipage}{0.35\linewidth}
		\includegraphics[width=1\linewidth]{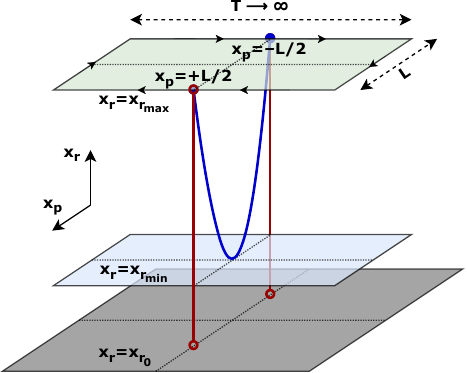}
	\end{minipage}
	\caption{
        The schematic representation for (1) the simple symmetric string profile and (2) the corresponding Wilson loop contour.
	}
    \label{fig:string-profile}
\end{figure}

To avoid ambiguities in the schematic representation of the arbitrary string shape, which generally is non-symmetric with respect to the string turning point, $x_{i_{\min}}$, and can have complex shape in several space directions, we show pictures only for a simple and clear case of the symmetric shape, where only one coordinate, $x_r$, depends on the string coordinate, $\sigma$, while others stay fixed, $dx_i/d\sigma=0$ at $i \neq r$.
For such cases, the string profile and the corresponding Wilson loop contour are shown in Fig.~\ref{fig:string-profile}.
We also show straight strings that run from the boundary, $x_{r_{\max}}$, up to the horizon or deep bulk limit, $x_{r_0}$.
These strings, representing free deconfined quarks, will be used later in the renormalization procedure, described in the corresponding section.
The simple symmetric string profile will be used for the calculations in Sec.~\ref{sec:brane} and Sec.~\ref{sec:rindler-ads}, while in the current section we do not restrict ourself by considering such cases and will show general results for an arbitrary string configuration.
For some example of the asymmetric string configuration, see the discussions about the parity violation later in this section.

\subsection*{The Equations of Motion}

The explicit view of the induced metric~\eqref{eq:induced-metric:general} in the terms of the background metric components~\eqref{eq:metric}~is
\begin{align}\label{eq:induced-metric:1}
	g_{\tau\tau}     & = G_{00},                                                             \\
	\label{eq:induced-metric:2}
	g_{\sigma\sigma} & = G_{pp} + 2\sum_{i} G_{pi} x_i' + \sum_{i}\sum_{j} G_{ij} x_i' x_j', \\
	\label{eq:induced-metric:3}
	g_{\tau\sigma}   & = g_{\sigma\tau} = G_{p0} + \sum_i G_{0i} x_i',
\end{align}
where we denoted $x_i' = \deriv{x_i}{\sigma}$.
Using~\eqref{eq:induced-metric:1}--\eqref{eq:induced-metric:3} one can write down the Lagrangian density~as
\begin{align}\label{eq:lagrangian}
	\L & = \sqrt{-g} = \sqrt{g_{\tau\sigma}^2 - g_{\tau\tau}g_{\sigma\sigma}} = \sqrt{-g_p - \sum_i x_i' \left[ 2h_{pi} + G_{00}\sum_{j} G_{ij} x_j' \right] + \left(\sum_i G_{0i} x_i'\right)^2}.
\end{align}
Here, we introduced following geometrical objects, which are the base terms in our work:
\begin{align}\label{eq:hab}
	h_{ab} & = G_{00} G_{ab} - G_{0a} G_{0b},     \\
	\label{eq:gp}
	g_a    & = h_{aa} = G_{00} G_{aa} - G_{0a}^2.
\end{align}
Quantity~\eqref{eq:hab} characterizes a geometric coupling between directions $a$ and $b$ as seen by the string \worldsheet{}.
For $a=b$, Eq.~\eqref{eq:gp} corresponds to the determinant of the \worldsheet{} metric for a string oriented purely along the $a$-direction.
Hence, $g_p$ in~\eqref{eq:lagrangian} defines the geometry of the background \spacetime{} swept by the unperturbed static string oriented along the $p$-direction.
When $a \neq b$, $h_{ab}$ describes how motions in directions $a$ and $b$ couple in the string dynamics, so $h_{pi}$ in~\eqref{eq:lagrangian} reflects a geometrical connection between $x_p$ and $x_i$ directions.
More precisely, $h_{ab}$ is a projection metric on the spatial sections with $x_0=\const$, that will be discussed later.

Using~\eqref{eq:lagrangian}, we find a first integral of the equations of motion (EoM), which corresponds to the conserved Hamiltonian density~\footnote{In fact, this is \textit{the reduced Hamiltonian density}, but for brevity we will simply call it the Hamiltonian density. For details see App.~\ref{sec:app:virasoro}}:

\begin{align}\label{eq:hamiltonian}
	\H & = \sum_i \P_i x_i' - \L = \frac{g_p + \sum_i h_{pi} x_i'}{\L} = \C,
\end{align}
where the quantities
\begin{align}\label{eq:canonical-mom}
	\P_i = \pderiv{\L}{x_i'} = -\frac{h_{pi} + G_{00} \sum_{j} G_{ij} x_j' - G_{0i} \sum_j G_{0j} x_j'}{\L}
\end{align}
are the canonical momenta.
For our purposes, it is sufficient to use only a set of $\P_i$, while other momenta are fixed via the Virasoro constraints, see App.~\ref{sec:app:virasoro}.
The constant $\C$ is defined from the boundary condition at the string turning point, where all the coordinate derivatives vanish: $x_i'\Big\vert_{x_i = x_{i_{\min}}}=0$.
Substituting this condition into~\eqref{eq:hamiltonian} and solving for $\C$ yields:
\begin{align}\label{eq:C}
	\C = - \sqrt{-g_p}\Big\vert_{x_i = x_{i_{\min}}} = -\sqrt{G_{0p}^2 - G_{00} G_{pp}} \Big\vert_{x_i = x_{i_{\min}}}.
\end{align}
One finds that the canonical momenta in Eq.~\eqref{eq:canonical-mom}, evaluated at the string turning point, are constant:
\begin{align}\label{eq:canonical-mom:turning-point}
	\P_i \Big \vert_{x_i = x_{i_{\min}}} = -\frac{h_{pi}}{\sqrt{-g_p}} \Big \vert_{x_i = x_{i_{\min}}} = \frac{1}{\C} h_{pi} \Big \vert_{x_i = x_{i_{\min}}}.
\end{align}
These vanish for spacetimes with certain symmetries satisfying $h_{pi}=0$ (the simplest case is the diagonal metric).

Now it's convenient to split coordinates $x_i$ into radial, $x_r$, and transverse, $x_i$ ($i \neq r$), directions to change the integration variable $\sigma$ to $x_r$ via $\dd \sigma~=~\dd x_r/x_r'$.
To solve for the derivative $x_r'$, we begin by squaring both sides of Eq.~\eqref{eq:hamiltonian} to eliminate the square root present in the Lagrangian~$\L$:
\begin{align*}
	\frac{(g_p^{(T)})^2 + 2g_p^{(T)} h_{pr} x_r' + h_{pr}^2 x_r'^2}{\L^2_T - g_r x_r'^2 - 2x_r' \left[h_{pr} - G_{0r}\sum_{i \neq r} G	_{0i} x_i'\right]} = \C^2,
\end{align*}
where we defined
\begin{align}\label{eq:gr}
	g_r       & = h_{rr} = G_{00}G_{rr} - G_{0r}^2,                                                                                                                             \\
	g_p^{(T)} & = g_p + \sum_{i \neq r} h_{pi} x_i',                                                                                                                            \\
	\L^2_T    & = \L^2 \vert_{x_r' = 0} = -g_p - \sum_{i \neq r} x_i' \left[ 2h_{pi} + G_{00} \sum_{j \neq r} G_{ij}x_j' \right] + \left( \sum_{i \neq r} G_{0i} x_i'\right)^2.
\end{align}
Here, $\L_T$ is the Lagrangian density for a purely transverse string profile with $x_r'=0$, $g_p^{(T)}$ is a modification of~\eqref{eq:gp} that takes into account the transverse corrections of the string shape, $g_r$ is the determinant of the induced metric for a string oriented purely along the radial direction.

Then, by collecting terms with the same powers of $x_r'$, we obtain a quadratic equation:
\begin{align}\label{eq:x_r':equation}
	a x_r'^2 + b x_r'+ c = 0
\end{align}
with following coefficients
\begin{align*}
	a & = h_{pr}^2 + \C^2 g_r,                                                                               \\
	b & = 2 \left[g_p^{(T)} h_{pr} + \C^2 \left( h_{pr} - G_{0r}\sum_{i \neq r} G_{0i} x_i' \right) \right], \\
	c & = (g_p^{(T)})^2 - \C^2 \L^2_T.
\end{align*}
The solution of Eq.~\eqref{eq:x_r':equation} is given by a standard expression:
\begin{align}\label{eq:x_r'}
	x_r' & = \frac{-b \pm \sqrt{b^2 - 4ac}}{2a}.
\end{align}
For instance, in the backgrounds family with $h_{pr}=0$ for string configurations with simple transverse shape $x_i'=0$ (see Fig.~\ref{fig:string-profile}) the general expression for $x_r'$ simplifies to
\begin{align}\label{eq:x_r':special}
	x_r'^2 & = -\frac{g_p (g_p + \C^2)}{\C^2 g_r},
\end{align}
while considering of diagonal metrics leads to
\begin{align}\label{eq:x_r':diagonal}
	x_r'^2 & = -\frac{G_{pp} (G_{00}G_{pp} + \C^2)}{\C^2 G_{rr}}.
\end{align}

\subsection*{Possible source of the parity violation}

For arbitrary stationary background $h_{pr} \neq 0$, so a consideration of the simple string ansatz with $x_i'=0$ leads to the following view of EoM~\eqref{eq:x_r'}:
\begin{align}\label{eq:x_r':hpr}
	x_r' = \frac{\sqrt{g_p + \C^2}}{h_{pr}^2 + \C^2 g_r} \left( - h_{pr}\sqrt{g_p + \C^2} \pm |\C| \sqrt{-h} \right),
\end{align}
where we defined $h = - (h_{pr}^2 - g_r g_p) = -(h_{pr}^2 - h_{rr} h_{pp})$, the meaning of which we will find below.
Note that the equation~\eqref{eq:x_r':hpr} is generally non-symmetric with respect to the string turning point.
Hence, a non-zero value of the quantity $h_{pr}$ signals a breakdown of symmetry of the string profile along the involved directions, which may be related to the parity violation.

As we noted earlier, $h_{pr}$ reflects space asymmetry in the $p-r$ plane.
The physical meaning of $h$ becomes especially clear when the stationary metric is written in the standard form with splitting along the time-like Killing field:
\begin{align}\label{eq:metric:canonical}
	\dd s^2 = G_{00} (\dd t + A_a \dd x^a)^2 + \frac{h_{ab}}{G_{00}} \dd x^a \dd x^b,
\end{align}
where $G_{00}$ is the norm of the time-like Killing field, $A_a = G_{0a} / G_{00}$ is the gravitomagnetic potential, so $h_{pr}$ defines the geometry of the spatial sections.
Hence,
\begin{align}
	h = -(h_{pr}^2 - h_{rr} h_{pp}) = \det \begin{pmatrix} h_{pp} & h_{pr}\\ h_{pr} & h_{rr} \end{pmatrix} = G_{00}\det \begin{pmatrix} G_{00} & G_{0p} & G_{0r} \\ G_{0p} & G_{pp} & G_{pr} \\ G_{0r} & G_{pr} & G_{rr} \end{pmatrix}
\end{align}
is the determinant of the projection metric on spatial sections with $x_0 = \tau = \const$.
In the parametrization~\eqref{eq:metric:canonical}, $h_{pr}$ combines two possible sources of parity violation: non-zero off-diagonal metric element $G_{pr} \neq 0$ and a presence of the gravitomagnetic fields $A_p\neq0$, $A_r\neq0$.
For instance, it may correspond to pseudoscalar condensates or some stationary flows.

Consequently, the asymmetry of the string profile, the source of which is $h_{pr}\neq0$, may probe a parity violation in the quark-\antiquark{} interaction.
A quantitative measure of this violation can be provided by a shift of the string turning point away from the midpoint between the quark and \antiquark{}, which is computed from the first integral of the equations of motion.

As an example of a background with $h_{pr} \neq 0$, one can consider the Kerr-AdS$_5$ solution (in the static-at-infinity frame~\cite{Gibbons:2004ai}):
\begin{align} 
    {\rm d} s^{2} \sim &-\left(1 + \frac{y^2}{\ell^2}\right) {\rm d} T^2 + \frac{{\rm d} y^2}{1 + \frac{y^2}{\ell^2} - \frac{2M}{\Delta^2 y^2}} + \frac{2M}{\Delta^3 y^2} \left({\rm d} T - a \sin^2 \Theta {\rm d} \Phi - b \cos^2 \Theta {\rm d} \Psi \right)^2 \nonumber\\
&+ y^2 \left({\rm d} \Theta^2 + \sin^2 \Theta {\rm d} \Phi^2 + \cos^2 \Theta {\rm d} \Psi^2\right),
\end{align}
with angles $0 \leq \Theta \leq \frac{\pi}{2}$, $0 \leq \Phi, \Psi \leq 2\pi$, and $\Delta = 1 - \frac{a^2}{\ell^2} \sin^2 \Theta - \frac{b^2}{\ell^2} \cos^2 \Theta$.
Here, $a$ and $b$ are two independent rotation parameters in orthogonal planes, which can give rise to a pseudoscalar (axial) structure in spacetime through the topological invariants of the curved geometry, such as $R\tilde{R}$, which change sign under parity reflection.
Indeed, parametrizing the string profile by $(x_p, x_r) = (\Phi, \Psi)$ or $(x_p, x_r) = (\Psi, \Phi)$, one obtains $h_{pr} \neq 0$:
\begin{equation}
h_{pr} = -\frac{a b M \sin^2 2\Theta}{2\Delta^3 y^2} \left(1 + \frac{y^2}{\ell^2} \right).
\end{equation}
As follows from the equation of motion~\eqref{eq:x_r':hpr}, this non-zero $h_{pr}$ generates two distinct branches for the string profile, which explicitly manifest parity breaking in the static gauge.

When $h_{pr}=0$, the profile becomes symmetric, however, this condition does not necessarily restore full parity invariance of the background: it may merely reflect a specific compensation or a configuration that is insensitive to the existing symmetry violations.
Moreover, one be mindful that $h_{pr}$ is not a diffeomorphism-invariant scalar: its non-zero value in the chosen parametrization signals parity violation, but the invariant physical content of the latter should be extracted from appropriate curvature invariants or conserved quantities.

\begin{figure}[!htb]\centering
    \includegraphics[width=0.2\textheight]{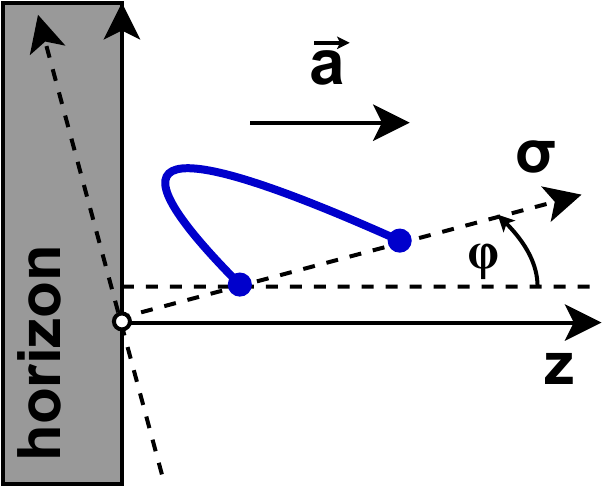}
	\caption{
        The asymmetric string configuration in the 4D Rindler space.
	}
    \label{fig:rindler:asymmetric-string}
\end{figure}

For instance, one can consider the usual 4D Rindler space
\begin{align}
    \dd s^2 = - a^2 z^2 \dd t^2 + \dd x^2 + \dd y^2 + \dd z^2.
\end{align}
This background belongs to the family with $h_{pr} = 0$ for any pair $(x_p, x_r)$.
We show an asymmetric string configuration in Fig.~\ref{fig:rindler:asymmetric-string}, where the \worldsheet{} coordinate $\sigma$ is chosen such that the string is oriented at some angle $\phi$ to the acceleration axis $z$.
Here the asymmetry is evident; however, even in an apparently "straight" configuration with $\sigma = z$ --- where the string is stretched purely along the acceleration axis without sagging --- the energy density remains non-homogeneous.
Although the spatial shape in $(x,y,z)$ coordinates is a straight line, the energy per unit coordinate length (or the effective local tension) decreases toward the horizon.
From the perspective of an observer constantly accelerating in the Rindler wedge, the parity is manifestly broken by the restriction to a non-inertial frame and the limited accessible region.

\subsection*{General expression for the static potential}

Expressing $\L$ via $\C$ from~\eqref{eq:hamiltonian}, one can simplify the action~\eqref{eq:action:general}:
\begin{align}\label{eq:action:C}
	S_\textrm{NG} & = \frac{\T}{2\pi\alpha' \C} \int_{-L/2}^{L/2} \dd\sigma\, \Big(g_p^{(T)} + h_{pr} x_r' \Big).
\end{align}
Taking into account that the EoM~\eqref{eq:x_r':hpr} generally is not symmetric with respect to the string turning point $x_{r_{\min}}$, we rewrite expressions~\eqref{eq:L}~and~\eqref{eq:action:C} as
\begin{align}\label{eq:L-and-action:x_r:L}
	L     & = \int_{x_{r_{\min}}}^{x_{r_{\max}}}\dd x_r \Big(\frac{1}{x_{r_+}'} +  \frac{1}{x_{r_-}'} \Big),                                                                     \\\label{eq:L-and-action:x_r:action}
	S_\textrm{NG} & = \frac{\T}{2\pi\alpha'\C} \left[ \int_{x_{r_{\min}}}^{x_{r_{\max}}} \dd x_r\, \Big(\frac{g_p^{(T)}}{x_{r_+}'} + \frac{g_p^{(T)}}{x_{r_-}'} + 2h_{pr} \Big) \right],
\end{align}
where we denoted two different branches by $x_{r_\pm}$.

The Nambu--Goto action~\eqref{eq:L-and-action:x_r:action} typically exhibits divergence that requires renormalization.
For asymptotically AdS spaces (and other geometries with similar ultraviolet behavior) this is accomplished by subtracting the action of two straight, disconnected strings extending from the boundary to the deep infrared (see red straight lines in Fig.~\ref{fig:string-profile}):
\begin{align}\label{eq:renorm:S0}
	S_0 & = \frac{\T}{\pi\alpha'} \int_{x_{r_0}}^{x_{r_{\max}}} \dd x_r\, \sqrt{-g_r}= \frac{\T}{\pi\alpha'} \left(\int_{x_{r_0}}^{x_{r_{\min}}} + \int_{x_{r_{\min}}}^{x_{r_{\max}}}\right) \dd x_r\, \sqrt{-g_r}
\end{align}
with $g_r$ defined in equation~\eqref{eq:gr}, and $x_{r_0}$ denoted the infrared endpoint: either a horizon radius or, in horizon-free geometries, the deep bulk limit.
Hence, the following expression for the renormalized action can be obtained:
\begin{align}
	S_\textrm{NG}^\ren = S_\textrm{NG} - S_0 = \frac{\T}{\pi\alpha'} \left[ \frac{1}{2\C} \int_{x_{r_{\min}}}^{x_{r_{\max}}} \dd x_r\, \Big(\frac{g_p^{(T)}}{x_{r_+}'} + \frac{g_p^{(T)}}{x_{r_-}'} + 2h_{pr} - 2\C\sqrt{-g_r} \Big) - \int_{x_{r_0}}^{x_{r_{\min}}} \dd x_r\, \sqrt{-g_r} \right].
\end{align}

Finally, associating the string with a heavy quark-\antiquark{} pair, the static potential energy~\eqref{eq:vqq-def} can be explicitly written as
\begin{align}\label{eq:result:V:general}
	\pi\alpha' V = \frac{1}{2\C} \int_{x_{r_{\min}}}^{x_{r_{\max}}} \dd x_r\, \Big(\frac{g_p^{(T)}}{x_{r_+}'} + \frac{g_p^{(T)}}{x_{r_-}'} + 2h_{pr} - 2\C\sqrt{-g_r} \Big) - \int_{x_{r_0}}^{x_{r_{\min}}} \dd x_r\, \sqrt{-g_r},
\end{align}
which is a general expression obtained within the string theory for a general static string configuration in arbitrary stationary $D$-dimensional background.
The obtained result does not depend on whether one works within the framework of the holographic models, or effective description of the static QCD fields between a heavy quark pair.

Note that without additional assumptions, confident simplifications of~\eqref{eq:result:V:general} are not justified.
For example, while one may postulate string symmetry about the string turning point $x_{r_{\min}}$, such a symmetric profile may not represent the minimum-energy configuration when the metric coupling $h_{pr}$ is non-zero (see Eq.~\eqref{eq:x_r':hpr} and discussions later).
Although symmetric solutions might exist in special cases even for $h_{pr} \neq 0$, rigorous confidence in this symmetry requires $h_{pr} = 0$.

\subsection*{Isolation of the linear-in-distance term for some background families}

It is easy to see that the general formulas~\eqref{eq:L-and-action:x_r:L}~and~\eqref{eq:result:V:general} are reduced to particular cases obtained in the literature earlier.
For instance, for diagonal metrics with account of~\eqref{eq:x_r':diagonal} it simplifies to
\begin{align}\label{eq:result:L:diagonal}
    L & = 2 |\C| \int_{x_{r_{\min}}}^{x_{r_{\max}}} dx_r \sqrt{-G_{00}G_{rr}} \frac{1}{\sqrt{G_{00}G_{pp} (G_{00}G_{pp} + \C^2)}} \\
	\label{eq:result:V:diagonal:0}
	\pi\alpha'V & = \int_{x_{r_{\min}}}^{x_{r_{\max}}} \dd x_r\, \sqrt{-G_{00} G_{rr}} \left(\frac{\sqrt{G_{00} G_{pp}}}{\sqrt{G_{00}G_{pp} + \C^2} } - 1\right) - \int_{x_{r_0}}^{x_{r_{\min}}} \dd x_r\, \sqrt{-G_{00} G_{rr}} \\
	\label{eq:result:V:diagonal}
	  & = -\frac{\C L}{2} + \int_{x_{r_{\min}}}^{x_{r_{\max}}} \dd x_r\, \sqrt{-G_{00} G_{rr}} \left( \sqrt{1 + \frac{\C^2}{G_{00}G_{pp}}}-1 \right) - \int_{x_{r_0}}^{x_{r_{\min}}} \dd x_r\, \sqrt{-G_{00} G_{rr}},
\end{align}
where we used the following identity:
\begin{align}\label{eq:L-isolation}
	\frac{x}{\sqrt{x^2+1}} = \frac{\sqrt{x^2+1}}{x} - \frac{1}{x\sqrt{x^2+1}}
\end{align}
The equation~\eqref{eq:result:V:diagonal} is exactly the same obtained in~\cite{Giataganas2012}, taking into account a relation between the constants definitions ($\C = - c$).

For the Schwarzschild-AdS$_5$ and Kerr-AdS$_5$ black holes considered in~\cite{PhysRevD.107.106017} we also receive the same expressions (see Eqs.~(3.12)~and~(3.32) in~\cite{PhysRevD.107.106017}).
In fact, these black holes solutions are the members of the backgrounds family with $h_{pr}=0$, and for the string configurations with fixed transverse shape, $x_i'=0$, shown in Fig.~\ref{fig:string-profile}, we can isolate the linear part in the potential using~\eqref{eq:x_r':special}~and~\eqref{eq:L-isolation}:
\begin{align}\label{eq:result:L:special}
	L & = 2|\C| \int_{x_{r_{\min}}}^{x_{r_{\max}}} \dd x_r \sqrt{-g_r} \frac{1}{\sqrt{g_p (g_p + \C^2)}}                                                                                                                    \\
	\label{eq:result:V:special:1}
	\pi \alpha' V & = \int_{x_{r_{\min}}}^{x_{r_{\max}}} \dd x_r\, \sqrt{-g_r} \Big(\frac{\sqrt{g_p}}{\sqrt{g_p + \C^2}} - 1 \Big) - \int_{x_{r_0}}^{x_{r_{\min}}} \dd x_r\, \sqrt{-g_r}\\
	\label{eq:result:V:special}
	  & = -\frac{\C L}{2} + \int_{x_{r_{\min}}}^{x_{r_{\max}}} \dd x_r\, \sqrt{-g_r} \left( \sqrt{1+\frac{\C^2}{g_p}} - 1 \right) - \int_{x_{r_0}}^{x_{r_{\min}}} \dd x_r\, \sqrt{-g_r}.
\end{align}

The result~\eqref{eq:result:V:special} also means that for such backgrounds (with $h_{pr}=0$) the Nambu--Goto action for a simple static symmetric string configuration ($x_i'=0$, $x_r' \neq 0$) can be initially written as:
\begin{align}\label{eq:result:action:hpr}
	S_\textrm{NG} = \frac{\T}{\pi\alpha'} \left( -\frac{\C L}{2} + \int_{x_{r_{\min}}}^{x_{r_{\max}}} \dd x_r\, \sqrt{-g_r} \sqrt{1+\frac{\C^2}{g_p}} \right)
\end{align}
with $\C$ and $L$ defined in~\eqref{eq:C}~and~\eqref{eq:result:L:special}, correspondingly.

\section{The quark-\antiquark{} potential via AdS/CFT at finite temperature}\label{sec:brane}

\subsection*{The potential and separation distance in the black brane background}

A straightforward generalization of the first calculations~\cite{PhysRevLett.80.4859, REY1998171} of the static potential energy of the quark-\antiquark{} pair, performed in the conformal background of the pure AdS space in the large-$N$ limit, to the finite temperature case was carried out in~\cite{BRANDHUBER199836}.
The background for such case consists of the $N$ coincident D3-branes (the black brane) and looks as
\begin{align}\label{eq:brane:metric}
    \dd s^2 &= -f(U)\frac{U^2}{\R^2}\, \dd t^2 + \frac{1}{f(U)} \frac{\R^2}{U^2}\, \dd U^2 + \frac{U^2}{\R^2}\, \dd x_p^2 + \R^2 \dd \Omega_{d-2},
\end{align}
with the blackening function $f(U) = 1 - U_T^4/U^4$ and AdS curvature $\R=\sqrt{4 \pi g N}$.
The parameter $U_T^4 = \frac{2^7}{3} \pi^4 g^2 \mu$, expressed via the energy density above extremality on the brane $\mu$, and the coupling constant $g$, is related to the Hawking temperature as $\TH=U_T/(\pi \R^2)$.

In this section, following our general formulas from Sec.~\ref{sec:general}, we obtain a precise analytic expressions for the separation, $L$, and static potential, $V$, for the quark-\antiquark{} pair in the large-$N$ limit and then analyze their different temperature limits.
We use the following string configuration:
\begin{align}
    \tau = t, \qquad \sigma = x_p, \qquad x_r = U=U(\sigma), \qquad U(\sigma \pm L/2) = U_{\max} = \infty,
\end{align}
schematic representation of which is shown in Fig.~\ref{fig:string-profile}.
The basic units for the calculations are
\begin{align}\label{eq:brane:basic}
    g_p &= G_{tt} G_{xx} = -f(U) \frac{U^4}{R^4} = \frac{1}{R^4} (U_T^4 - U^4),\\
    g_r &= G_{tt} G_{rr} = -1,\\
    \C &= - \sqrt{-g_p} \Big\vert_{U=U_m} = - \frac{1}{R^2} \sqrt{U_m^4 - U_T^4},
\end{align}
where $U_m$ is the coordinate of the string turning point.
Using these units and Eqs.~\eqref{eq:result:L:special}-\eqref{eq:result:V:special}, one can find the following expressions for $L$ and $V$ (see details of the calculations in App.~\ref{sec:app:brane}):
\begin{align}
      \label{eq:brane:distance}
      L &=  \frac{\R^2}{U_m} \frac{\pi}{\sqrt{2} \beta} \sqrt{1-z}\, {_2F_1}\left(\frac{1}{2}, \frac{3}{4}; \frac{5}{4}; z\right),\\
    \label{eq:brane:potential:hypergeometric-view}
    \pi\alpha'V &= U_T + U_m (1 - z) \frac{\pi}{\sqrt{2} \beta} \left[\frac{1}{2}\, {_2F_1}\left(\frac{1}{2}, \frac{3}{4}; \frac{5}{4}; z\right) - {_2F_1}\left(\frac{3}{4}, \frac{3}{2}; \frac{5}{4}; z\right)\right]\\
    \label{eq:brane:potential:distance-view}
                &= U_T - \frac{U_m^2}{\R^2} \sqrt{1-z} \left( \frac{1+z}{1-z} \frac{L}{2} + 2z \frac{\partial L}{\partial z} \right).
\end{align}
Here $_2F_1(a, b; c; x)$ is the Gauss hypergeometric function, and we introduced a convenient shorthand notation
\begin{align}
    z=U_T^4/U_m^4 
\end{align}
and coefficient
\begin{align}\label{eq:brane:beta}
    \beta = \frac{\Gamma\left(\frac{1}{4}\right)^2}{4 \sqrt{\pi}} = \frac{\pi}{2} \, _2F_1\left(\frac{1}{2}, \frac{1}{2}; 1; \frac{1}{2}\right) = K\left(\frac{1}{2}\right),
\end{align}
which is actually the elliptic integral of the first kind
\begin{align}
    K(m)=\int_{0}^{1} \dd x \frac{1}{\sqrt{(1-x^2)(1 - m x^2)}}   
\end{align}
with elliptic modulus $k = \sqrt{m} = 1/\sqrt{2}$. 
The formulas~\eqref{eq:brane:distance}~and~\eqref{eq:brane:potential:hypergeometric-view} were also obtained in Ref.~\cite{Avramis:2006em}, however, the authors did not recognize the expression~\eqref{eq:brane:potential:distance-view}.

It is exciting that the potential~\eqref{eq:brane:potential:distance-view} is expressed completely in terms of the quarks separation, $L$, and its derivative, $\frac{\partial L}{\partial z}$.
The derivative with respect to $z$ carries a direct physical meaning when one recalls that the temperature at a given bulk depth $U_m$ is redshifted.
The local temperature at some point $U$ is given by the Tolman--Ehrenfest law:
\begin{align}
    T(U) = \frac{\TH}{\sqrt{-G_{tt}(U)}} = \frac{1}{\pi \R} \frac{U_T}{U} \frac{1}{\sqrt{f(U)}}.
\end{align}
In turn, the local temperature at the string turning point, $U_m$, is
\begin{align}
    T(U_m) = \frac{1}{\pi \R} \frac{z^{1/4}}{\sqrt{1-z}}.
\end{align}
Consequently, the derivative $\frac{\partial L}{\partial z}$ quantifies how the quark-\antiquark{} separation changes as the local temperature at the string turning point varies.
Finally, rewriting $U_T=\pi\R^2\TH$ and $U_m=\frac{\R}{\sqrt{2}} \frac{\TH}{T_m}\sqrt{1 + \sqrt{1 + 4 \pi ^4 \R^4 T_m^4}}$, it is possible to express Eqs.~\eqref{eq:brane:distance}-\eqref{eq:brane:potential:distance-view} in terms of the Hawking, $\TH$, and local, $T_m=T(U_m)$, temperatures as
\begin{align}
      \label{eq:brane:distance:T}
      L &= \frac{1}{\sqrt{2}\pi \beta} \frac{\sqrt{z}}{\R T_m\TH} \, {_2F_1}\left(\frac{1}{2}, \frac{3}{4}; \frac{5}{4}; z\right),\\
    \label{eq:brane:potential:hypergeometric-view:T}
    \pi\alpha'V &= \pi\R^2 \TH + \frac{\pi}{\sqrt{2} \beta} \frac{\R\TH}{T_m} \sqrt{1-z} \left[\frac{1}{2}\, {_2F_1}\left(\frac{1}{2}, \frac{3}{4}; \frac{5}{4}; z\right) - {_2F_1}\left(\frac{3}{4}, \frac{3}{2}; \frac{5}{4}; z\right)\right]\\
    \label{eq:brane:potential:distance-view:T}
                &= \pi \R^2 \TH - \frac{\TH^2}{T_m^2} \frac{1}{\sqrt{1-z}} \left( \frac{1+z}{1-z} \frac{L}{2} + 2z \frac{\partial L}{\partial z} \right).
\end{align}

\subsection*{The screening length and critical separation distance}

\begin{figure}[!htb]\centering
	\begin{minipage}{0.325\linewidth}
		\includegraphics[width=1\linewidth]{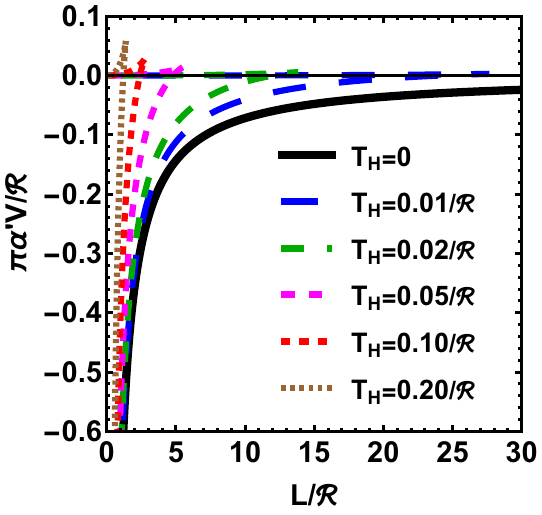}
	\end{minipage}
	\begin{minipage}{0.325\linewidth}
		\includegraphics[width=1\linewidth]{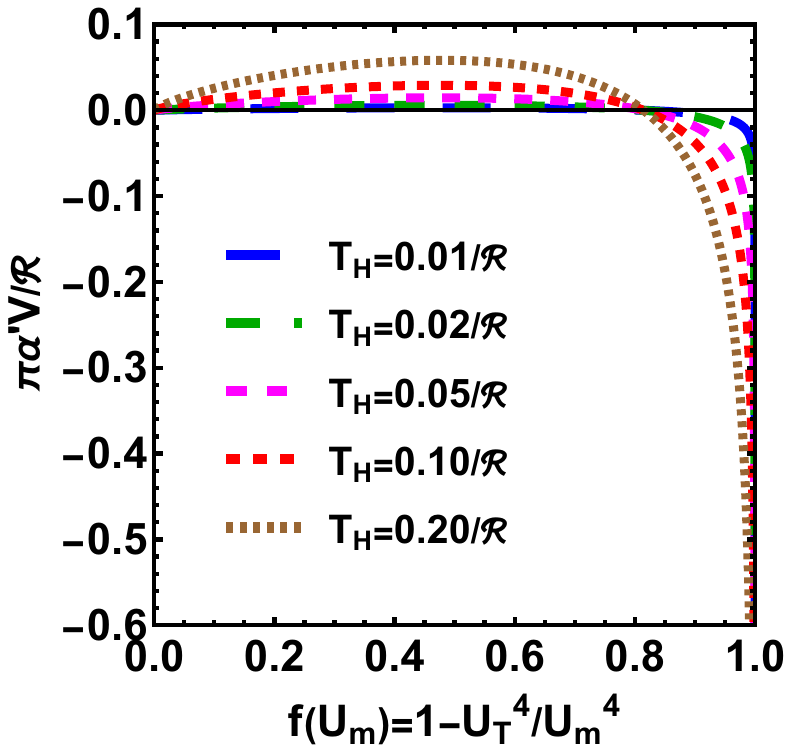}
	\end{minipage}
	\begin{minipage}{0.325\linewidth}
		\includegraphics[width=1\linewidth]{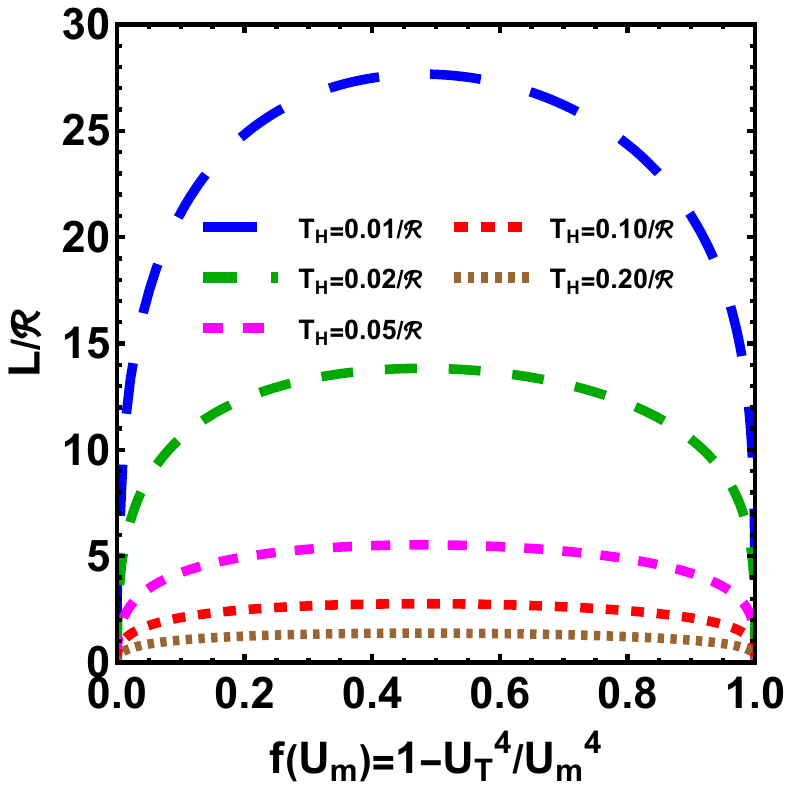}
	\end{minipage}
	\caption{
		Numerical calculations for (1) the quark-\antiquark{} potential $V$~\eqref{eq:brane:potential:hypergeometric-view} depending on the distance $L$~\eqref{eq:brane:distance} between quarks; (2) the potential $V$ as a function of $f(U_m)$; (3) the distance $L$ as a function of $f(U_m)$.
		Different line colors and styles, corresponding to different values of the Hawking temperature, are shown in legends.
        All the quantities are expressed in terms of the AdS curvature $\R$ to avoid the dimensions.
        For the plot (1) we also show the zero-temperature results by a black solid curve.
	}
	\label{fig:brane:compl}
\end{figure}

We show dependencies of the potential and distance between quarks in Fig.~\ref{fig:brane:compl} for several values of the Hawking temperature $\TH$.
To avoid dimensions we express all the quantities in terms of the curvature $\R$, so it is easy to restore all dimensions setting the dimension for $\R$.
Also the potential is multiplied by $\pi\alpha'$ to keep arbitrariness of $\alpha'$.

As one can see, in opposite to the conformal case at zero temperature, where the potential and distance have only one branch, a presence of the finite temperature leads to the double-valued behavior for such quantities.
From the physical point of view, the maximum attainable separation between quark and \antiquark{} is the screening length, beyond which the interaction is practically turned off.
The value of the string turning point at the screening length, $U_s$, is simply related to the parameter $U_T$ as
\begin{align}
    U_s = 2^{1/4} U_T,  
\end{align}
at which point the blackening function takes a middle value, $f(U_s) = 1/2$, and the local temperature depends only on the curvature $\R$:
\begin{align}
    T_s = T(U_s) = \frac{2^{1/4}}{\pi \R}.
\end{align}

\begin{figure}[!htb]\centering
	\begin{minipage}{0.3\linewidth}
        \includegraphics[width=1\linewidth]{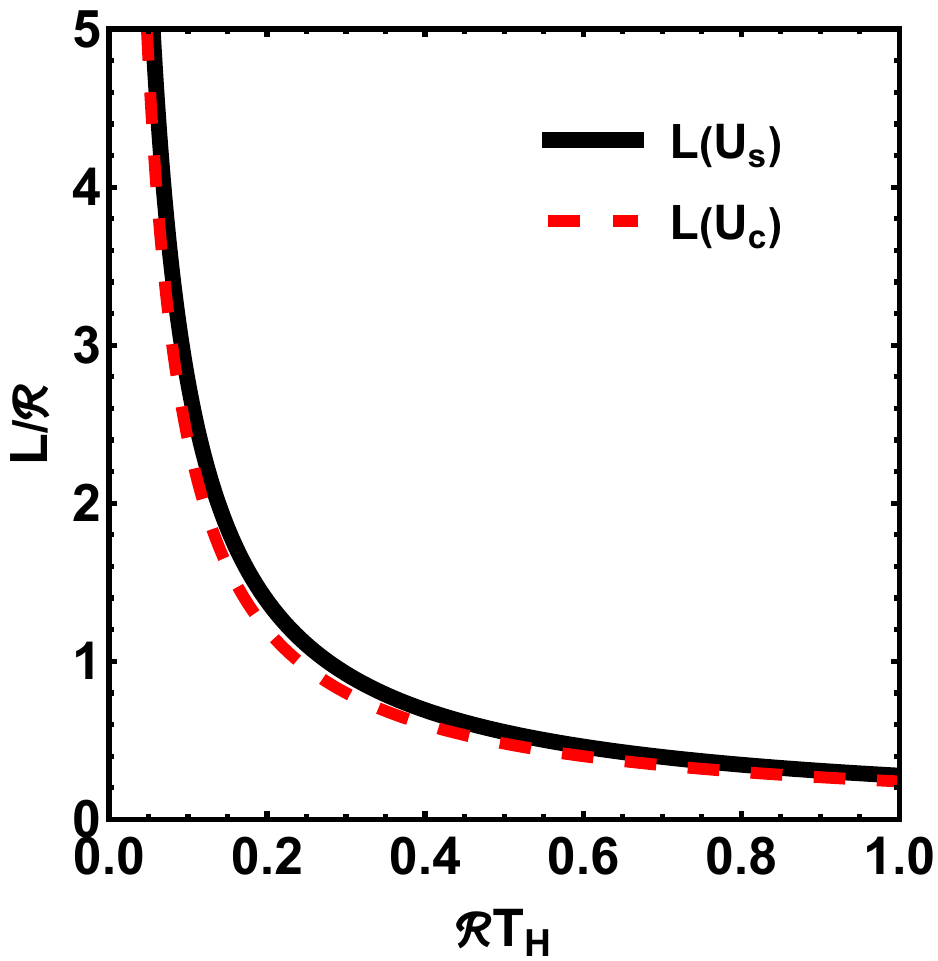}
    \end{minipage}
	\caption{
        The screening length, $L(U_s)$, depicted by solid black line, and the critical distance, $L(U_c)$, depicted by red dashed line, as functions of the Hawking temperature, $\TH$.
        All the quantities are expressed in terms of the AdS curvature $\R$ to avoid the dimensions.
	}
	\label{fig:brane:Lc-Ls}
\end{figure}

The lower branch of the potential corresponds to the case of $U_m > U_s$ and at $V < 0$ reflects the bound state for the quark-\antiquark{} pair --- quarkonium.
Starting from a some point $U_c$, where the static potential vanishes, $V(U_c) = 0$, there is a metastable bound state of the system for a parameters in range of $U_c \geq U_m > U_s$.
For such cases, the U-shaped string configuration, corresponding to a confined pair of quarks, is energetically disfavored compared with a pair of disconnected straight strings that run from the boundary down to the horizon.
The latter representing two independent, deconfined quarks, on which the quarkonium can dissociate via tunneling or thermal fluctuations.
The value of $U_c$ may be well estimated as
\begin{align}
    U_c^{\rm approx} = \left(\frac{4 \sqrt{2} \sqrt{\pi}}{\beta} \right)^{1/4} U_T, \qquad \qquad T_c = T(U_c) \approx \frac{1}{\pi \R} \frac{\left(\frac{4 \sqrt{2} \sqrt{\pi}}{\beta} \right)^{1/4}}{\sqrt{1 - \frac{4 \sqrt{2} \sqrt{\pi}}{\beta}}}.
\end{align}
however, it is not a precise value of the solution for $V(U_c) = 0$ and the ratio of our estimation, $U_c^{\rm approx}$, to a precise solution, $U_c$, is $U_c^{\rm approx}/U_c \approx 0.9991$, but practically the difference is insignificant.
Moreover, the quarks separation at this point, $L(U_c)$, is slightly smaller than the screening length, see Fig.~\ref{fig:brane:Lc-Ls}.
The upper branch with $U_m < U_s$ corresponds to completely unstable configuration in which the string turning point lies too close to the horizon, $U_T$.
In this regime, any small fluctuation causes the middle area of the string to be irreversibly pulled into the horizon, effectively dissociating the quarkonium.

\subsection*{The small temperature limit}

In the zero-temperature limit ($\TH = U_T = z = 0$), our calculation yields exact agreement with the results of Refs.~\cite{PhysRevLett.80.4859, REY1998171, BRANDHUBER199836}:
\begin{align}
    \label{eq:brane:distance:zero-temperature}
    L_0 &= \frac{\pi}{\sqrt{2} \beta} \frac{\R^2}{U_m} = \frac{2 \sqrt{2} \pi^{3/2}}{\Gamma \left(\frac{1}{4}\right)^2} \frac{\R^2}{U_m} ,\\
    \label{eq:brane:potential:zero-temperature}
    V_0 &= -\frac{\pi}{4 \beta^2} \frac{\R^2}{L_0} = - \frac{4 \pi ^2}{\Gamma \left(\frac{1}{4}\right)^4} \frac{\R^2}{L_0} = - \frac{1}{2\sqrt{2} \beta} U_m = - \frac{\sqrt{2\pi}}{\Gamma \left(\frac{1}{4}\right)^2} U_m,
\end{align}
where we set $\alpha' = 1$ for clarity.
The series expansion at the small temperatures looks as
\begin{align}
    \label{eq:brane:distance:small-T-expansion}
    L_{\TH \approx 0} &= L_0  - \frac{4 \beta^4}{5} (L_0 \TH)^4 - \frac{8 \beta^8}{5} (L_0 \TH)^8 - \frac{160 \beta^{12}}{39} (L_0 \TH)^{12} + O\left(\TH^{16}\right),\\
    \label{eq:brane:potential:small-T-expansion}
    V_{\TH \approx 0} &= V_0 - \frac{4 \beta^2}{\pi}(L_0 \TH) + 2 \beta^4 (L_0 \TH)^4 +\frac{6 \beta^8}{5} (L_0 \TH)^{8} + \frac{28 \beta^{12}}{15} (L_0 \TH)^{12} + O\left(\TH^{16}\right),
\end{align}
where the coefficient $\beta$ is given by~\eqref{eq:brane:beta}.
The conformal nature of the theory reveals itself in the fact that $V L$ can depend on $\TH$ only through the combination $L \TH$, as it was noted in Ref.~\cite{BRANDHUBER199836}.
However, actual combination is $L_0 \TH$, due a fact that $L$ itself depends on the temperature.
Also in~\cite{BRANDHUBER199836} a lowest correction term $\sim L_0 \TH$ was missed (see Eq.~(8) in~\cite{BRANDHUBER199836}), but it comes from $U_T = \pi \R^2 \TH$ after the renormalization procedure.

\begin{figure}[!htb]\centering
	\begin{minipage}{0.4\linewidth}
        \includegraphics[width=1\linewidth]{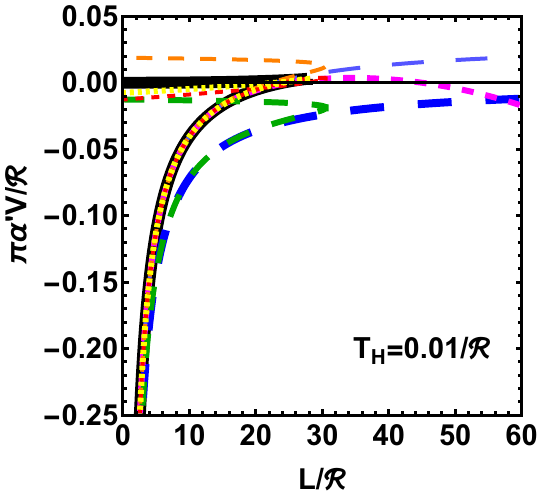}
    \end{minipage}\hspace*{0.05\linewidth}
	\begin{minipage}{0.3\linewidth}
		\includegraphics[width=1\linewidth]{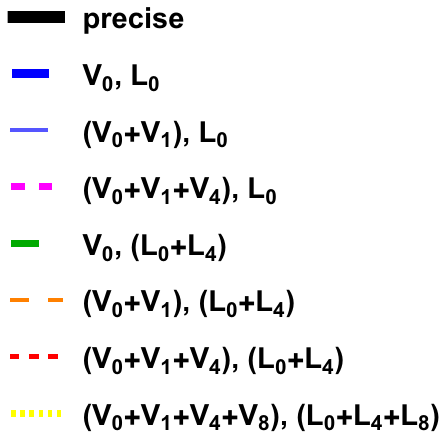}
	\end{minipage}
	\caption{
        The dependence of the potential from the series expansion~\eqref{eq:brane:distance:small-T-expansion} as a function of the separation $L$ from the expansion~\eqref{eq:brane:potential:small-T-expansion} at the fixed Hawking temperature $\TH=0.01/\R$.
        Different terms of the expansions, indexed in terms of the temperature powers, are depicted in the legend.
        All the quantities are expressed in terms of the AdS curvature $\R$ to avoid the dimensions.
	}
	\label{fig:brane:series}
\end{figure}

To analyze the series expansions~\eqref{eq:brane:distance:small-T-expansion},~\eqref{eq:brane:potential:small-T-expansion}, we plot different terms with respect to the temperature powers in Fig.~\ref{fig:brane:series}.
At small separations, the string worldsheet remains close to the conformal boundary, $U_m \gg U_T$, where the blackening function $f(U_m) \approx 1$.
In this region, the metric~\eqref{eq:brane:metric} is close to the pure AdS, and the system exhibits approximate conformal invariance.
Consequently, the leading behavior of the potential is universal and coincides with the zero-temperature Coulomb form, $V \approx V_0 \sim \frac{1}{L_0}$.
As the separation increases, the string probes deeper into the bulk, eventually reaching the region, where $f(U_m)$ deviates significantly from unity. 
The full non-linear dependence on temperature, encoded in the metric, then becomes essential, and the precise forms of the potential and distance are sensitive to higher-order terms in the temperature expansions.
Note that the "swallowtail" behavior of the potential and separation distance is caused by temperature corrections for the separation only.
The universal small-$L$ behavior is thus a robust UV feature, while the near-critical region reflects the detailed IR deformation of the conformal theory by the thermal scale.

\section{The quark-\antiquark{} potential in the Rindler-AdS space}\label{sec:rindler-ads}

\subsection*{The Rindler-AdS background}

In this section we will clarify the influence of acceleration on the quarkonium dissociation within the holographic approach.
One need to consider the Rindler-AdS space, the dual of which is the $\mathcal{N} = 4$ super Yang-Mills theory.
The Rindler-AdS \spacetime{} has been widely investigated in~\cite{Deser:1997ri, Czech:2012be, Almheiri:2014lwa, Parikh2018, Arenas-Henriquez:2023hur}.
The heavy quark-\antiquark{} potential in the presence of acceleration was previously calculated for the $\mathcal{N} = 2$ confining SYM theory in~\cite{Ghoroku:2010sp}, and for $\mathcal{N} = 4$ SYM in~\cite{Hirayama:2010xi}, however, here we present more comprehensive study, including first precise analytic expressions.

The Rindler-AdS metric is given by the following expression:
\begin{align}\label{eq:rindler-ads:metric}
	\dd s^2 = - a_c^2\xi^2\, \dd t^2 + \frac{\dd\xi^2}{1+a_c^2\xi^2} + (1+a_c^2\xi^2) \left[\dd\chi^2 + \frac{1}{a_c^2} \sinh^2 (a_c\chi)\, \dd\Omega_{d-2}^2\right].
\end{align}
For a qualitative analysis, one often sets the coordinate Rindler acceleration equal to the inverse AdS radius, $a_c=1/\R$ (see, for instance, Ref.~\cite{Parikh2018}).
The numerical value of $a_c$ itself is not a physical invariant (it can be changed by rescaling the coordinates), but the AdS radius $\R$ provides the intrinsic geometric scale of the \spacetime{}.
The quantity $a_c=1/\R$ then carries a correct dimension of the acceleration and serves as a natural reference value.

In the limit $a_c \xi \longrightarrow 0$ the metric~\eqref{eq:rindler-ads:metric} reduces to the line element of flat Rindler space with a horizon at $\xi=0$:
\begin{align}
	\label{eq:rindler-ads:metric:limit}
	\dd s^2 = - a_c^2\xi^2 \dd t^2 + \dd\xi^2 + \dd\chi^2 + \chi^2 \dd\Omega_{d-2}^2.
\end{align}
The coordinates~\eqref{eq:rindler-ads:metric} describe a region satisfying $(X^1)^2-(X^0)^2>0$ and $(X^1)^2-(X^{d+1})^2>0$ in a $(d+2)$-dimensional embedding Minkowski space with two time-like directions:
\begin{align}
	\label{eq:rindler-ads:metric:embedding}
	-(X^0)^2 + (X^1)^2 + ... + (X^d)^2 - (X^{d+1})^2 = -\R^2,
\end{align}
which possesses the symmetry group $O(2,d)$, corresponding to AdS$_{d+1}$.
This embedding shows that the Rindler observers in AdS follow hyperbolic trajectories in the ambient Minkowski space, i.e., they are also Rindler observers with respect to the embedding space.

At $\xi \longrightarrow \infty$ the metric~\eqref{eq:rindler-ads:metric} asymptotically approaches the conformal boundary of AdS:
\begin{align}
	\dd s^2_{\rm boundary} = -\dd t^2 + \dd\chi^2 + \frac{1}{a_c^2} \sinh^2 (a_c\chi)\, \dd \Omega_{d-2}^2,
\end{align}
which is the metric on $\mathbb{R} \times \mathbb{H}^{d-1}$, where the hyperbolic spatial section $\mathbb{H}^{d-1}$ has a curvature radius of $1/a_c = \R$.
Moreover, for any finite $\xi$, the induced metric on a constant-$\xi$ hypersurface is also conformally equivalent to $\mathbb{R} \times \mathbb{H}^{d-1}$.
This reflects the fact that the Rindler-AdS foliation preserves the conformal structure of the boundary at every finite $\xi$, even though the holographic dual CFT in the AdS/CFT sense is usually defined only on the ideal conformal boundary attached at $\xi\longrightarrow\infty$.

As it known from the general relativity, the temperature seen by an observer moving with constant acceleration
is not always proportional to the proper acceleration.
For the Rindler-AdS space the proper acceleration of an observer at constant $\xi$ is given by
\begin{align}\label{eq:rindler-ads:acceleration}
	a_{\rm prop}^2 = a^2_{\rm loc} + a^2_c = \frac{1}{\xi^2} + \frac{1}{\R^2},
\end{align}
where the local acceleration is defined as $a_{\rm loc} = \frac{1}{\xi}$.
In work~\cite{Deser:1997ri} a general formula for the local temperature in (A)dS$_{d+1}$ was obtained:
\begin{align}\label{eq:rindler-ads:temperature:general}
	T_{\rm loc} = \frac{1}{2\pi} \sqrt{\frac{2\Lambda}{d(d-1)} + a^2_{\rm prop}} = \frac{a_{\rm embed}}{2\pi},
\end{align}
where $a_{\rm embed}$ is the proper acceleration of the Rindler observer in the flat embedding space and $\Lambda$ is the cosmological constant.

For AdS$_{d+1}$ the cosmological constant looks as $\Lambda = - \frac{1}{\R^2} \frac{d(d-1)}{2}$, therefore the condition for a real value of the temperature requires the proper acceleration to satisfy $a_{\rm prop} \geq \frac{1}{\R} = a_{\rm crit}$, where we defined a critical acceleration $a_{\rm crit}$.
The observers with $a_{\rm prop} > a_{\rm crit}$ possess a non-degenerate Rindler-type horizon and measure a non-zero local temperature~\eqref{eq:rindler-ads:temperature:general}, which can be rewritten as
\begin{align}
	T(\xi) = T_{\rm loc} = \frac{1}{2\pi} \sqrt{a^2_{\rm prop} - a^2_{\rm crit}}.
\end{align}
At $a_{\rm prop} = a_{\rm crit}$ the horizon becomes extremal (zero surface gravity) and the local temperature vanishes.
As an example, such observer is located at $\xi \longrightarrow \infty$.
Finally, observers with $a_{\rm prop} < a_{\rm crit}$ have no causal horizon.
Their trajectories are globally extendible and they do not perceive a thermal spectrum.

Remarkably, the critical acceleration coincides exactly with the parameter that appears in the Rindler-AdS metric~\eqref{eq:rindler-ads:metric}, i.e. $a_{\rm crit} = a_c$.
The proper acceleration~\eqref{eq:rindler-ads:acceleration} is always greater than or equal to $a_c$.
Thus, the Rindler-AdS coordinate system naturally covers only the worldlines of super-critical (and, asymptotically, critical) observers.
The parameter $a_c$ in the metric therefore has a clear physical interpretation: it sets both the AdS curvature scale $\R$ and the minimal acceleration required for an observer to possess a horizon and a well-defined temperature in AdS.

Using~\eqref{eq:rindler-ads:acceleration}, one can obtain a final expression for the local temperature:
\begin{align}\label{eq:rindler-ads:temperature}
	T(\xi) = T_{\rm loc} = \frac{1}{2\pi\xi}.
\end{align}
Remarkably, this is precisely the same expression as for an accelerated detector in flat Rindler space, despite the non-zero curvature of AdS.
The horizon at $\xi=0$ possesses the Hawking temperature $\TH$, defined via its surface gravity.
For the Rindler-AdS metric, one finds $\TH = a_c/(2\pi)= 1/(2\pi\R)$, so the local temperature~\eqref{eq:rindler-ads:temperature} can then be written as
\begin{align}
	T(\xi) = \frac{\TH}{\sqrt{-G_{tt}(\xi)}} = \frac{1}{2\pi\xi},
\end{align}
which is exactly the Tolman--Ehrenfest law for temperature distribution in a static gravitational field.
Thus, while the horizon itself has a fixed temperature $\TH$, an accelerated observer at finite $\xi$ experiences a higher local temperature due to gravitational redshift.

\subsection*{Influence of the acceleration on the static quark-\antiquark{} potential}

Choosing the following string ansatz,
\begin{align}\label{eq:rindler-ads:configuration:1}
	\tau = t, \qquad \sigma=\chi, \qquad \xi=\xi(\sigma),\qquad \xi(\sigma=\pm L/2) = \xi_{\max} = \infty,
\end{align}
corresponding to a simple symmetric string shape depicted in Fig.~\ref{fig:string-profile}, one can find our main quantities for the potential and distance expressions:
\begin{align}\label{eq:rindler-ads:determinants}
	g_p & = G_{tt} G_{\chi\chi} = - a_c^2\xi^2 (1+a_c^2\xi^2),    \\
	g_r & = G_{tt} G_{\xi\xi} = -\frac{a_c^2\xi^2}{1+a_c^2\xi^2}.
\end{align}
The first integral takes the following value
\begin{align}
	\C & = - a_c \xi_m \sqrt{1+a_c^2\xi_m^2},
\end{align}
where the physical value of the string turning point, $\xi_m$, satisfying $\xi \geq 0$, $\C^2 \geq 0$, $a_c>0$, is given by
\begin{align}\label{eq:rindler-ads:xi-min}
	\xi_m & = \frac{\sqrt{\sqrt{4\C^2+1}-1}}{\sqrt{2}a_c}.
\end{align}
In the special case of an acceleration $a_c=1/\xi_m$, the first integral is a universal constant $\C=-\sqrt{2}$.

Now, using~\eqref{eq:rindler-ads:determinants}-\eqref{eq:rindler-ads:xi-min} one can analytically calculate the integrals in the expressions for the distance~\eqref{eq:result:L:special} and potential~\eqref{eq:result:V:special} between quarks.
Direct calculations (see Appendix~\ref{sec:app:rindler-ads} for details) lead to the following expressions for the distance $L$ and potential $V$:
\begin{align}
	\label{eq:result:L:rindler-ads}
	L           & = 2\xi_m\, \Re\left[\tilde{\Pi}(n, m) - \frac{K(1-m)}{1-n}\right],                                                     \\
	\label{eq:result:V:rindler-ads:1}
	\pi\alpha'V & = \frac{1}{a_c} + \sqrt{\frac{1}{a_c^2}+\xi_m^2}\, \Re\big[i\,K(m) -\tilde{E}(m)\big]                                      \\
	\label{eq:result:V:rindler-ads:2}
	            & =\frac{1}{a_c} -\frac{\C L}{2} + \frac{\xi_m \C}{n}\, \Re \big[\tilde{K}(m) - n\, \tilde{\Pi}(n, m) - \tilde{E}(m)\big],
\end{align}
where we introduced linear combinations of the complete elliptic integrals
\begin{align}
	\tilde{\Pi}(n, m) & = m\,\Pi(1-n, 1-m) + i\,\Pi(n, m), \\
	\tilde{K}(m)      & = m\,K(1-m) + i\,K(m),             \\
	\tilde{E}(m)      & = E(1-m) + i\,E(m)
\end{align}
with the characteristic $n$, elliptic modulus $k$ and parameter $m$ given by
\begin{align}
	n=-a_c^2\xi_m^2, \qquad m=k^2=\frac{n}{1-n}.
\end{align}
It was tested that the expressions~\eqref{eq:result:L:rindler-ads}-\eqref{eq:result:V:rindler-ads:2} give the same result as direct numerical integrations of~\eqref{eq:result:L:rindler-ads}~and~\eqref{eq:result:V:special} in the Rindler-AdS case.
Notably, the elliptic integrals in the context of the static interquark potential are also arisen in different setups, for instance~\cite{Dorn:2007zy, Drukker:2011za}.

Calculations of the asymptotic behavior lead to the following expressions:
\begin{align}
    \TH \sim a_c = \frac{1}{\R} \longrightarrow 0&:\qquad L \sim \hat{L} = 2 \xi_m\, \log \left(\frac{4}{a_c \xi_m}\right), \qquad V \sim \frac{a_c \xi_m}{4} (\hat{L} - 3 \xi_m) = \frac{\sqrt{\pi}}{2\sqrt{2}} \sqrt{\frac{|V_0|}{L_0}} (\hat{L} + 6 \sqrt{2} \beta V_0);\\
    \TH \sim a_c = \frac{1}{\R} \longrightarrow \infty &:\qquad L \sim L_0, \qquad V \sim V_0 \sim - \frac{1}{L_0},
\end{align}
where $L_0$ and $V_0$ are the conformal results in pure AdS defined in~\eqref{eq:brane:distance:zero-temperature}~and~\eqref{eq:brane:potential:zero-temperature}, correspondingly, and $\beta$ is given by~\eqref{eq:brane:beta}.
Interestingly, that the high-temperature (large acceleration or small curvature) limit in the Rindler-AdS asymptotically comes to the results for a conformal case of pure AdS.
The opposite limit is related to the logarithmic divergencies caused by a singularity in the metric~\eqref{eq:rindler-ads:metric}.

To compare different AdS geometries (different radii $\R$) while keeping a fixed reference length $\ell$, it is convenient to introduce a dimensionless scaling parameter $a_0$ via the relation $a_c=a_0/\ell$, where $\ell$ is a fixed reference length, for example the AdS radius of a chosen ``reference'' geometry (so that $a_0=1$ in that case).
This parametrization emphasizes that, when $\ell$ is held fixed, the dimensionless number $a_0$ distinguishes one geometry from another, whereas $a_c$ (or equivalently $1/\R$) absorbs the dimensional dependence.

Consequently, physical observables should depend on acceleration only through dimensionless combinations.
For example, in the final part of this section we will see that the quark-\antiquark{} potential, written as dimensionless function, $a_c V$, of the dimensionless distance between quarks, $a_c L$, is independent of the specific numerical choice of $a_0$, which is a direct consequence of the unphysical nature of $a_c$.
On the other hand, the dimensionless ratio $V/\ell$ expressed as a function of $L/\ell$ retains an explicit dependence on $a_0$, thereby reflecting how the underlying geometry changes with the rescaled acceleration parameter.

\begin{figure}[!htb]\centering
	\begin{minipage}{0.325\linewidth}
		\includegraphics[width=1\linewidth]{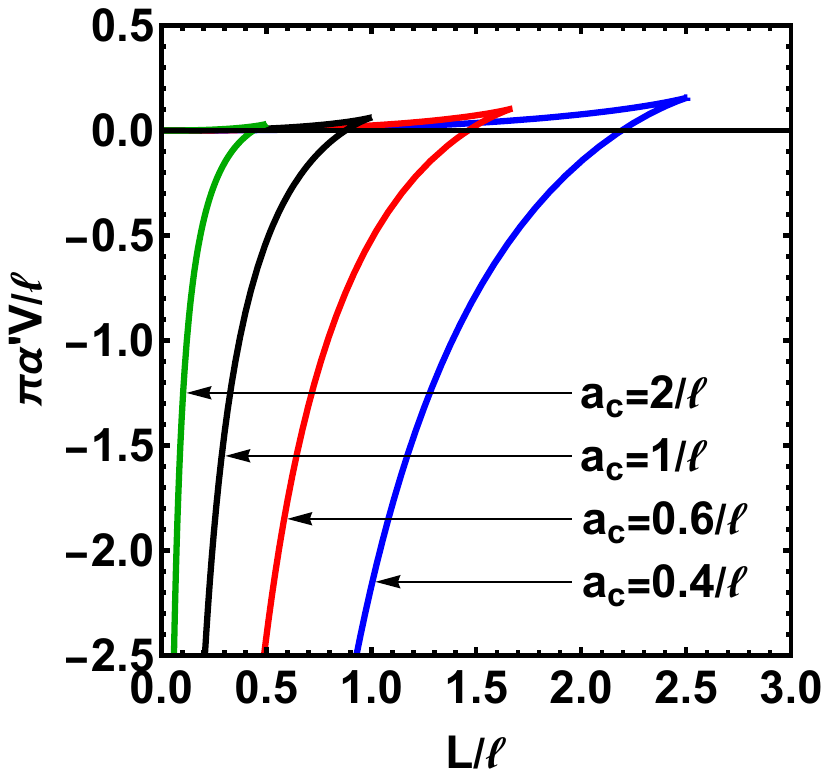}
	\end{minipage}
	\begin{minipage}{0.325\linewidth}
		\includegraphics[width=1\linewidth]{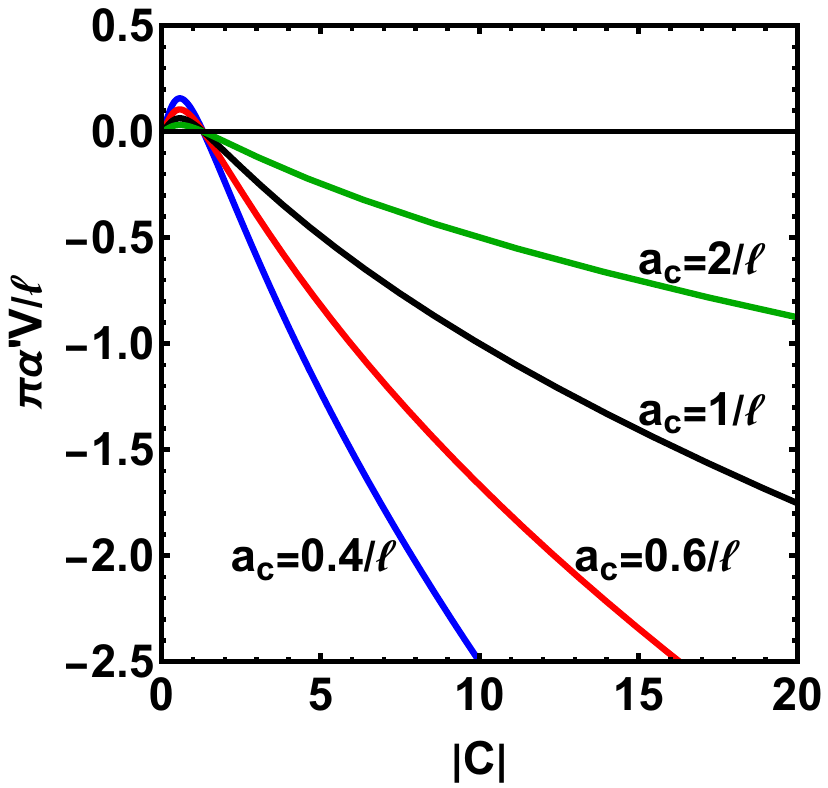}
	\end{minipage}
	\begin{minipage}{0.325\linewidth}
		\includegraphics[width=1\linewidth]{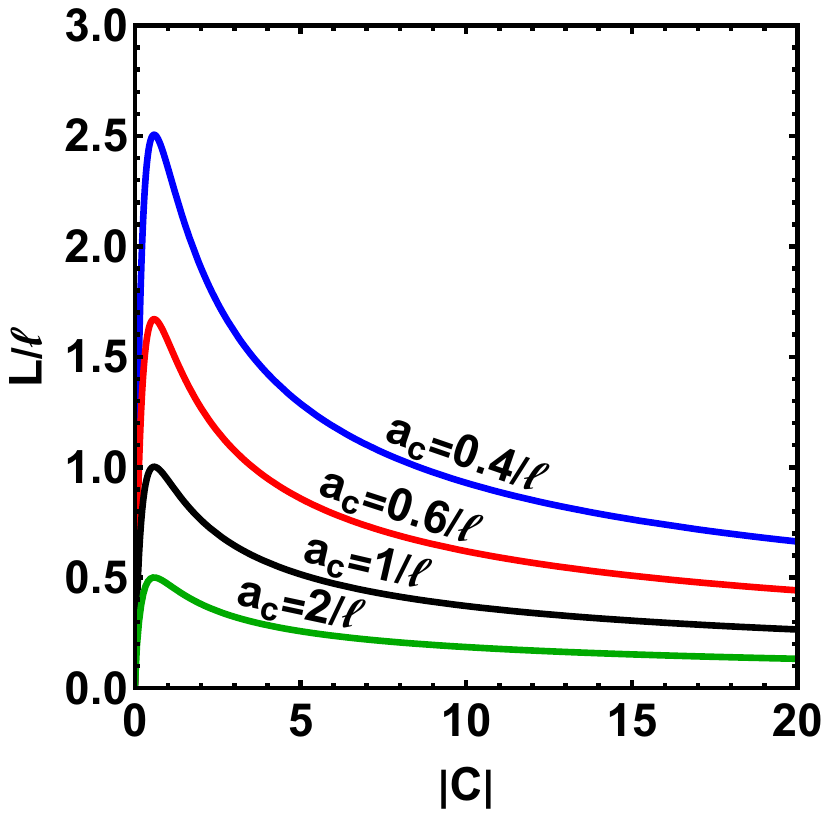}
	\end{minipage}
	\caption{
		Numerical calculations in the Rindler-AdS background~\eqref{eq:rindler-ads:metric} for (1) the quark-\antiquark{} potential $V$~\eqref{eq:result:V:special} depending on the distance $L$~\eqref{eq:result:L:special} between quarks; (2) the potential $V$ as a function of the absolute value of constant $\C$; (3) the distance $L$ as a function of the absolute value of constant $\C$.
		Different line colors correspond to different values of the acceleration: blue for $a_c=0.4/\ell$, red for $a=0.6/\ell$, black for $a_c=1/\ell$, and green for $a_c=2/\ell$.
	}
	\label{fig:Rindler:1D}
\end{figure}

We show dependencies of the potential and distance between quarks in the Rindler-AdS background at in Fig.~\ref{fig:Rindler:1D} for several values of the acceleration parameter $a_0=a_c\ell=0.4,\, 0.6,\, 1,\, 2$.
To avoid dimensions we express all the quantities in terms of $\ell$.
Also the potential is multiplied by $\pi\alpha'$ to keep arbitrariness of $\alpha'$.

\begin{figure}[!htb]\centering
	\begin{minipage}{0.32\linewidth}
		\includegraphics[width=1\linewidth]{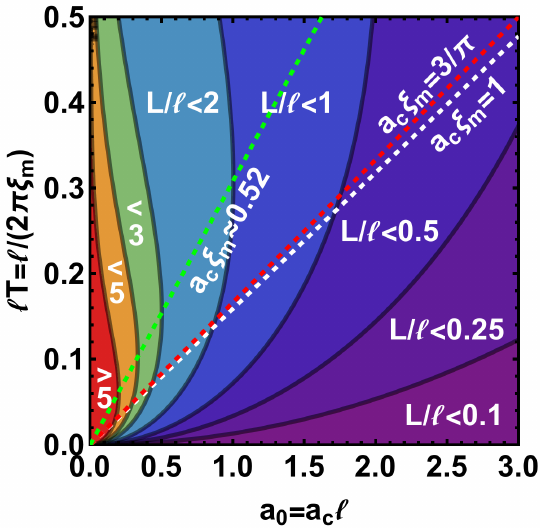}
	\end{minipage}\hspace*{1em}
	\begin{minipage}{0.32\linewidth}
		\includegraphics[width=1\linewidth]{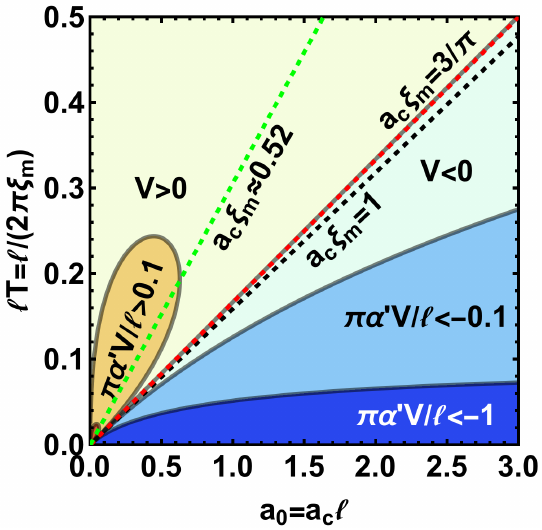}
	\end{minipage}
	\caption{
		Contours plots of the distance $L$ (first pane) and potential $V$ (second pane) between quarks as functions of the dimensionless acceleration parameter $a_0 = a_c \ell$ and of the dimensionless temperature at the string turning point $\ell T = \ell/(2\pi\xi_m)$.
		Different values of $V$ or $L$, corresponding certain contours, are depicted by colors and labels.
		The characteristic temperature $T_c = (\pi/3) \TH = a_c/6$ is shown by a red dashed line, while the Hawking temperature $\TH$ is drawn by a rest dashed line.
	}
	\label{fig:Rindler:2D}
\end{figure}

Again, one can observe a double-valued behavior for the potential and separation distance, so we refer to Sec.~\ref{sec:brane} for the explanation.
However, now a critical value of the string turning point, $\xi_c$, at which the potential takes a zero value, $V(\xi_c)$, is related to the acceleration/curvature as
\begin{align}
    \xi_c = \frac{3}{\pi a_c} = \frac{3\R}{\pi},
\end{align}
where the temperature is 
\begin{align}
    T_c = \frac{\pi}{3} \TH = \frac{a_c}{6},
\end{align}
that is depicted by a red dashed line in the contour plots in Fig.~\ref{fig:Rindler:2D}.
Note that it is slightly higher (the factor is $\approx 1.05$) than the horizon temperature $\TH$, which is depicted by a black/white line in Fig.~\ref{fig:Rindler:2D}.

The parameters relation for the separation length can be find from the condition $\frac{\partial V(a_c, \xi_m)}{\partial \xi_m} \Big\vert_{\xi_m = \xi_s} = 0$, from which one finds
\begin{align}\label{eq:rindler-ads:screening-length}
    E \left(\frac{1 + a_c^2 \xi_s^2}{1 + 2 a_c^2 \xi_s^2}\right) = \frac{1}{2} K \left(\frac{1 + a_c^2 \xi_s^2}{1 + 2 a_c^2 \xi_s^2}\right).
\end{align}
The approximate solution of~\eqref{eq:rindler-ads:screening-length} and corresponding temperature are
\begin{align}
    \xi_s \approx \frac{0.5163}{a_c}, \qquad \qquad T_s = T(\xi_s) \approx \frac{a_c}{2\pi \cdot 0.5163},
\end{align}
which are shown in Fig.~\ref{fig:Rindler:2D} by a green dashed line.

\begin{figure}[!htb]\centering
	\begin{minipage}{0.35\linewidth}
		\includegraphics[width=1\linewidth]{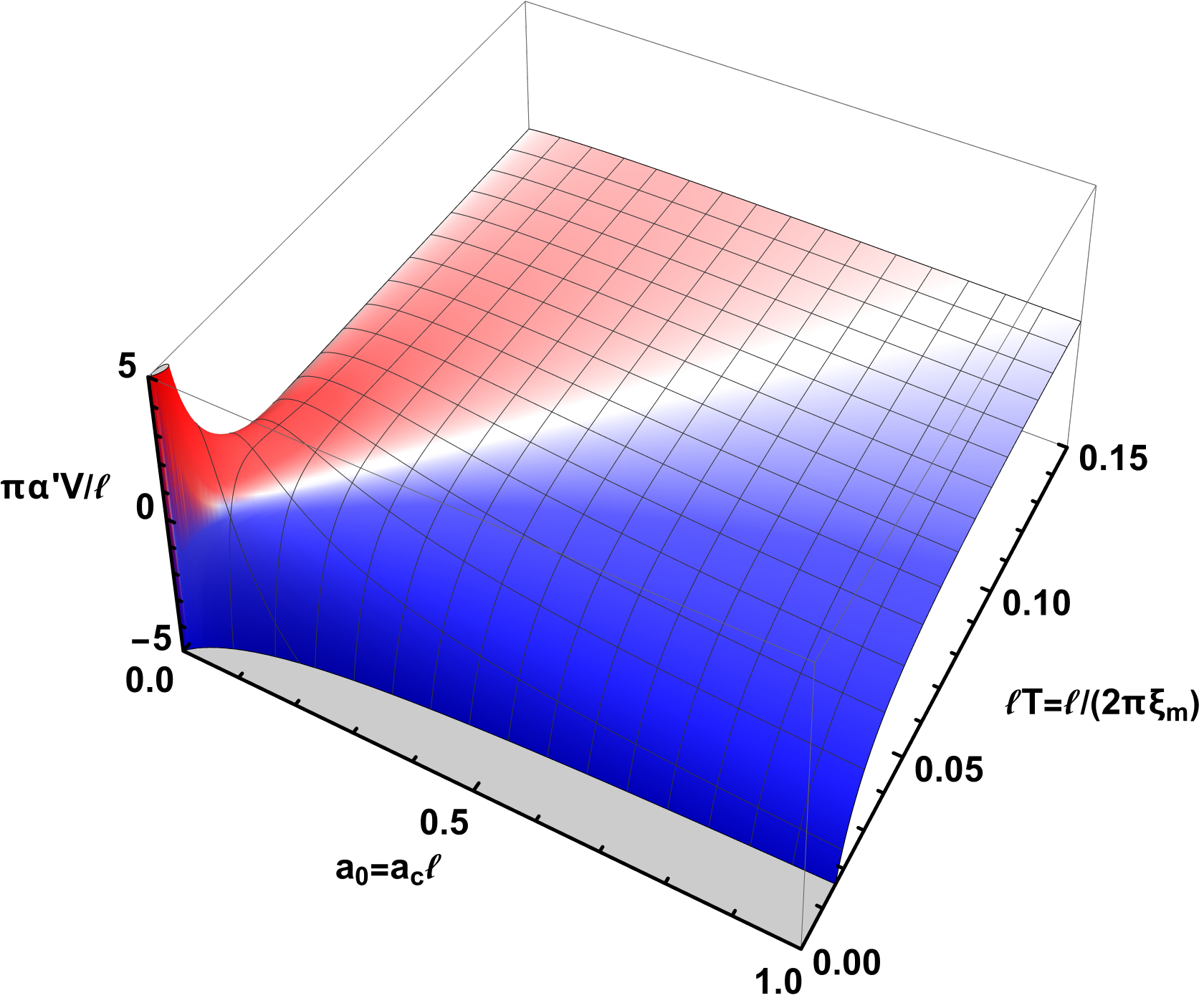}
	\end{minipage}\hspace*{1em}
	\begin{minipage}{0.32\linewidth}
		\includegraphics[width=1\linewidth]{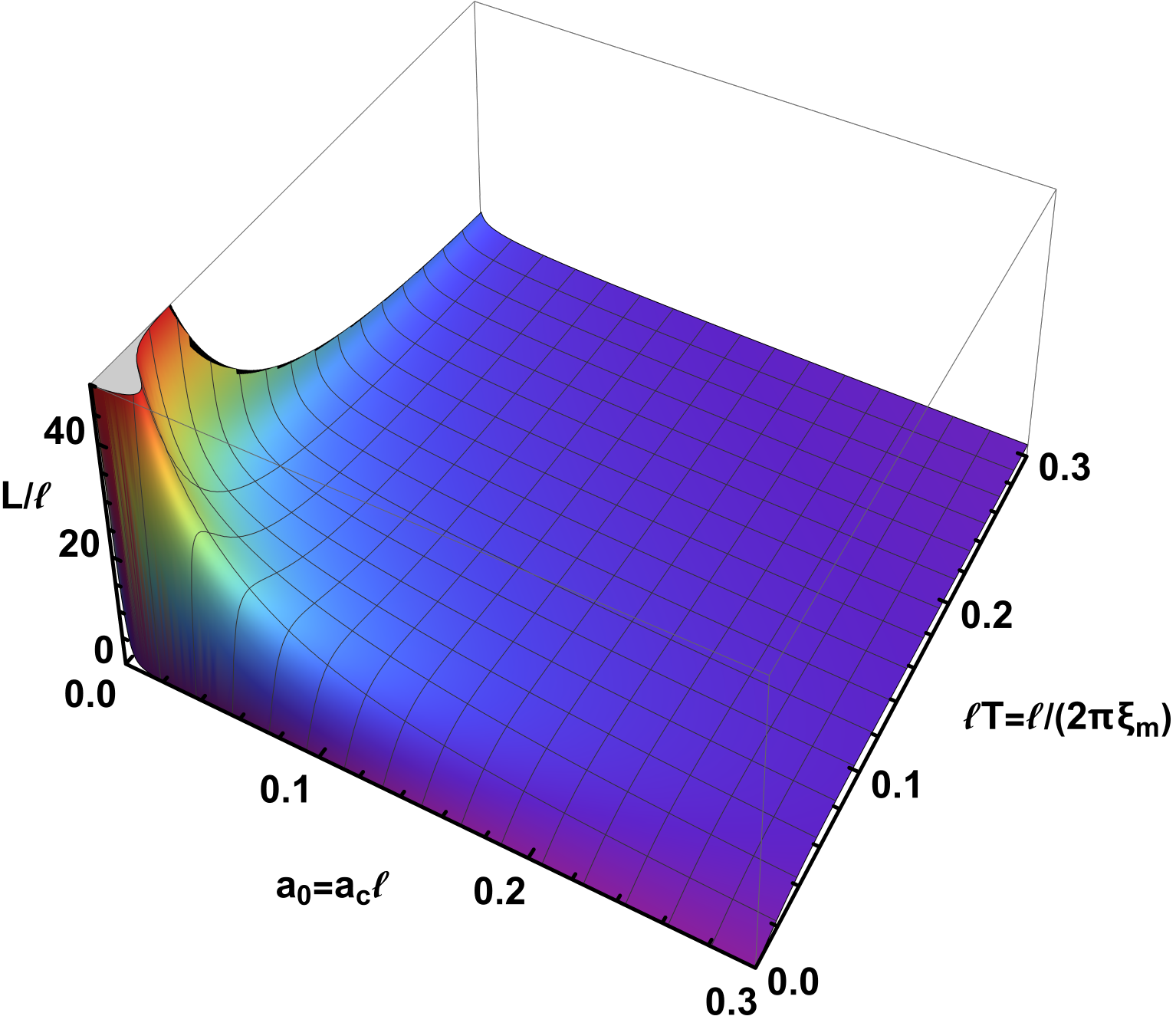}
	\end{minipage}\hspace*{1em}
	\begin{minipage}{0.32\linewidth}
		\includegraphics[width=1\linewidth]{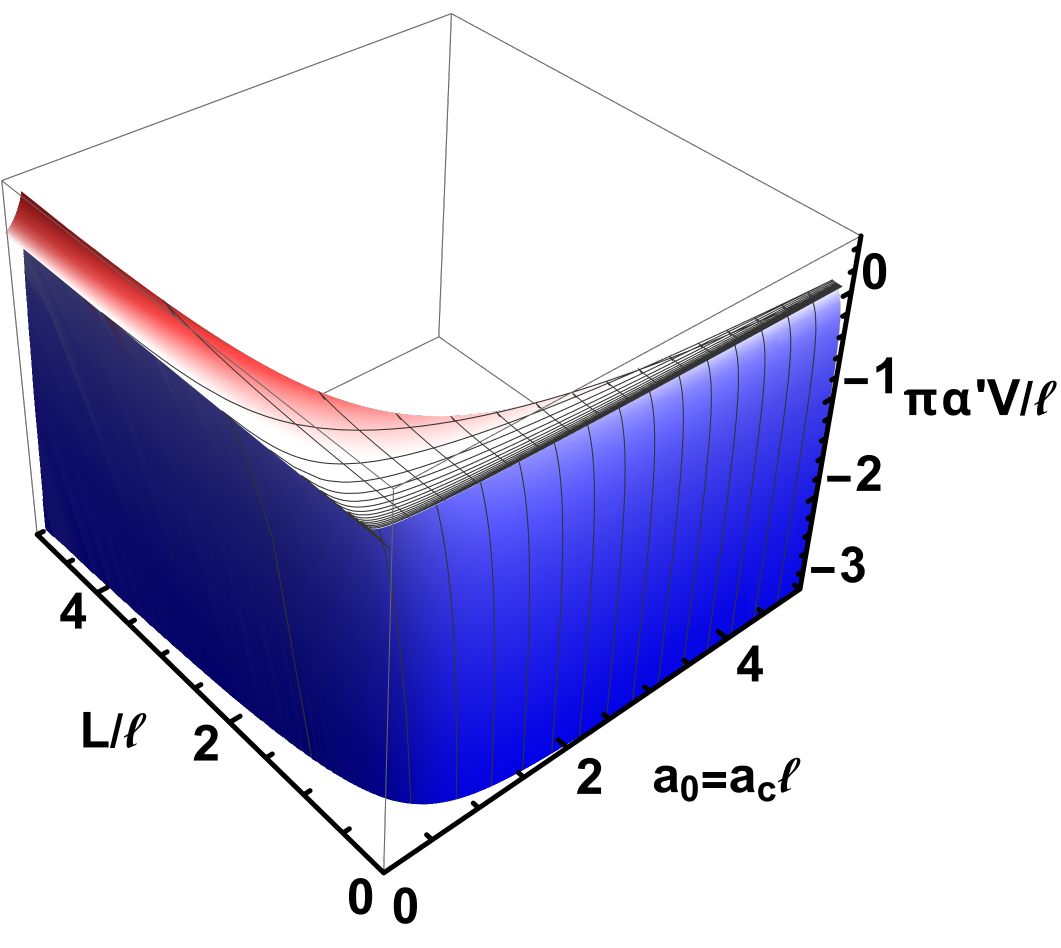}
	\end{minipage}
	\caption{
		3-dimensional plots of the distance $L$ (first pane) and potential $V$ (second pane) between quarks as functions of the dimensionless acceleration parameter $a_0 = a_c \ell$ and of the dimensionless temperature at the string turning point $\ell T = \ell/(2\pi\xi_m)$; the potential $V$ as a function of the distance $L$ and acceleration parameter $a_0$ (third pane).
	}
	\label{fig:Rindler:3D}
\end{figure}

In Fig.~\ref{fig:Rindler:3D} we also shown 3-dimensional plots of the distance and potential in the acceleration--temperature at the string turning point plane.
As one can see, at low acceleration $a_0$ and temperatures $T(\xi_m)$ the distance and potential undergo faster growth and diverge in the limit of $a_c \longrightarrow 0$ and $T(\xi_m) \longrightarrow 0$.

An accelerated observer in a Rindler wedge has access only to the part of spacetime limited by its horizon.
As acceleration increases, the observer moves closer to the horizon.
Two accelerated quarks, in order to remain in causal connection with each other, must be closer to each other, otherwise the horizon will break the string between them.
The critical distance at which the string can exist decreases with increasing acceleration, see Fig.~\ref{fig:Rindler:1D}.
The shorter string is in a more ''compressed'' state.
The energy stored in the string per unit length (tension) is constant, but due to the curvature of spacetime and the horizon influence, the effective energy that must be expended to separate the quarks increases.
This manifests itself as an increase in the interquark potential, which can be seen in Fig.~\ref{fig:Rindler:1D}.

Hence, the characteristic temperature increases with the acceleration $a_c=a_0/\ell = 1/\R$.
This occurs either by increasing the dimensionless parameter $a_0$ at fixed curvature scale $\ell$, or by decreasing $\ell$ at fixed $a_0$, which is the same as the AdS horizon $\R$ decrease.
However, one can argue that the acceleration parameter $a_c$ is unphysical, so a certain value of this sets only the scale between time and space coordinates.
One can perform a transformation of the coordinates in the Rindler-AdS metric~\eqref{eq:rindler-ads:metric} and redefine the originally chosen acceleration value $a_c$ to another value or even get rid of the acceleration by the coordinate transformation $\tilde{x}_i = a_c x_i$.

\begin{figure}[!htb]\centering
	\begin{minipage}{0.4\linewidth}
		\includegraphics[width=1\linewidth]{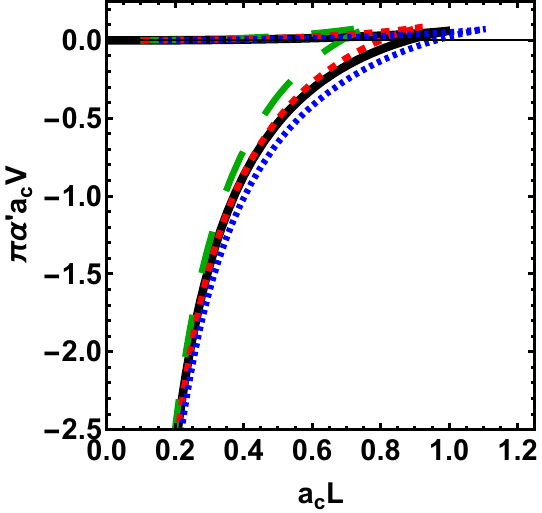}
    \end{minipage}\hspace*{0.05\linewidth}
	\begin{minipage}{0.3\linewidth}
		\includegraphics[width=1\linewidth]{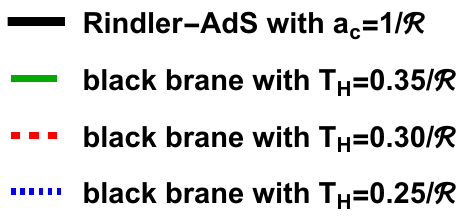}
	\end{minipage}
	\caption{
        The scaled onto acceleration potential, $\tilde{V} = a_c V$, as a function of the acceleration-scaled distance, $\tilde{L} = a_c L$.
        Results in the black brane background~\eqref{eq:brane:distance}-\eqref{eq:brane:potential:hypergeometric-view} are also shown at different Hawking temperatures, $\R \TH = \TH / a_c = 0.35, 0.30, 0.25$, depicted in the legend.
    }
	\label{fig:Rindler:aV}
\end{figure}

Indeed, the potential scaled onto acceleration value, $\tilde{V} = a_c V$, as a function of the acceleration-scaled distance, $\tilde{L} = a_c L$, shows no dependence on the certain acceleration value $a_c$, see Fig.~\ref{fig:Rindler:aV}.
Similar scaling is also observed in the lattice QCD calculations~\cite{Braguta:2026nfy, Braguta:2026kuz}.
In such terms the Hawking temperature is a universal constant $\tilde{T}_H = 1/2\pi$, which does not depend on the acceleration or AdS radius values.
Hence, one have the characteristic temperatures are also constants: $\tilde{T}_{c} = (\pi/3) \tilde{T}_H = 1/6$ and $\tilde{T}_s \approx 1/(2\pi \cdot 0.5163)$.

For a comparison we also show results obtained in Sec.~\ref{sec:brane} at several values of the Hawking temperatures.
As one can see, scaled results in the Rindler-AdS space are very close to the results in the black brane background at $\R \TH = \TH / a_c \approx 0.3$.

\section{Conclusions}\label{sec:summary}

In this paper, we have presented calculations of the separation distance and static potential energy for a heavy quark-\antiquark{} pair within the string theory for an arbitrary stationary background.
We established that for a simple U-shaped string with only radial dependence on the space string coordinate, $x_r'(\sigma) \neq 0$, the string is generally not symmetric with respect to the turning point, and the symmetry (at least for the EoM) restores only for backgrounds with a single constraint: $h_{pr} = G_{00} G_{pr} - G_{0p} G_{0r} = 0$.
Consequently, such asymmetric strings may probe a possibility of the parity violation in the quark-\antiquark{} interaction, taking into account that confident evidences of the violation should be extracted from other appropriate invariant quantities.

Nevertheless, we obtained general expressions for the distance~\eqref{eq:L-and-action:x_r:L} and potential~\eqref{eq:result:V:general} within a non-symmetric arbitrary string profile.
Then, we shown that for a simple string ansatz (only radial dependency $x_i'=0$, $x_r' \neq 0$ and symmetry with respect to the turning point, see~Fig.~\ref{fig:string-profile}) one can directly isolate a linear-in-distance term even for complicated non-diagonal metrics with the only one constraint ($h_{pr} = 0$), see~Eq.~\eqref{eq:result:V:special}.
It is a wide class of the backgrounds, including simple diagonal (or semi-diagonal with respect to $p$ and $r$ as for the AdS family solutions) cases, stationary metrics, special cases with $G_{pr}=\frac{G_{0p}G_{0r}}{G_{00}}$, where one can diagonalize this metric part by a simple time shift depending on $r$ and $p$, etc.
It also means that one can initially work with a simplified version of the Nambu--Goto action, expressed in terms of the first integral and the distance between string endpoints (quarks), see~\eqref{eq:result:action:hpr}.

As a first example, we apply our general formulas to the AdS/CFT at finite temperature formulated in terms of $N$ coincident D3-branes (the black brane), which is in large-$N$ limit dual to the finite-temperature $\mathcal{N}=4$ SYM plasma.
We obtained precise analytic expressions for the separation distance~\eqref{eq:brane:distance} and static potential~\eqref{eq:brane:potential:hypergeometric-view} in terms of the hypergeometric functions.
We noted that the potential may be expressed in terms of the separation and its derivative by temperature, see~Eq.~\eqref{eq:brane:potential:distance-view} or Eq.~\eqref{eq:brane:potential:distance-view:T} for an expression obtained completely in temperature terms.
We found values of the parameters and temperatures at the critical separation and screening length and analyze their dependence on the Hawking temperature.
Performing a small-temperature expansion, it was shown that the series coefficients are expressed in terms of the dimensionless product of the separation distance in conformal limit and Hawking temperature, $L_0 \TH$, and the "swallowtail" behavior of the potential and distance are caused by temperature corrections for the separation distance.

In the last part of the work, within the framework of the holographic duality between the Rindler-AdS space and the accelerated $\mathcal{N}=4$ SYM plasma, we find analytic expressions for the distance~\eqref{eq:result:L:rindler-ads} and static potential~\eqref{eq:result:V:rindler-ads:1},~\eqref{eq:result:V:rindler-ads:2} between quarks in terms of the elliptic integrals.
Analyzing their asymptotics, we observed that the high Hawking temperature (high acceleration or small curvature) limits of the Rindler-AdS results come to the conformal expressions obtained for a pure AdS.
We shown that the distance between quarks decreases, the static potential between them increases, and the characteristic temperatures, $T_c = (\pi/3) T_H = a_c/6$ and $T_s \approx a_c/ (2\pi \cdot 0.5163)$, increase with an acceleration, $a_c$.
However, we observe that an acceleration-scaled potential, $\tilde{V}=a_c V$, as a function of the acceleration-scaled distance, $\tilde{L}=a_c L$, does not depend on the certain value of the acceleration, $a_c$.
This result, reflecting self-similarity of the holographic setup, can be also obtained in the dimensionless metric after rescaling of the coordinates onto the acceleration, $\tilde{x}_i = a_c x_i$, for which one obtains universal values of the characteristic temperatures.

Finally, we want to note that the interquark potential obtained in the Rindler-AdS background has a similar dependence on the distance as in the case of the pure AdS~\cite{PhysRevLett.80.4859}, $D$-brane~\cite{BRANDHUBER199836} and $D$-instanton~\cite{Chen:2022obe} in the AdS space, the Schwarzschild-AdS and Kerr-AdS black holes~\cite{PhysRevD.107.106017}, and other cases~\cite{Avramis:2006em, White:2007tu, Dorn:2007zy, Chen:2023yug, Cheng:2025inr}.
Therefore, we can conclude that this is a common feature of the simple string ansatz~\eqref{eq:rindler-ads:configuration:1} in the AdS-like backgrounds.

\appendix
\numberwithin{equation}{section}
\makeatletter

\section{The Virasoro constraints}\label{sec:app:virasoro}

The Nambu--Goto action has invariance under reparametrization ($\tau, \sigma \longrightarrow \tilde{\tau}, \tilde{\sigma}$), but it will be lost after fixing of the gauge.
To restore this symmetry and eliminate unphysical degrees of freedom, one must impose the Virasoro constraints, which require the vanishing of the \worldsheet{} stress-energy tensor.

Let's start from the general Lagrangian density without any constraints on the background and string configuration:
\begin{align}
    \tilde{\L} = \sqrt{-(\dot{X}^\mu \dot{X}_\mu) (X'^\nu X_\nu') + (\dot{X}^\mu X_\mu')^2},
\end{align}
where we denoted $\dot{x} = \partial x / \partial \tau$ and $x' = \partial x / \partial \sigma$.
In the Polyakov approach, the Virasoro constraints can be found by nullifying of the Hilbert stress–energy tensor, but for the Nambu--Goto action the tensor is trivial because the induced metric $g_{ab} = G_{MN} \partial_a X^M \partial_b X^N$ is not an independent variable.
In such case, we must use the Noether's theorem, where from one finds the Canonical stress-energy tensor:
\begin{align}
    \Theta^a_b = \frac{\partial \tilde{\L}}{\partial (\partial_a X^\mu)} \partial_b X^\mu - \delta^a_b \tilde{\L}.  
\end{align}
The explicit view of the components is
\begin{align}
    \Theta^\tau_\tau = \frac{\partial \tilde{\L}}{\partial \dot{X}^\mu} \dot{X}^\mu - \tilde{\L} = \P^{(\tau)}_\mu \dot{X}^\mu - \tilde{\L}, \qquad & \qquad
    \Theta^\tau_\sigma = \frac{\partial \tilde{\L}}{\partial \dot{X}^\mu} X'^\mu = \P^{(\tau)}_\mu X'^\mu,
    \\
    \Theta^\sigma_\sigma = \frac{\partial \tilde{\L}}{\partial X'^\mu} X'^\mu - \tilde{\L} = \P^{(\sigma)}_\mu X'^\mu - \tilde{\L}, \qquad            & \qquad
    \Theta^\sigma_\tau = \frac{\partial \tilde{\L}}{\partial X'^\mu} \dot{X}^\mu = \P^{(\sigma)}_\mu \dot{X}^\mu.
\end{align}

Now, we come back to the stationary background~\eqref{eq:metric}, fix the gauge and nullify the components of $\Theta^a_b$:
\begin{align}\label{eq:virasoro-1}
    \Theta^\tau_\tau = \P^{(\tau)}_\mu \dot{X}^\mu - \tilde{\L} = \P^{(\tau)}_0 - \L = 0, \qquad   & \qquad
    \Theta^\tau_\sigma = \P^{(\tau)}_\mu X'^\mu = \P^{(\tau)}_i x'^i = 0,                                   \\
    \Theta^\sigma_\sigma = \P^{(\sigma)}_\mu X'^\mu - \tilde{\L} = \P_p + \P_i x'^i - \L = 0, \qquad & \qquad
    \Theta^\sigma_\tau = \P^{(\sigma)}_\mu \dot{X}^\mu = \P_0 = 0.
\end{align}
Here, $\P^{(\sigma)}_\mu \longrightarrow \P_\mu$, $\tilde{\L} \longrightarrow \L$, $X_i \longrightarrow x_i$ are our common notation in the paper.

Finally, we obtain the Virasoro constraints for our general string configuration in arbitrary stationary background:
\begin{align}\label{eq:virasoro-2}
    \P^{(\tau)}_0 = \L, \qquad                    & \qquad \P^{(\tau)}_i x'^i = 0, \\
    \P_p = - \P_i x'^i + \L = - \H = - \C, \qquad & \qquad \P_0 = 0,
\end{align}
where we used the expression for the (reduced) Hamiltonian density with the constant $\C$ fixed by the string turning point in the configuration, see~\eqref{eq:string-config},~\eqref{eq:hamiltonian}, and~\eqref{eq:C}.
As one can see, total Hamiltonian density is zero~\eqref{eq:virasoro-1}, but the reduced one~\eqref{eq:virasoro-2} is constant, which is more convenient quantity for calculations.
Hence, in any case of the string configuration like~\eqref{eq:string-config} in a stationary background one can simply use our general expressions obtained in Sec.~\ref{sec:general} without any problems with the Virasoro constraints.

\section{Analytic calculations in the AdS/CFT at finite temperature}\label{sec:app:brane}

Inserting~\eqref{eq:brane:basic} into~\eqref{eq:result:L:special} leads to
\begin{align}
    \nonumber
    L &= 2 |\C| \int_{U_m}^{\infty} dU \sqrt{\frac{-g_r}{g_p (g_p + C^2)}} \\
      \nonumber
      &= 2 \R^2 \sqrt{U_m^4 - U_T^4} \int_{U_m}^{\infty} dU \frac{1}{\sqrt{ (U_T^4 - U^4) (U_m^4 - U^4)}} \\
      \nonumber
      &= \sslash x = \frac{U_T}{U};\qquad dU = -U_T \frac{dx}{x^2};\qquad x_0 = \frac{U_T}{U_m} \sslash \\
      \nonumber
      &= 2 \frac{\R^2}{U_T^3} \sqrt{U_m^4 - U_T^4}\, \frac{x^3}{3}\, F_1\left(\frac{3}{4};\frac{1}{2},\frac{1}{2};\frac{7}{4};x^4,\frac{x^4}{x_0^4}\right) \Big\vert_{0}^{x_0} \\
      &=  \frac{\R^2}{U_m} \frac{\pi}{\sqrt{2} \beta} \sqrt{1-z}\, {_2F_1}\left(\frac{1}{2}, \frac{3}{4}; \frac{5}{4}; z\right)
\end{align}
where $F_1(a; b_1, b_1; c; x, y)$ is the Appell hypergeometric function of two variables, $_2F_1(a, b; c; x)$ is the Gauss hypergeometric function, and we shown a temporary variable transformation inside the double slashes.
Here we also introduced a convenient shorthand notation
\begin{align}
    z=U_T^4/U_m^4 
\end{align}
and coefficient
\begin{align}
    \beta = \frac{\Gamma\left(\frac{1}{4}\right)^2}{4 \sqrt{\pi}} = \frac{\pi}{2} \, _2F_1\left(\frac{1}{2}, \frac{1}{2}; 1; \frac{1}{2}\right) = K\left(\frac{1}{2}\right),
\end{align}
which is actually the elliptic integral of the first kind
\begin{align}
    K(m)=\int_{0}^{1} \dd x \frac{1}{\sqrt{(1-x^2)(1 - m x^2)}}   
\end{align}
with the elliptic modulus $k = \sqrt{m} = 1/\sqrt{2}$. 

The expression for the potential may be calculated in a similar way.
After a paste of~\eqref{eq:brane:basic} into~\eqref{eq:result:V:special:1} one obtains
\begin{align}
      \nonumber
    \pi\alpha'V &= \int_{U_m}^{\infty} dU \sqrt{-g_r} \left( \sqrt{\frac{g_p}{g_p + C^2}} - 1\right) -  \int_{U_m}^{U_T} dU \sqrt{-g_r} \\
      \nonumber
      \nonumber
    &= \int_{U_m}^{\infty} dU \left(\sqrt{\frac{ U_T^4 - U^4}{U_m^4 - U^4}} - 1\right) - U_m + U_T \\
      \nonumber
      &= \sslash x = \frac{U_T}{U};\qquad dU = -U_T \frac{dx}{x^2};\qquad x_0 = \frac{U_T}{U_m} \sslash \\
      \nonumber
      &= U_T \int_{0}^{x_0} \frac{dx}{x^2} \left(\sqrt{\frac{1 - x^4}{1 - \frac{x^4}{x_0^4}}} - 1\right) - U_m + U_T \\  
      \nonumber
      &= U_T - U_m + U_T \Bigg[ \frac{3 x^7}{7 x_0^4} F_1\left(\frac{7}{4};\frac{1}{2},\frac{1}{2};\frac{11}{4};x^4,\frac{x^4}{x_0^4}\right) - \frac{x^3}{3} \left(1 + \frac{2}{x_0^4}\right) F_1\left(\frac{3}{4};\frac{1}{2},\frac{1}{2};\frac{7}{4};x^4,\frac{x^4}{x_0^4}\right)\\
        \nonumber
      &\qquad\qquad\qquad\qquad+ \frac{1}{x} \left( 1 + \sqrt{1 - x^4} \sqrt{1 - \frac{x^4}{x_0^4}} \right) \Bigg] \Bigg\vert_{0}^{x_0} \\
      &= U_T + U_m (1 - z) \frac{\pi}{\sqrt{2} \beta} \left[\frac{1}{2}\, {_2F_1}\left(\frac{1}{2},\frac{3}{4};\frac{5}{4};z\right) - {_2F_1}\left(\frac{3}{4},\frac{3}{2};\frac{5}{4};z\right)\right].
\end{align}

One can recognize a linear-in-distance term in the potential due to the first hypergeometric function.
Moreover, it is possible to express the second hypergeometric function in terms of the first one, hence, express the potential in terms of the distance explicitly.
Using the following identity
\begin{align}
    {_2F_1}\left(\frac{3}{4},\frac{3}{2};\frac{5}{4};z\right) = \, {_2F_1}\left(\frac{1}{2},\frac{3}{4};\frac{5}{4};z\right) + 2 z \frac{d}{dz}\, {_2F_1}\left(\frac{1}{2},\frac{3}{4};\frac{5}{4};z\right),
\end{align}
we obtain the final view of the potential:
\begin{align}\nonumber
    \pi \alpha' V &= U_T - \frac{U_m^2}{\R^2} \sqrt{1-z}  \left[\frac{L}{2} + \frac{2 \sqrt{1-z}}{U_m} z \frac{d}{dz} \left( \frac{U_m L}{\sqrt{1-z}} \right)\right],\\
                  &= U_T - \frac{U_m^2}{\R^2} \sqrt{1-z} \left( \frac{1+z}{1-z} \frac{L}{2} + 2z \frac{\partial L}{\partial z} \right).
\end{align}
It is easy to check that an application of the second formula~\eqref{eq:result:V:special} leads to the same result as above.

\section{Analytic calculations in the Rindler-AdS background}\label{sec:app:rindler-ads}

First, inserting~\eqref{eq:rindler-ads:determinants}-\eqref{eq:rindler-ads:xi-min} into~\eqref{eq:result:L:special} leads to the following expression for the distance between string endpoints $L$:
\begin{align}
	\nonumber
	L & = 2 |\C| \int_{\xi_m}^{\infty} \frac{\dd \xi}{1+a_c^2\xi^2} \frac{1}{\sqrt{a_c^2\xi^2 (1 + a_c^2\xi^2) - \C^2}}                                              \\
	\nonumber
	  & = \sslash x=a_c\xi; x_m=a_c \xi_m \sslash = \frac{2 |\C|}{a_c} \int_{a_c\xi_m}^{\infty} \frac{\dd x}{1+x^2} \frac{1}{\sqrt{(x^2-x_m^2)(x^2+x_m^2+1)}}          \\
	\nonumber
	  & = \sslash y=\frac{x}{x_m}; n=-x_m^2; m=k^2=\frac{n}{1-n} \sslash = 2i\xi_m \int_{1}^{\infty} \frac{\dd y}{1-n y^2} \frac{1}{\sqrt{(1-y^2)(1 - m y^2)}} \\
	\label{eq:app:L:1}
	  & = 2i\xi_m \Pi\left(\arcsin y; n, m\right) \Big\vert_{y=1}^{y\longrightarrow\infty},
\end{align}
where for clarity we shown auxiliary variable redefinitions inside the double slashes.
Here we also introduced
\begin{align}
	\Pi(z; n, m) & = \int_{0}^{z}\dd x \frac{1}{(1-nx^2)\sqrt{(1-x^2)(1-m x^2)}}. \\
	\Pi(n, m)    & = \Pi(z=\pi/2; n, m),
\end{align}
which are the incomplete and complete elliptic integrals of the third kind, correspondingly, with the characteristic $n$ and elliptic modulus $k=\sqrt{m}$.

One can note that $n=-a_c^2\xi_m^2 < 0$ and $m=k^2=n/(1-n) < 0$, so the modulus $k$ is complex-valued.
While this representation is mathematically well-defined, the analytic continuation of elliptic integrals to complex parameters introduces branch ambiguities, so to avoid this problem we take the real part of the analytically continued expression.
Using the identity for the asymptotic behavior of the elliptic integral of the third kind with imaginary argument,
\begin{align}
	\nonumber
	\lim_{\alpha\longrightarrow\infty}\Pi\left(-i\alpha; n, m\right) & =i\frac{F(-\pi/2; 1-m) - n \Pi(-\pi/2; 1-n, 1-m)}{1-n} \\
	                                                                 & = -i\frac{n \Pi(1-n, 1-m) - K(1-m)}{1-n},
\end{align}
we can evaluate the upper limit explicitly.
Substituting this into our expression for the proper length yields to
\begin{align}\nonumber
	L & = 2\xi_m\, \Re \Big[\frac{n \Pi(1-n, 1-m) - K(1-m)}{1-n} + i\Pi(n, m)\Big] \\
        \label{eq:app:L:final}
      & = 2\xi_m\, \Re \Big[\tilde{\Pi}(n, m) - \frac{K(1-m)}{1-n} \Big],
\end{align}
where we introduced the following shorthand notation:
\begin{align}
	\tilde{\Pi}(n, m) & = m\,\Pi(1-n, 1-m) + i\,\Pi(n, m).
\end{align}

Similarly, one can calculate an analytic formula for the potential.
After direct substitution of~\eqref{eq:rindler-ads:determinants}-\eqref{eq:rindler-ads:xi-min} into~\eqref{eq:result:V:special:1} we obtain
\begin{align}
	\nonumber
	\pi\alpha'\, V & = \int_{\xi_m}^{\infty} \dd \xi\, \frac{a_c^2\xi^2}{\sqrt{\C^2 - a_c^2\xi^2 (1+a_c^2\xi^2)}} - \int_{0}^{\infty} \dd \xi \frac{a_c\xi}{\sqrt{1+a_c^2\xi^2}},
\end{align}
where the calculations of the second integral is straightforward:
\begin{align}\label{eq:app:V:int2:divergence}
	\int_{0}^{\infty} \dd \xi \frac{a_c\xi}{\sqrt{1+a_c^2\xi^2}} & = \sqrt{\frac{1}{a_c^2}+\xi^2}\Big\vert_{\xi=0}^{\xi\longrightarrow\infty}.
\end{align}
The first integral is calculated in a similar way as~\eqref{eq:app:L:1}, namely
\begin{align}
	\nonumber
	\int_{\xi_m}^{\infty} \dd \xi\, & \frac{a_c^2\xi^2}{\sqrt{\C^2 - a_c^2\xi^2 (1+a_c^2\xi^2)}}  = \sslash y=\xi/\xi_m, \quad n = -a_c^2\xi_m^2, \quad m=k^2 = \frac{n}{1-n} \sslash                                                                                     \\
	\nonumber
	& =i \sqrt{\frac{1}{a_c^2}+\xi_m^2} \left[ \int_{1}^{\infty} \dd y\, \frac{\sqrt{1-my^2}}{\sqrt{1-y^2}} - \int_{1}^{\infty} \dd y\,\frac{1}{\sqrt{(1-y^2)(1-my^2)}} \right] \\
	\nonumber
	                               & = i \sqrt{\frac{1}{a_c^2}+\xi_m^2} \left[ E(\arcsin y; m) - F(\arcsin y; m) \right] \Big\vert_{y=1}^{y\longrightarrow\infty},
\end{align}
where
\begin{align}
	F(z;m)  & = \int_{0}^{z} \dd x \frac{1}{\sqrt{(1-x^2)(1-m x^2)}},  \\
	E(z; m) & = \int_{0}^{z} \dd x \frac{\sqrt{1-m x^2}}{\sqrt{1-x^2}}
\end{align}
are the incomplete elliptic integrals of the first and second kind, correspondingly.
In turn, at the lower bound $y=1$ one come to the complete integrals of the first, $K(m)$, and second, $E(m)$, kinds:
\begin{align}
    K(m) &= F(z=\pi/2;m),\\
    E(m) &= E(z=\pi/2;m).
\end{align}

At the upper integration bound $y\longrightarrow\infty$ one can use the following limit for $F(z; m)$:
\begin{align}
	\lim_{y\longrightarrow\infty} F(\arcsin y; m) = -i\, K(1-m),
\end{align}
while for $E(z; m)$ one must explicitly perform the imaginary-argument transformation:
\begin{align}\label{eq:app:V:int1:divergence}
	E(i\, \phi; m) = i\, \Big(F(\psi; 1-m) - E(\psi; 1-m) + \tan \psi \sqrt{1-(1-m)\sin^2\psi}\Big).
\end{align}
At $y\longrightarrow\infty$ the two first terms transform to the complete integrals with minus sign.
The last term in~\eqref{eq:app:V:int1:divergence} diverges in the same way as the upper bound in~\eqref{eq:app:V:int2:divergence}, so we can get rid of them both.
Collecting~\eqref{eq:app:V:int2:divergence}-\eqref{eq:app:V:int1:divergence}, we come to the following formula for the potential:
\begin{align}\nonumber
	\pi \alpha' V & = \frac{1}{a_c} + \sqrt{\frac{1}{a^2}+\xi_m^2} \Re \Big[ i\,K(m) - E(1-m) - i\,E(m) \Big], \\
	              & = \frac{1}{a_c} + \sqrt{\frac{1}{a^2}+\xi_m^2} \Re \Big[ i\,K(m) - \tilde{E}(m) \Big],
\end{align}
where we also explicitly used only a real part of the expression as it was done for $L$ in~\eqref{eq:app:L:final} and defined common shorthand notation
\begin{align}
	\tilde{E}(m) & = E(1-m) + i\,E(m).
\end{align}

Similarly, starting from~\eqref{eq:result:V:special}, one can obtain the second form of the potential:
\begin{align}
	\pi \alpha' V & = \frac{1}{a_c} -\frac{\C L}{2} + \frac{\xi_m \C}{n}\, \Re \Big[\tilde{K}(m) - n\, \tilde{\Pi}(n, m) - \tilde{E}(m)\Big],
\end{align}
where
\begin{align}
	\tilde{K}(m) & = m\,K(1-m) + i\,K(m).
\end{align}

\begin{acknowledgments}
    We are grateful to V.~Braguta, E.~Kolomeitsev, and A.~Golubtsova for the valuable discussions.
    The work was supported by the grant of the Foundation for the Advancement of Theoretical Physics and Mathematics
"BASIS" No.~25-1-5-80-1.
\end{acknowledgments}

\bibliography{stationary-potential.bib}

@article{Voskre-mag,
  author = {Voskresensky, D. N. and Anisimov, N. Yu.},
  title = {Properties of a pion condensate in a magnetic field},
  journal = {Zh. Eksp. Teor. Fiz.},
  volume = {78},
  pages = {2845-2853},
  year = {1980},
  note = {[Sov. Phys. JETP \textbf{51}, 13 (1980)]}
}

@article{SIT-2009,
  author = {Skokov, V. and Illarionov, A. Y. and Toneev, V. D.},
  title = {Estimate of the magnetic field strength in heavy-ion collisions},
  journal = {International Journal of Modern Physics A},
  volume = {24},
  pages = {5925-5932},
  year = {2009},
  doi = {10.1142/S0217751X09047570}
}

@article{Bzdak:2011yy,
    author = "Bzdak, Adam and Skokov, Vladimir",
    title = "{Event-by-event fluctuations of magnetic and electric fields in heavy ion collisions}",
    eprint = "1111.1949",
    archivePrefix = "arXiv",
    primaryClass = "hep-ph",
    reportNumber = "BNL-96541-2011-JA, RBRC-927",
    doi = "10.1016/j.physletb.2012.02.065",
    journal = "Phys. Lett. B",
    volume = "710",
    pages = "171--174",
    year = "2012"
}

@article{Deng:2012pc,
    author = "Deng, Wei-Tian and Huang, Xu-Guang",
    title = "{Event-by-event generation of electromagnetic fields in heavy-ion collisions}",
    eprint = "1201.5108",
    archivePrefix = "arXiv",
    primaryClass = "nucl-th",
    doi = "10.1103/PhysRevC.85.044907",
    journal = "Phys. Rev. C",
    volume = "85",
    pages = "044907",
    year = "2012"
}

@article{Bloczynski:2012en,
    author = "Bloczynski, John and Huang, Xu-Guang and Zhang, Xilin and Liao, Jinfeng",
    title = "{Azimuthally fluctuating magnetic field and its impacts on observables in heavy-ion collisions}",
    eprint = "1209.6594",
    archivePrefix = "arXiv",
    primaryClass = "nucl-th",
    doi = "10.1016/j.physletb.2012.12.030",
    journal = "Phys. Lett. B",
    volume = "718",
    pages = "1529--1535",
    year = "2013"
}

@article{Gursoy:2014aka,
    author = "Gursoy, Umut and Kharzeev, Dmitri and Rajagopal, Krishna",
    title = "{Magnetohydrodynamics, charged currents and directed flow in heavy ion collisions}",
    eprint = "1401.3805",
    archivePrefix = "arXiv",
    primaryClass = "hep-ph",
    doi = "10.1103/PhysRevC.89.054905",
    journal = "Phys. Rev. C",
    volume = "89",
    number = "5",
    pages = "054905",
    year = "2014"
}

@article{Inghirami:2016iru,
    author = "Inghirami, Gabriele and Del Zanna, Luca and Beraudo, Andrea and Moghaddam, Mohsen Haddadi and Becattini, Francesco and Bleicher, Marcus",
    title = "{Numerical magneto-hydrodynamics for relativistic nuclear collisions}",
    eprint = "1609.03042",
    archivePrefix = "arXiv",
    primaryClass = "hep-ph",
    doi = "10.1140/epjc/s10052-016-4516-8",
    journal = "Eur. Phys. J. C",
    volume = "76",
    number = "12",
    pages = "659",
    year = "2016"
}

@article{Gursoy:2018yai,
    author = {G{\"u}rsoy, Umut and Kharzeev, Dmitri and Marcus, Eric and Rajagopal, Krishna and Shen, Chun},
    title = "{Charge-dependent Flow Induced by Magnetic and Electric Fields in Heavy Ion Collisions}",
    eprint = "1806.05288",
    archivePrefix = "arXiv",
    primaryClass = "hep-ph",
    doi = "10.1103/PhysRevC.98.055201",
    journal = "Phys. Rev. C",
    volume = "98",
    number = "5",
    pages = "055201",
    year = "2018"
}

@article{Galilo:2011nh,
    author = "Galilo, Bogdan V. and Nedelko, Sergei N.",
    title = "{Impact of the strong electromagnetic field on the QCD effective potential for homogeneous Abelian gluon field configurations}",
    eprint = "1107.4737",
    archivePrefix = "arXiv",
    primaryClass = "hep-ph",
    doi = "10.1103/PhysRevD.84.094017",
    journal = "Phys. Rev. D",
    volume = "84",
    pages = "094017",
    year = "2011"
}

@article{Bali:2012zg,
    author = "Bali, G. S. and Bruckmann, F. and Endrodi, G. and Fodor, Z. and Katz, S. D. and Schafer, A.",
    title = "{QCD quark condensate in external magnetic fields}",
    eprint = "1206.4205",
    archivePrefix = "arXiv",
    primaryClass = "hep-lat",
    doi = "10.1103/PhysRevD.86.071502",
    journal = "Phys. Rev. D",
    volume = "86",
    pages = "071502",
    year = "2012"
}

@article{Ayala:2014iba,
    author = "Ayala, Alejandro and Loewe, M. and Mizher, Ana Julia and Zamora, R.",
    title = "{Inverse magnetic catalysis for the chiral transition induced by thermo-magnetic effects on the coupling constant}",
    eprint = "1406.3885",
    archivePrefix = "arXiv",
    primaryClass = "hep-ph",
    doi = "10.1103/PhysRevD.90.036001",
    journal = "Phys. Rev. D",
    volume = "90",
    number = "3",
    pages = "036001",
    year = "2014"
}

@article{Mueller:2015fka,
    author = "Mueller, Niklas and Pawlowski, Jan M.",
    title = "{Magnetic catalysis and inverse magnetic catalysis in QCD}",
    eprint = "1502.08011",
    archivePrefix = "arXiv",
    primaryClass = "hep-ph",
    doi = "10.1103/PhysRevD.91.116010",
    journal = "Phys. Rev. D",
    volume = "91",
    number = "11",
    pages = "116010",
    year = "2015"
}

@article{Mao:2016fha,
    author = "Mao, Shijun",
    title = "{Inverse magnetic catalysis in Nambu{\textendash}Jona-Lasinio model beyond mean field}",
    eprint = "1602.06503",
    archivePrefix = "arXiv",
    primaryClass = "hep-ph",
    doi = "10.1016/j.physletb.2016.05.018",
    journal = "Phys. Lett. B",
    volume = "758",
    pages = "195--199",
    year = "2016"
}

@article{Toneev:2016tgj,
    author = "Toneev, V. and Rogachevsky, O. and Voronyuk, V.",
    title = "{Evidence for creation of strong electromagnetic fields in relativistic heavy-ion collisions}",
    eprint = "1604.06231",
    archivePrefix = "arXiv",
    primaryClass = "hep-ph",
    doi = "10.1140/epja/i2016-16264-1",
    journal = "Eur. Phys. J. A",
    volume = "52",
    number = "8",
    pages = "264",
    year = "2016"
}

@article{Farias:2016gmy,
    author = "Farias, R. L. S. and Timoteo, V. S. and Avancini, S. S. and Pinto, M. B. and Krein, G.",
    title = "{Thermo-magnetic effects in quark matter: Nambu--Jona-Lasinio model constrained by lattice QCD}",
    eprint = "1603.03847",
    archivePrefix = "arXiv",
    primaryClass = "hep-ph",
    doi = "10.1140/epja/i2017-12320-8",
    journal = "Eur. Phys. J. A",
    volume = "53",
    number = "5",
    pages = "101",
    year = "2017"
}

@article{Braguta:2019yci,
    author = "Braguta, V. V. and Chernodub, M. N. and Kotov, A. Yu and Molochkov, A. V. and Nikolaev, A. A.",
    title = "{Finite-density QCD transition in a magnetic background field}",
    eprint = "1909.09547",
    archivePrefix = "arXiv",
    primaryClass = "hep-lat",
    doi = "10.1103/PhysRevD.100.114503",
    journal = "Phys. Rev. D",
    volume = "100",
    number = "11",
    pages = "114503",
    year = "2019"
}

@article{Taya:2024wrm,
    author = "Taya, Hidetoshi and Nishimura, Toru and Ohnishi, Akira",
    title = "{Estimation of electric field in intermediate-energy heavy-ion collisions}",
    eprint = "2402.17136",
    archivePrefix = "arXiv",
    primaryClass = "hep-ph",
    reportNumber = "RIKEN-iTHEMS-Report-24",
    doi = "10.1103/PhysRevC.110.014901",
    journal = "Phys. Rev. C",
    volume = "110",
    number = "1",
    pages = "014901",
    year = "2024"
}

@article{Schwinger:1951nm,
    author = "Schwinger, Julian S.",
    editor = "Milton, K. A.",
    title = "{On gauge invariance and vacuum polarization}",
    doi = "10.1103/PhysRev.82.664",
    journal = "Phys. Rev.",
    volume = "82",
    pages = "664--679",
    year = "1951"
}

@article{Baur:2007zz,
    author = "Baur, Gerhard and Hencken, Kai and Trautmann, Dirk",
    title = "{Electron-Positron Pair Production in Relativistic Heavy Ion Collisions}",
    eprint = "0706.0654",
    archivePrefix = "arXiv",
    primaryClass = "nucl-th",
    doi = "10.1016/j.physrep.2007.09.002",
    journal = "Phys. Rept.",
    volume = "453",
    pages = "1--27",
    year = "2007"
}

@article{Gelis:2015kya,
    author = "Gelis, Francois and Tanji, Naoto",
    title = "{Schwinger mechanism revisited}",
    eprint = "1510.05451",
    archivePrefix = "arXiv",
    primaryClass = "hep-ph",
    doi = "10.1016/j.ppnp.2015.11.001",
    journal = "Prog. Part. Nucl. Phys.",
    volume = "87",
    pages = "1--49",
    year = "2016"
}

@article{PhysRevC.95.054902,
  title = {Global hyperon polarization at local thermodynamic equilibrium with vorticity, magnetic field, and feed-down},
  author = {Becattini, Francesco and Karpenko, Iurii and Lisa, Michael Annan and Upsal, Isaac and Voloshin, Sergei A.},
  journal = {Phys. Rev. C},
  volume = {95},
  issue = {5},
  pages = {054902},
  numpages = {13},
  year = {2017},
  publisher = {American Physical Society},
  doi = {10.1103/PhysRevC.95.054902},
  url = {https://link.aps.org/doi/10.1103/PhysRevC.95.054902}
}

@article{Guo:2019joy,
    author = "Guo, Yu and Shi, Shuzhe and Feng, Shengqin and Liao, Jinfeng",
    title = "{Magnetic Field Induced Polarization Difference between Hyperons and Anti-hyperons}",
    eprint = "1905.12613",
    archivePrefix = "arXiv",
    primaryClass = "nucl-th",
    doi = "10.1016/j.physletb.2019.134929",
    journal = "Phys. Lett. B",
    volume = "798",
    pages = "134929",
    year = "2019"
}

@article{Liu:2021nyg,
    author = "Liu, Yu-Chen and Huang, Xu-Guang",
    title = "{Spin polarization formula for Dirac fermions at local equilibrium}",
    eprint = "2109.15301",
    archivePrefix = "arXiv",
    primaryClass = "nucl-th",
    doi = "10.1007/s11433-022-1903-8",
    journal = "Sci. China Phys. Mech. Astron.",
    volume = "65",
    number = "7",
    pages = "272011",
    year = "2022"
}

@article{Xu:2022hql,
    author = "Xu, Kun and Lin, Fan and Huang, Anping and Huang, Mei",
    title = "{{\ensuremath{\Lambda}}/{\ensuremath{\Lambda}}{\textasciimacron} polarization and splitting induced by rotation and magnetic field}",
    eprint = "2205.02420",
    archivePrefix = "arXiv",
    primaryClass = "hep-ph",
    doi = "10.1103/PhysRevD.106.L071502",
    journal = "Phys. Rev. D",
    volume = "106",
    number = "7",
    pages = "L071502",
    year = "2022"
}

@article{Buzzegoli:2022qrr,
    author = "Buzzegoli, Matteo",
    title = "{Spin polarization induced by magnetic field and the relativistic Barnett effect}",
    eprint = "2211.04549",
    archivePrefix = "arXiv",
    primaryClass = "nucl-th",
    doi = "10.1016/j.nuclphysa.2023.122674",
    journal = "Nucl. Phys. A",
    volume = "1036",
    pages = "122674",
    year = "2023"
}

@article{Liu:2024hii,
    author = "Liu, Zhiwei and Bai, Yunfan and Zheng, Shiqi and Huang, Anping and Chen, Baoyi",
    title = "{Exploring spin polarization of heavy quarks in magnetic fields and hot medium}",
    eprint = "2404.02032",
    archivePrefix = "arXiv",
    primaryClass = "nucl-th",
    doi = "10.1103/PhysRevC.110.034910",
    journal = "Phys. Rev. C",
    volume = "110",
    number = "3",
    pages = "034910",
    year = "2024"
}

@article{Fukushima:2008xe,
    author = "Fukushima, Kenji and Kharzeev, Dmitri E. and Warringa, Harmen J.",
    title = "{The Chiral Magnetic Effect}",
    eprint = "0808.3382",
    archivePrefix = "arXiv",
    primaryClass = "hep-ph",
    doi = "10.1103/PhysRevD.78.074033",
    journal = "Phys. Rev. D",
    volume = "78",
    pages = "074033",
    year = "2008"
}

@article{Sadofyev:2010is,
    author = "Sadofyev, A. V. and Shevchenko, V. I. and Zakharov, V. I.",
    title = "{Notes on chiral hydrodynamics within effective theory approach}",
    eprint = "1012.1958",
    archivePrefix = "arXiv",
    primaryClass = "hep-th",
    doi = "10.1103/PhysRevD.83.105025",
    journal = "Phys. Rev. D",
    volume = "83",
    pages = "105025",
    year = "2011"
}

@article{Toneev:2012zx,
    author = "Toneev, V. D. and Konchakovski, V. P. and Voronyuk, V. and Bratkovskaya, E. L. and Cassing, W.",
    title = "{Event-by-event background in estimates of the chiral magnetic effect}",
    eprint = "1208.2519",
    archivePrefix = "arXiv",
    primaryClass = "nucl-th",
    doi = "10.1103/PhysRevC.86.064907",
    journal = "Phys. Rev. C",
    volume = "86",
    pages = "064907",
    year = "2012"
}

@article{Kharzeev:2013ffa,
	author = "Kharzeev, Dmitri E.",
	title = "{The Chiral Magnetic Effect and Anomaly-Induced Transport}",
	eprint = "1312.3348",
	archivePrefix = "arXiv",
	primaryClass = "hep-ph",
	doi = "10.1016/j.ppnp.2014.01.002",
	journal = "Prog. Part. Nucl. Phys.",
	volume = "75",
	pages = "133--151",
	year = "2014"
}

@article{PhysRevC.77.024906,
  title = {Angular momentum conservation in heavy ion collisions at very high energy},
  author = {Becattini, F. and Piccinini, F. and Rizzo, J.},
  journal = {Phys. Rev. C},
  volume = {77},
  issue = {2},
  pages = {024906},
  numpages = {8},
  year = {2008},
  publisher = {American Physical Society},
  doi = {10.1103/PhysRevC.77.024906},
  url = {https://link.aps.org/doi/10.1103/PhysRevC.77.024906}
}

@article{Gao:2007bc,
  author        = {Gao, Jian-Hua and Chen, Shou-Wan and Deng, Wei-Tian and Liang, Zuo-Tang and Wang, Qun and Wang, Xin-Nian},
  title         = {{Global quark polarization in non-central A+A collisions}},
  eprint        = {0710.2943},
  archiveprefix = {arXiv},
  primaryclass  = {nucl-th},
  reportnumber  = {LBNL-63515},
  doi           = {10.1103/PhysRevC.77.044902},
  journal       = {Phys. Rev. C},
  volume        = {77},
  pages         = {044902},
  year          = {2008},
  language      = {english}
}

@article{Jiang:2016woz,
  author        = {Jiang, Yin and Lin, Zi-Wei and Liao, Jinfeng},
  title         = {Rotating quark-gluon plasma in relativistic heavy ion collisions},
  eprint        = {1602.06580},
  archiveprefix = {arXiv},
  primaryclass  = {hep-ph},
  doi           = {10.1103/PhysRevC.94.044910},
  journal       = {Phys. Rev. C},
  volume        = {94},
  number        = {4},
  pages         = {044910},
  year          = {2016},
  note          = {[Erratum: Phys.Rev.C 95, 049904 (2017)]},
  language      = {english}
}

@article{Sass:2022ucj,
    author = {Sass, Nils and M{\"u}ller, Marco and Garcia-Montero, Oscar and Elfner, Hannah},
    collaboration = "SMASH",
    title = "{Global angular momentum generation in heavy-ion reactions within a hadronic transport approach}",
    eprint = "2212.14385",
    archivePrefix = "arXiv",
    primaryClass = "nucl-th",
    doi = "10.1103/PhysRevC.108.044903",
    journal = "Phys. Rev. C",
    volume = "108",
    number = "4",
    pages = "044903",
    year = "2023"
}

@article{Tsegelnik:2022eoz,
    author = "Tsegelnik, N. S. and Kolomeitsev, E. E. and Voronyuk, V.",
    title = "{Helicity and vorticity in heavy-ion collisions at energies available at the JINR Nuclotron-based Ion Collider facility}",
    eprint = "2211.09219",
    archivePrefix = "arXiv",
    primaryClass = "nucl-th",
    doi = "10.1103/PhysRevC.107.034906",
    journal = "Phys. Rev. C",
    volume = "107",
    number = "3",
    pages = "034906",
    year = "2023"
}

@article{PhysRevLett.47.229,
  title = {$\ensuremath{\Lambda}$ Production near Threshold in Central Nucleus-Nucleus Collisions},
  author = {Harris, J. W. and Sandoval, A. and Stock, R. and Stroebele, H. and Renfordt, R. E. and Geaga, J. V. and Pugh, H. G. and Schroeder, L. S. and Wolf, K. L. and Dacal, A.},
  journal = {Phys. Rev. Lett.},
  volume = {47},
  issue = {4},
  pages = {229--232},
  numpages = {0},
  year = {1981},
  publisher = {American Physical Society},
  doi = {10.1103/PhysRevLett.47.229},
  url = {https://link.aps.org/doi/10.1103/PhysRevLett.47.229}
}

@article{Anikina:1984,
  author   = {M. Anikina and others},
  title    = {Characteristics of $\Lambda$ and $K^0$ particles produced in central nucleus-nucleus Collisions at a 4.5\,GeV/$c$ momentum per incident nucleon},
  journal  = {Z. Phys. C},
  volume   = {25},
  pages    = {1-11},
  year     = {1984},
  language = {english}
}

@article{STAR:2007ccu,
    author = "Abelev, B. I. and others",
    collaboration = "STAR",
    title = "{Global polarization measurement in Au+Au collisions}",
    eprint = "0705.1691",
    archivePrefix = "arXiv",
    primaryClass = "nucl-ex",
    reportNumber = "STAR-05-11-2007",
    doi = "10.1103/PhysRevC.76.024915",
    journal = "Phys. Rev. C",
    volume = "76",
    pages = "024915",
    year = "2007",
    note = "[Erratum: Phys.Rev.C 95, 039906 (2017)]"
}

@article{STAR:2017ckg,
    author = "Adamczyk, L. and others",
    collaboration = "STAR",
    title = "{Global $\Lambda$ hyperon polarization in nuclear collisions: evidence for the most vortical fluid}",
    eprint = "1701.06657",
    archivePrefix = "arXiv",
    primaryClass = "nucl-ex",
    doi = "10.1038/nature23004",
    journal = "Nature",
    volume = "548",
    pages = "62--65",
    year = "2017"
}

@article{Petersen2017,
    title = {The fastest-rotating fluid},
    author = {Petersen, H.},
    journal = {Nature},
    year = {2017},
    volume = {548},
    pages = {34--35},
    doi = {10.1038/548034a},
    url = {https://doi.org/10.1038/548034a}
}

@article{STAR:2020xbm,
    author = "Adam, J. and others",
    collaboration = "STAR",
    title = "{Global Polarization of $\Xi$ and $\Omega$ Hyperons in Au+Au Collisions at $\sqrt {s_{NN}}$ = 200  GeV}",
    eprint = "2012.13601",
    archivePrefix = "arXiv",
    primaryClass = "nucl-ex",
    doi = "10.1103/PhysRevLett.126.162301",
    journal = "Phys. Rev. Lett.",
    volume = "126",
    number = "16",
    pages = "162301",
    year = "2021",
    note = "[Erratum: Phys.Rev.Lett. 131, 089901 (2023)]"
}

@article{HADES:2022enx,
    author = "Abou Yassine, R. and others",
    collaboration = "HADES",
    title = "{Measurement of global polarization of {\ensuremath{\Lambda}} hyperons in few-GeV heavy-ion collisions}",
    eprint = "2207.05160",
    archivePrefix = "arXiv",
    primaryClass = "nucl-ex",
    doi = "10.1016/j.physletb.2022.137506",
    journal = "Phys. Lett. B",
    volume = "835",
    pages = "137506",
    year = "2022"
}

@article{PhysRevC.33.1999,
  title = {${\ensuremath{\Lambda}}^{0}$ nonpolarization: Possible signature of quark matter},
  author = {Panagiotou, Apostolos D.},
  journal = {Phys. Rev. C},
  volume = {33},
  issue = {6},
  pages = {1999--2002},
  numpages = {0},
  year = {1986},
  publisher = {American Physical Society},
  doi = {10.1103/PhysRevC.33.1999},
  url = {https://link.aps.org/doi/10.1103/PhysRevC.33.1999}
}

@article{PhysRevLett.109.232301,
  title = {Chiral Anomaly and Local Polarization Effect from the Quantum Kinetic Approach},
  author = {Gao, Jian-Hua and Liang, Zuo-Tang and Pu, Shi and Wang, Qun and Wang, Xin-Nian},
  journal = {Phys. Rev. Lett.},
  volume = {109},
  issue = {23},
  pages = {232301},
  numpages = {5},
  year = {2012},
  publisher = {American Physical Society},
  doi = {10.1103/PhysRevLett.109.232301},
  url = {https://link.aps.org/doi/10.1103/PhysRevLett.109.232301}
}

@article{Becattini:2013fla,
  author        = {Becattini, F. and Chandra, V. and Del Zanna, L. and Grossi, E.},
  title         = {{Relativistic distribution function for particles with spin at local thermodynamical equilibrium}},
  eprint        = {1303.3431},
  archiveprefix = {arXiv},
  primaryclass  = {nucl-th},
  doi           = {10.1016/j.aop.2013.07.004},
  journal       = {Annals Phys.},
  volume        = {338},
  pages         = {32--49},
  year          = {2013},
  language      = {english}
}

@article{Sorin:2016smp,
    author = "Sorin, Alexander and Teryaev, Oleg",
    title = "{Axial anomaly and energy dependence of hyperon polarization in Heavy-Ion Collisions}",
    eprint = "1606.08398",
    archivePrefix = "arXiv",
    primaryClass = "nucl-th",
    doi = "10.1103/PhysRevC.95.011902",
    journal = "Phys. Rev. C",
    volume = "95",
    number = "1",
    pages = "011902",
    year = "2017"
}

@article{Teryaev:2017wlm,
    author = "Teryaev, Oleg V. and Zakharov, Valentin I.",
    title = "{From the chiral vortical effect to polarization of baryons: A model}",
    doi = "10.1103/PhysRevD.96.096023",
    journal = "Phys. Rev. D",
    volume = "96",
    number = "9",
    pages = "096023",
    year = "2017"
}

@article{Csernai:2018yok,
    author = "Csernai, L. P. and Kapusta, J. I. and Welle, T.",
    title = "{$\Lambda$ and $\bar{\Lambda}$ spin interaction with meson fields generated by the baryon current in high energy nuclear collisions}",
    eprint = "1807.11521",
    archivePrefix = "arXiv",
    primaryClass = "nucl-th",
    doi = "10.1103/PhysRevC.99.021901",
    journal = "Phys. Rev. C",
    volume = "99",
    number = "2",
    pages = "021901",
    year = "2019"
}

@article{BECATTINI2021136519,
    title = {Spin-thermal shear coupling in a relativistic fluid},
    journal = {Physics Letters B},
    volume = {820},
    pages = {136519},
    year = {2021},
    issn = {0370-2693},
    doi = {https://doi.org/10.1016/j.physletb.2021.136519},
    url = {https://www.sciencedirect.com/science/article/pii/S0370269321004597},
    author = {F. Becattini and M. Buzzegoli and A. Palermo},
}

@article{Liu:2021uhn,
    author = "Liu, Shuai Y. F. and Yin, Yi",
    title = "{Spin polarization induced by the hydrodynamic gradients}",
    eprint = "2103.09200",
    archivePrefix = "arXiv",
    primaryClass = "hep-ph",
    doi = "10.1007/JHEP07(2021)188",
    journal = "JHEP",
    volume = "07",
    pages = "188",
    year = "2021"
}

@article{Yi:2021ryh,
    author = "Yi, Cong and Pu, Shi and Yang, Di-Lun",
    title = "{Reexamination of local spin polarization beyond global equilibrium in relativistic heavy ion collisions}",
    eprint = "2106.00238",
    archivePrefix = "arXiv",
    primaryClass = "hep-ph",
    doi = "10.1103/PhysRevC.104.064901",
    journal = "Phys. Rev. C",
    volume = "104",
    number = "6",
    pages = "064901",
    year = "2021"
}

@article{PhysRevD.104.054043,
  title = {Spin Hall effect in heavy-ion collisions},
  author = {Liu, Shuai Y. F. and Yin, Yi},
  journal = {Phys. Rev. D},
  volume = {104},
  issue = {5},
  pages = {054043},
  numpages = {8},
  year = {2021},
  publisher = {American Physical Society},
  doi = {10.1103/PhysRevD.104.054043},
  url = {https://link.aps.org/doi/10.1103/PhysRevD.104.054043}
}

@article{Lin:2022tma,
    author = "Lin, Shu and Wang, Ziyue",
    title = "{Shear induced polarization: collisional contributions}",
    eprint = "2206.12573",
    archivePrefix = "arXiv",
    primaryClass = "hep-ph",
    doi = "10.1007/JHEP12(2022)030",
    journal = "JHEP",
    volume = "12",
    pages = "030",
    year = "2022"
}

@article{Unruh:1974bw,
    author = "Unruh, W. G.",
    title = "{Second quantization in the Kerr metric}",
    doi = "10.1103/PhysRevD.10.3194",
    journal = "Phys. Rev. D",
    volume = "10",
    pages = "3194--3205",
    year = "1974"
}

@article{Vilenkin:1979ui,
    author = "Vilenkin, A.",
    title = "{Macroscopic parity violating effects: neutrino fluxes from rotating black holes and in rotating thermal radiation}",
    doi = "10.1103/PhysRevD.20.1807",
    journal = "Phys. Rev. D",
    volume = "20",
    pages = "1807--1812",
    year = "1979"
}

@article{Letaw:1979wy,
    author = "Letaw, John R. and Pfautsch, Jonathan D.",
    title = "{The Quantized Scalar Field in Rotating Coordinates}",
    reportNumber = "ORO-3992-382",
    doi = "10.1103/PhysRevD.22.1345",
    journal = "Phys. Rev. D",
    volume = "22",
    pages = "1345",
    year = "1980"
}

@article{Iyer:1982ah,
    author = "Iyer, Bala R.",
    title = "{Dirac field theory in rotating coordinates}",
    doi = "10.1103/PhysRevD.26.1900",
    journal = "Phys. Rev. D",
    volume = "26",
    pages = "1900--1905",
    year = "1982"
}

@article{Davies:1996ks,
    author = "Davies, Paul C. W. and Dray, Tevian and Manogue, Corinne A.",
    title = "{The Rotating quantum vacuum}",
    eprint = "gr-qc/9601034",
    archivePrefix = "arXiv",
    reportNumber = "ADP-95-43-M-36-REV, ADP-95-43-M-36",
    doi = "10.1103/PhysRevD.53.4382",
    journal = "Phys. Rev. D",
    volume = "53",
    pages = "4382--4387",
    year = "1996"
}

@article{Duffy:2002ss,
    author = "Duffy, Gavin and Ottewill, Adrian C.",
    title = "{The Rotating quantum thermal distribution}",
    eprint = "hep-th/0211096",
    archivePrefix = "arXiv",
    doi = "10.1103/PhysRevD.67.044002",
    journal = "Phys. Rev. D",
    volume = "67",
    pages = "044002",
    year = "2003"
}

@article{Fischer:2003zz,
    author = "Fischer, Uwe R. and Baym, Gordon",
    title = "{Vortex states of rapidly rotating dilute Bose-Einstein condensates}",
    eprint = "cond-mat/0111443",
    archivePrefix = "arXiv",
    doi = "10.1103/PhysRevLett.90.140402",
    journal = "Phys. Rev. Lett.",
    volume = "90",
    pages = "140402",
    year = "2003"
}

@article{Ambrus:2014uqa,
    author = "Ambru\c{s}, Victor E. and Winstanley, Elizabeth",
    title = "{Rotating quantum states}",
    eprint = "1401.6388",
    archivePrefix = "arXiv",
    primaryClass = "hep-th",
    doi = "10.1016/j.physletb.2014.05.031",
    journal = "Phys. Lett. B",
    volume = "734",
    pages = "296--301",
    year = "2014"
}

@article{Ebihara:2016fwa,
    author = "Ebihara, Shu and Fukushima, Kenji and Mameda, Kazuya",
    title = "{Boundary effects and gapped dispersion in rotating fermionic matter}",
    eprint = "1608.00336",
    archivePrefix = "arXiv",
    primaryClass = "hep-ph",
    doi = "10.1016/j.physletb.2016.11.010",
    journal = "Phys. Lett. B",
    volume = "764",
    pages = "94--99",
    year = "2017"
}

@article{Chernodub:2020qah,
    author = "Chernodub, M. N.",
    title = "{Inhomogeneous confining-deconfining phases in rotating plasmas}",
    eprint = "2012.04924",
    archivePrefix = "arXiv",
    primaryClass = "hep-ph",
    doi = "10.1103/PhysRevD.103.054027",
    journal = "Phys. Rev. D",
    volume = "103",
    number = "5",
    pages = "054027",
    year = "2021"
}

@article{Chen:2022smf,
    author = "Chen, Shi and Fukushima, Kenji and Shimada, Yusuke",
    title = "{Perturbative Confinement in Thermal Yang-Mills Theories Induced by Imaginary Angular Velocity}",
    eprint = "2207.12665",
    archivePrefix = "arXiv",
    primaryClass = "hep-ph",
    doi = "10.1103/PhysRevLett.129.242002",
    journal = "Phys. Rev. Lett.",
    volume = "129",
    number = "24",
    pages = "242002",
    year = "2022"
}

@article{Ambrus:2023bid,
    author = "Ambru{\c{s}}, Victor E. and Chernodub, Maxim N.",
    title = "{Rigidly rotating scalar fields: Between real divergence and imaginary fractalization}",
    eprint = "2304.05998",
    archivePrefix = "arXiv",
    primaryClass = "hep-th",
    doi = "10.1103/PhysRevD.108.085016",
    journal = "Phys. Rev. D",
    volume = "108",
    number = "8",
    pages = "085016",
    year = "2023"
}

@article{Voskresensky:2023znr,
    author = "Voskresensky, D. N.",
    title = "{Pion-sigma meson vortices in rotating systems}",
    eprint = "2311.06804",
    archivePrefix = "arXiv",
    primaryClass = "nucl-th",
    doi = "10.1103/PhysRevD.109.034030",
    journal = "Phys. Rev. D",
    volume = "109",
    number = "3",
    pages = "034030",
    year = "2024"
}

@article{Chernodub:2025jwy,
    author = "Chernodub, Maxim",
    title = "{Negative moment of inertia of large-$N_c$ gluons on a ring}",
    eprint = "2506.06230",
    archivePrefix = "arXiv",
    primaryClass = "hep-th",
    month = "6",
    year = "2025"
}

@article{Unruh:1983ac,
    author = "Unruh, William G. and Weiss, Nathan",
    title = "{Acceleration Radiation in Interacting Field Theories}",
    reportNumber = "Print-83-1008 (BRITISH COLUMBIA)",
    doi = "10.1103/PhysRevD.29.1656",
    journal = "Phys. Rev. D",
    volume = "29",
    pages = "1656",
    year = "1984"
}

@article{Ohsaku:2004rv,
    author = "Ohsaku, Tadafumi",
    title = "{Dynamical chiral symmetry breaking and its restoration for an accelerated observer}",
    eprint = "hep-th/0407067",
    archivePrefix = "arXiv",
    doi = "10.1016/j.physletb.2004.08.019",
    journal = "Phys. Lett. B",
    volume = "599",
    pages = "102--110",
    year = "2004"
}

@article{Korsbakken:2004bv,
    author = "Korsbakken, Jan Ivar and Leinaas, Jon Magne",
    title = "{The Fulling-Unruh effect in general stationary accelerated frames}",
    eprint = "hep-th/0406080",
    archivePrefix = "arXiv",
    reportNumber = "OSLO-TP-1-04",
    doi = "10.1103/PhysRevD.70.084016",
    journal = "Phys. Rev. D",
    volume = "70",
    pages = "084016",
    year = "2004"
}

@article{Ebert:2006bh,
    author = "Ebert, D. and Zhukovsky, V. Ch.",
    title = "{Restoration of Dynamically Broken Chiral and Color Symmetries for an Accelerated Observer}",
    eprint = "hep-th/0612009",
    archivePrefix = "arXiv",
    reportNumber = "HU-EP-06-45",
    doi = "10.1016/j.physletb.2006.12.013",
    journal = "Phys. Lett. B",
    volume = "645",
    pages = "267--274",
    year = "2007"
}

@article{Castorina:2012yg,
    author = "Castorina, P. and Finocchiaro, M.",
    title = "{Symmetry Restoration By Acceleration}",
    eprint = "1207.3677",
    archivePrefix = "arXiv",
    primaryClass = "hep-th",
    doi = "10.4236/jmp.2012.311209",
    journal = "J. Mod. Phys.",
    volume = "3",
    pages = "1703",
    year = "2012"
}

@article{Takeuchi:2015nga,
    author = "Takeuchi, Shingo",
    title = "{Bose{\textendash}Einstein condensation in the Rindler space}",
    eprint = "1501.07471",
    archivePrefix = "arXiv",
    primaryClass = "hep-th",
    doi = "10.1016/j.physletb.2015.09.013",
    journal = "Phys. Lett. B",
    volume = "750",
    pages = "209--217",
    year = "2015"
}

@article{Benic:2015qha,
	author = "Benic, Sanjin and Fukushima, Kenji",
	title = "{Unruh effect and condensate in and out of an accelerated vacuum}",
	eprint = "1503.05790",
	archivePrefix = "arXiv",
	primaryClass = "hep-th",
	year = "2015"
}

@article{Becattini:2017ljh,
    author = "Becattini, F.",
    title = "{Thermodynamic equilibrium with acceleration and the Unruh effect}",
    eprint = "1712.08031",
    archivePrefix = "arXiv",
    primaryClass = "gr-qc",
    doi = "10.1103/PhysRevD.97.085013",
    journal = "Phys. Rev. D",
    volume = "97",
    number = "8",
    pages = "085013",
    year = "2018"
}

@article{Prokhorov:2018bql,
    author = "Prokhorov, George Y. and Teryaev, Oleg V. and Zakharov, Valentin I.",
    title = "{Effects of rotation and acceleration in the axial current: density operator vs Wigner function}",
    eprint = "1807.03584",
    archivePrefix = "arXiv",
    primaryClass = "hep-th",
    doi = "10.1007/JHEP02(2019)146",
    journal = "JHEP",
    volume = "02",
    pages = "146",
    year = "2019"
}

@article{Prokhorov:2019cik,
    author = "Prokhorov, George Y. and Teryaev, Oleg V. and Zakharov, Valentin I.",
    title = "{Unruh effect for fermions from the Zubarev density operator}",
    eprint = "1903.09697",
    archivePrefix = "arXiv",
    primaryClass = "hep-th",
    doi = "10.1103/PhysRevD.99.071901",
    journal = "Phys. Rev. D",
    volume = "99",
    number = "7",
    pages = "071901",
    year = "2019"
}

@article{Prokhorov:2018qhq,
    author = "Prokhorov, G. and Teryaev, O. and Zakharov, V.",
    title = "{Axial current in rotating and accelerating medium}",
    eprint = "1805.12029",
    archivePrefix = "arXiv",
    primaryClass = "hep-th",
    doi = "10.1103/PhysRevD.98.071901",
    journal = "Phys. Rev. D",
    volume = "98",
    number = "7",
    pages = "071901",
    year = "2018"
}

@article{Scardigli:2018jlm,
    author = "Scardigli, Fabio and Blasone, Massimo and Luciano, Gaetano and Casadio, Roberto",
    title = "{Modified Unruh effect from Generalized Uncertainty Principle}",
    eprint = "1804.05282",
    archivePrefix = "arXiv",
    primaryClass = "hep-th",
    doi = "10.1140/epjc/s10052-018-6209-y",
    journal = "Eur. Phys. J. C",
    volume = "78",
    number = "9",
    pages = "728",
    year = "2018"
}

@article{Palermo:2021hlf,
    author = "Palermo, Andrea and Buzzegoli, Matteo and Becattini, Francesco",
    title = "{Exact equilibrium distributions in statistical quantum field theory with rotation and acceleration: Dirac field}",
    eprint = "2106.08340",
    archivePrefix = "arXiv",
    primaryClass = "hep-th",
    doi = "10.1007/JHEP10(2021)077",
    journal = "JHEP",
    volume = "10",
    pages = "077",
    year = "2021"
}

@article{Akhmedov:2021agm,
    author = "Akhmedov, E. T. and Bazarov, K. V. and Diakonov, D. V.",
    title = "{Quantum fields in the future Rindler wedge}",
    eprint = "2106.01791",
    archivePrefix = "arXiv",
    primaryClass = "hep-th",
    doi = "10.1103/PhysRevD.104.085008",
    journal = "Phys. Rev. D",
    volume = "104",
    number = "8",
    pages = "085008",
    year = "2021"
}

@article{Chernodub:2024wis,
    author = "Chernodub, M. N. and Goy, V. A. and Molochkov, A. V. and Stepanov, D. V. and Pochinok, A. S.",
    title = "{Extreme Softening of QCD Phase Transition under Weak Acceleration: First-Principles Monte~Carlo Results for Gluon Plasma}",
    eprint = "2409.01847",
    archivePrefix = "arXiv",
    primaryClass = "hep-lat",
    doi = "10.1103/PhysRevLett.134.111904",
    journal = "Phys. Rev. Lett.",
    volume = "134",
    number = "11",
    pages = "111904",
    year = "2025"
}

@article{Chernodub:2025ovo,
	author = "Chernodub, Maxim N.",
	title = "{Acceleration as refrigeration: Acceleration-induced spontaneous symmetry breaking in thermal medium}",
	eprint = "2501.16129",
	archivePrefix = "arXiv",
	primaryClass = "hep-th",
	year = "2025"
}

@article{Bordag:2025zbt,
    author = "Bordag, M. and Voskresensky, D. N.",
    title = "{Charged scalar bosons under rotation and acceleration}",
    eprint = "2510.07438",
    archivePrefix = "arXiv",
    primaryClass = "hep-ph",
    month = "10",
    year = "2025"
}

@article{Prokhorov:2025vak,
    author = "Prokhorov, G. Yu. and Shohonov, D. A. and Teryaev, O. V. and Tsegelnik, N. S. and Zakharov, V. I.",
    title = "{Modeling of acceleration in heavy-ion collisions: Occurrence of temperature below the Unruh temperature}",
    eprint = "2502.10146",
    archivePrefix = "arXiv",
    primaryClass = "nucl-th",
    doi = "10.1103/7m17-n41m",
    journal = "Phys. Rev. C",
    volume = "112",
    number = "6",
    pages = "064907",
    year = "2025"
}

@article{Prokhorov:2026swu,
    author = "Prokhorov, G. Yu.",
    title = "{Entanglement Viscosity: from Unitarity to Irreversibility in Accelerated Frames}",
    eprint = "2601.02083",
    archivePrefix = "arXiv",
    primaryClass = "hep-th",
    month = "1",
    year = "2026"
}

@article{Braguta:2026nfy,
    author = "Braguta, Viktor and Goy, Vladimir and Dey, Jayanta and Roenko, Artem",
    title = "{Spatial confinement-deconfinement transition in accelerated gluodynamics within lattice simulation}",
    eprint = "2602.20970",
    archivePrefix = "arXiv",
    primaryClass = "hep-lat",
    month = "2",
    year = "2026"
}

@inproceedings{Braguta:2026kuz,
    author = "Braguta, Victor V. and Goy, Vladimir A. and Dey, Jayanta and Roenko, Artem A.",
    title = "{Spatially inhomogeneous confinement-deconfinement phase transition in accelerated gluodynamics}",
    booktitle = "{42th International Symposium on Lattice Field Theory}",
    eprint = "2603.01754",
    archivePrefix = "arXiv",
    primaryClass = "hep-lat",
    doi = "10.22323/1.518.0112",
    month = "3",
    year = "2026"
}

@article{Yamamoto:2013zwa,
	author = "Yamamoto, Arata and Hirono, Yuji",
	title = "{Lattice QCD in rotating frames}",
	eprint = "1303.6292",
	archivePrefix = "arXiv",
	primaryClass = "hep-lat",
	reportNumber = "RIKEN-QHP-80",
	doi = "10.1103/PhysRevLett.111.081601",
	journal = "Phys. Rev. Lett.",
	volume = "111",
	pages = "081601",
	year = "2013"
}

@article{Braguta:2020biu,
    author = "Braguta, V. V. and Kotov, A. Yu. and Kuznedelev, D. D. and Roenko, A. A.",
    title = "{Study of the Confinement/Deconfinement Phase Transition in Rotating Lattice SU(3) Gluodynamics}",
    doi = "10.31857/S1234567820130029",
    journal = "Pisma Zh. Eksp. Teor. Fiz.",
    volume = "112",
    number = "1",
    pages = "9--16",
    year = "2020"
}

@article{Braguta:2021jgn,
	author = "Braguta, V. V. and Kotov, A. Yu. and Kuznedelev, D. D. and Roenko, A. A.",
	title = "{Influence of relativistic rotation on the confinement-deconfinement transition in gluodynamics}",
	eprint = "2102.05084",
	archivePrefix = "arXiv",
	primaryClass = "hep-lat",
	doi = "10.1103/PhysRevD.103.094515",
	journal = "Phys. Rev. D",
	volume = "103",
	number = "9",
	pages = "094515",
	year = "2021"
}

@article{Braguta:2021ucr,
    author = "Braguta, Victor and Kotov, A. Yu. and Kuznedelev, Denis and Roenko, Artem",
    title = "{Lattice study of the confinement/deconfinement transition in rotating gluodynamics}",
    eprint = "2110.12302",
    archivePrefix = "arXiv",
    primaryClass = "hep-lat",
    doi = "10.22323/1.396.0125",
    journal = "PoS",
    volume = "LATTICE2021",
    pages = "125",
    year = "2022"
}

@article{Chernodub:2022veq,
    author = "Chernodub, M. N. and Goy, V. A. and Molochkov, A. V.",
    title = "{Inhomogeneity of a rotating gluon plasma and the Tolman-Ehrenfest law in imaginary time: Lattice results for fast imaginary rotation}",
    eprint = "2209.15534",
    archivePrefix = "arXiv",
    primaryClass = "hep-lat",
    doi = "10.1103/PhysRevD.107.114502",
    journal = "Phys. Rev. D",
    volume = "107",
    number = "11",
    pages = "114502",
    year = "2023"
}

@article{Braguta:2022str,
    author = "Braguta, V. V. and Kotov, Andrey and Roenko, Artem and Sychev, Dmitry",
    title = "{Thermal phase transitions in rotating QCD with dynamical quarks}",
    eprint = "2212.03224",
    archivePrefix = "arXiv",
    primaryClass = "hep-lat",
    doi = "10.22323/1.430.0190",
    journal = "PoS",
    volume = "LATTICE2022",
    pages = "190",
    year = "2023"
}

@article{Braguta:2023kwl,
    author = "Braguta, V. V. and Kudrov, I. E. and Roenko, A. A. and Sychev, D. A. and Chernodub, M. N.",
    title = "{Lattice Study of the Equation of State of a Rotating Gluon Plasma}",
    doi = "10.1134/S0021364023600830",
    journal = "JETP Lett.",
    volume = "117",
    number = "9",
    pages = "639--644",
    year = "2023"
}

@article{Braguta:2023aio,
    author = "Braguta, V. V. and Chernodub, M. N. and Kudrov, I. E. and Roenko, A. A. and Sychev, D. A.",
    title = "{Influence of Relativistic Rotation on QCD Properties}",
    doi = "10.1134/S1063778824010150",
    journal = "Phys. Atom. Nucl.",
    volume = "86",
    number = "6",
    pages = "1249--1255",
    year = "2023"
}

@article{Yang:2023vsw,
    author = "Yang, Ji-Chong and Huang, Xu-Guang",
    title = "{QCD on Rotating Lattice with Staggered Fermions}",
    eprint = "2307.05755",
    archivePrefix = "arXiv",
    primaryClass = "hep-lat",
    year = "2023"
}

@article{Braguta:2023yjn,
    author = "Braguta, Victor V. and Chernodub, Maxim N. and Roenko, Artem A. and Sychev, Dmitrii A.",
    title = "{Negative moment of inertia and rotational instability of gluon plasma}",
    eprint = "2303.03147",
    archivePrefix = "arXiv",
    primaryClass = "hep-lat",
    doi = "10.1016/j.physletb.2024.138604",
    journal = "Phys. Lett. B",
    volume = "852",
    pages = "138604",
    year = "2024"
}

@article{Braguta:2023qex,
    author = "Braguta, Victor V. and Chernodub, Maxim N. and Kudrov, Ilya E. and Roenko, Artem A. and Sychev, Dmitrii A.",
    title = "{Moment of inertia and supervortical temperature of gluon plasma}",
    eprint = "2311.03947",
    archivePrefix = "arXiv",
    primaryClass = "hep-lat",
    doi = "10.22323/1.453.0181",
    journal = "PoS",
    volume = "LATTICE2023",
    pages = "181",
    year = "2024"
}

@article{Braguta:2023tqz,
    author = "Braguta, Victor V. and Chernodub, Maxim N. and Kudrov, Ilya E. and Roenko, Artem A. and Sychev, Dmitrii A.",
    title = "{Negative Barnett effect, negative moment of inertia of the gluon plasma, and thermal evaporation of the chromomagnetic condensate}",
    eprint = "2310.16036",
    archivePrefix = "arXiv",
    primaryClass = "hep-ph",
    doi = "10.1103/PhysRevD.110.014511",
    journal = "Phys. Rev. D",
    volume = "110",
    number = "1",
    pages = "014511",
    year = "2024"
}

@article{Braguta:2023iyx,
    author = "Braguta, Victor V. and Chernodub, Maxim N. and Roenko, Artem A.",
    title = "{New mixed inhomogeneous phase in vortical gluon plasma: First-principle results from rotating SU(3) lattice gauge theory}",
    eprint = "2312.13994",
    archivePrefix = "arXiv",
    primaryClass = "hep-lat",
    doi = "10.1016/j.physletb.2024.138783",
    journal = "Phys. Lett. B",
    volume = "855",
    pages = "138783",
    year = "2024"
}

@article{Braguta:2025yud,
    author = "Braguta, V. and Chernodub, M. and Eremeev, E. and Kudrov, I. and Roenko, A. and Sychev, D.",
    title = "{On the angular momentum and free energy of rotating gluon plasma}",
    eprint = "2512.04070",
    archivePrefix = "arXiv",
    primaryClass = "hep-lat",
    month = "12",
    year = "2025"
}

@article{Jiang:2016wvv,
	author = "Jiang, Yin and Liao, Jinfeng",
	title = "{Pairing Phase Transitions of Matter under Rotation}",
	eprint = "1606.03808",
	archivePrefix = "arXiv",
	primaryClass = "hep-ph",
	doi = "10.1103/PhysRevLett.117.192302",
	journal = "Phys. Rev. Lett.",
	volume = "117",
	number = "19",
	pages = "192302",
	year = "2016"
}

@article{Chernodub:2016kxh,
    author = "Chernodub, M. N. and Gongyo, Shinya",
    title = "{Interacting fermions in rotation: chiral symmetry restoration, moment of inertia and thermodynamics}",
    eprint = "1611.02598",
    archivePrefix = "arXiv",
    primaryClass = "hep-th",
    reportNumber = "RIKEN-QHP-255",
    doi = "10.1007/JHEP01(2017)136",
    journal = "JHEP",
    volume = "01",
    pages = "136",
    year = "2017"
}

@article{Wang:2018sur,
	author = "Wang, Xinyang and Wei, Minghua and Li, Zhibin and Huang, Mei",
	title = "{Quark matter under rotation in the NJL model with vector interaction}",
	eprint = "1808.01931",
	archivePrefix = "arXiv",
	primaryClass = "hep-ph",
	doi = "10.1103/PhysRevD.99.016018",
	journal = "Phys. Rev. D",
	volume = "99",
	number = "1",
	pages = "016018",
	year = "2019"
}

@article{Zhang:2018ome,
    author = "Zhang, Hui and Hou, Defu and Liao, Jinfeng",
    title = "{Mesonic Condensation in Isospin Matter under Rotation}",
    eprint = "1812.11787",
    archivePrefix = "arXiv",
    primaryClass = "hep-ph",
    doi = "10.1088/1674-1137/abae4d",
    journal = "Chin. Phys. C",
    volume = "44",
    number = "11",
    pages = "111001",
    year = "2020"
}

@article{Fujimoto:2021xix,
	author = "Fujimoto, Yuki and Fukushima, Kenji and Hidaka, Yoshimasa",
	title = "{Deconfining Phase Boundary of Rapidly Rotating Hot and Dense Matter and Analysis of Moment of Inertia}",
	eprint = "2101.09173",
	archivePrefix = "arXiv",
	primaryClass = "hep-ph",
	reportNumber = "KEK-TH-2290, J-PARC-TH-0236, RIKEN-iTHEMS-Report-21",
	doi = "10.1016/j.physletb.2021.136184",
	journal = "Phys. Lett. B",
	volume = "816",
	pages = "136184",
	year = "2021"
}

@article{Chen:2023cjt,
    author = "Chen, Hao-Lei and Zhu, Zhi-Bin and Huang, Xu-Guang",
    title = "{Quark-meson model under rotation: A functional renormalization group study}",
    eprint = "2306.08362",
    archivePrefix = "arXiv",
    primaryClass = "hep-ph",
    doi = "10.1103/PhysRevD.108.054006",
    journal = "Phys. Rev. D",
    volume = "108",
    number = "5",
    pages = "054006",
    year = "2023"
}

@article{Sun:2023kuu,
    author = "Sun, Fei and Xu, Kun and Huang, Mei",
    title = "{Splitting of chiral and deconfinement phase transitions induced by rotation}",
    eprint = "2307.14402",
    archivePrefix = "arXiv",
    primaryClass = "hep-ph",
    doi = "10.1103/PhysRevD.108.096007",
    journal = "Phys. Rev. D",
    volume = "108",
    number = "9",
    pages = "096007",
    year = "2023"
}

@article{Sun:2024anu,
    author = "Sun, Fei and Shao, Jingdong and Wen, Rui and Xu, Kun and Huang, Mei",
    title = "{Chiral phase transition and spin alignment of vector mesons in the polarized-Polyakov-loop Nambu{\textendash}Jona-Lasinio model under rotation}",
    eprint = "2402.16595",
    archivePrefix = "arXiv",
    primaryClass = "hep-ph",
    doi = "10.1103/PhysRevD.109.116017",
    journal = "Phys. Rev. D",
    volume = "109",
    number = "11",
    pages = "116017",
    year = "2024"
}

@article{Kiefer:2025xdp,
    author = "Kiefer, Lutz and Dash, Ashutosh and Rischke, Dirk H.",
    title = "{Magnetization by Rotation: Spin and Chiral Condensates in the NJL Model}",
    eprint = "2509.18881",
    archivePrefix = "arXiv",
    primaryClass = "nucl-th",
    month = "9",
    year = "2025"
}

@article{Maldacena:1997re,
    author = "Maldacena, Juan Martin",
    title = "{The Large $N$ limit of superconformal field theories and supergravity}",
    eprint = "hep-th/9711200",
    archivePrefix = "arXiv",
    reportNumber = "HUTP-97-A097, HUTP-98-A097",
    doi = "10.4310/ATMP.1998.v2.n2.a1",
    journal = "Adv. Theor. Math. Phys.",
    volume = "2",
    pages = "231--252",
    year = "1998"
}

@article{DeWolfe:2013cua,
    author = "DeWolfe, Oliver and Gubser, Steven S. and Rosen, Christopher and Teaney, Derek",
    title = "{Heavy ions and string theory}",
    eprint = "1304.7794",
    archivePrefix = "arXiv",
    primaryClass = "hep-th",
    reportNumber = "COLO-HEP-579, PUPT-2446, CCTP-2013-06",
    doi = "10.1016/j.ppnp.2013.11.001",
    journal = "Prog. Part. Nucl. Phys.",
    volume = "75",
    pages = "86--132",
    year = "2014"
}

@book{Casalderrey-Solana:2011dxg,
    author = "Casalderrey-Solana, Jorge and Liu, Hong and Mateos, David and Rajagopal, Krishna and Achim Wiedemann, Urs",
    title = "{Gauge/String Duality, Hot QCD and Heavy Ion Collisions}",
    eprint = "1101.0618",
    archivePrefix = "arXiv",
    primaryClass = "hep-th",
    reportNumber = "CERN-PH-TH-2010-316, MIT-CTP-4198, ICCUB-10-202",
    doi = "10.1017/9781009403504",
    isbn = "978-1-009-40350-4, 978-1-009-40349-8, 978-1-009-40352-8, 978-1-139-13674-7",
    publisher = "Cambridge University Press",
    year = "2014"
}

@article{Arefeva:2014kyw,
    author = "Aref'eva, I. Ya",
    title = "{Holographic approach to quark{\textendash}gluon plasma in heavy ion collisions}",
    doi = "10.3367/UFNe.0184.201406a.0569",
    journal = "Phys. Usp.",
    volume = "57",
    pages = "527--555",
    year = "2014"
}

@article{PhysRevLett.80.4859,
  title = {Wilson Loops in Large $\mathit{N}$ Field Theories},
  author = {Maldacena, Juan},
  journal = {Phys. Rev. Lett.},
  volume = {80},
  issue = {22},
  pages = {4859--4862},
  numpages = {0},
  year = {1998},
  month = {Jun},
  publisher = {American Physical Society},
  doi = {10.1103/PhysRevLett.80.4859},
  url = {https://link.aps.org/doi/10.1103/PhysRevLett.80.4859}
}

@article{REY1998171,
title = {Wilson-Polyakov loop at finite temperature in large-N gauge theory and anti-de Sitter supergravity},
journal = {Nuclear Physics B},
volume = {527},
number = {1},
pages = {171-186},
year = {1998},
issn = {0550-3213},
doi = {https://doi.org/10.1016/S0550-3213(98)00471-4},
url = {https://www.sciencedirect.com/science/article/pii/S0550321398004714},
author = {Soo-Jong Rey and Stefan Theisen and Jung-Tay Yee},
}

@article{BRANDHUBER199836,
title = {Wilson loops in the large N limit at finite temperature},
journal = {Physics Letters B},
volume = {434},
number = {1},
pages = {36-40},
year = {1998},
issn = {0370-2693},
doi = {https://doi.org/10.1016/S0370-2693(98)00730-8},
url = {https://www.sciencedirect.com/science/article/pii/S0370269398007308},
author = {A Brandhuber and N Itzhaki and J Sonnenschein and S Yankielowicz},
}

@article{Brandhuber:1999jr,
    author = "Brandhuber, A. and Sfetsos, K.",
    title = "{Wilson loops from multicenter and rotating branes, mass gaps and phase structure in gauge theories}",
    eprint = "hep-th/9906201",
    archivePrefix = "arXiv",
    reportNumber = "CERN-TH-99-191",
    doi = "10.4310/ATMP.1999.v3.n4.a4",
    journal = "Adv. Theor. Math. Phys.",
    volume = "3",
    pages = "851--887",
    year = "1999"
}

@article{KINAR2000103,
title = {$Q\bar{Q}$ potential from strings in curved space-time -- classical results},
journal = {Nuclear Physics B},
volume = {566},
number = {1},
pages = {103-125},
year = {2000},
issn = {0550-3213},
doi = {https://doi.org/10.1016/S0550-3213(99)00652-5},
url = {https://www.sciencedirect.com/science/article/pii/S0550321399006525},
author = {Y. Kinar and E. Schreiber and J. Sonnenschein},
}

@article{Rey:1998ik,
    author = "Rey, Soo-Jong and Yee, Jung-Tay",
    title = "{Macroscopic strings as heavy quarks in large N gauge theory and anti-de Sitter supergravity}",
    eprint = "hep-th/9803001",
    archivePrefix = "arXiv",
    reportNumber = "SNUTP-98-016",
    doi = "10.1007/s100520100799",
    journal = "Eur. Phys. J. C",
    volume = "22",
    pages = "379--394",
    year = "2001"
}

@article{Boschi-Filho:2006hfm,
    author = "Boschi-Filho, Henrique and Braga, Nelson R. F. and Ferreira, Cristine N.",
    title = "{Heavy quark potential at finite temperature from gauge/string duality}",
    eprint = "hep-th/0607038",
    archivePrefix = "arXiv",
    doi = "10.1103/PhysRevD.74.086001",
    journal = "Phys. Rev. D",
    volume = "74",
    pages = "086001",
    year = "2006"
}

@article{Avramis:2006em,
    author = "Avramis, Spyros D. and Sfetsos, Konstadinos and Zoakos, Dimitrios",
    title = "{On the velocity and chemical-potential dependence of the heavy-quark interaction in N=4 SYM plasmas}",
    eprint = "hep-th/0609079",
    archivePrefix = "arXiv",
    doi = "10.1103/PhysRevD.75.025009",
    journal = "Phys. Rev. D",
    volume = "75",
    pages = "025009",
    year = "2007"
}

@article{Liu:2006he,
    author = "Liu, Hong and Rajagopal, Krishna and Wiedemann, Urs Achim",
    title = "{Wilson loops in heavy ion collisions and their calculation in AdS/CFT}",
    eprint = "hep-ph/0612168",
    archivePrefix = "arXiv",
    reportNumber = "MIT-CTP-3794, CERN-PH-TH-2006-257",
    doi = "10.1088/1126-6708/2007/03/066",
    journal = "JHEP",
    volume = "03",
    pages = "066",
    year = "2007"
}

@article{White:2007tu,
    author = "White, C D",
    title = "{The Cornell potential from general geometries in AdS / QCD}",
    eprint = "hep-ph/0701157",
    archivePrefix = "arXiv",
    reportNumber = "NIKHEF-2007-001",
    doi = "10.1016/j.physletb.2007.07.006",
    journal = "Phys. Lett. B",
    volume = "652",
    pages = "79--85",
    year = "2007"
}

@article{Dorn:2007zy,
    author = "Dorn, Harald and Ngo, Thanh Hai",
    title = "{On the internal space dependence of the static quark-antiquark potential in N = 4 SYM plasma wind}",
    eprint = "0707.2754",
    archivePrefix = "arXiv",
    primaryClass = "hep-th",
    reportNumber = "HU-EP-07-26",
    doi = "10.1016/j.physletb.2007.08.039",
    journal = "Phys. Lett. B",
    volume = "654",
    pages = "41--45",
    year = "2007"
}

@Article{Giataganas2012,
	author={Giataganas, Dimitrios},
	title={Probing strongly coupled anisotropic plasma},
	journal={Journal of High Energy Physics},
	year={2012},
	day={05},
	volume={2012},
	number={7},
	pages={31},
	issn={1029-8479},
	doi={10.1007/JHEP07(2012)031},
	url={https://doi.org/10.1007/JHEP07(2012)031}
}

@article{Drukker:2011za,
    author = "Drukker, Nadav and Forini, Valentina",
    title = "{Generalized quark-antiquark potential at weak and strong coupling}",
    eprint = "1105.5144",
    archivePrefix = "arXiv",
    primaryClass = "hep-th",
    reportNumber = "IMPERIAL-TP-2011-ND-02, NSF-KITP-11-073, AEI-2011-027",
    doi = "10.1007/JHEP06(2011)131",
    journal = "JHEP",
    volume = "06",
    pages = "131",
    year = "2011"
}

@article{Giataganas:2011nz,
    author = "Giataganas, Dimitrios and Irges, Nikos",
    title = "{Flavor Corrections in the Static Potential in Holographic QCD}",
    eprint = "1104.1623",
    archivePrefix = "arXiv",
    primaryClass = "hep-th",
    reportNumber = "WITS-CTP-69",
    doi = "10.1103/PhysRevD.85.046001",
    journal = "Phys. Rev. D",
    volume = "85",
    pages = "046001",
    year = "2012"
}

@article{Witten:1998zw,
    author = "Witten, Edward",
    editor = "Bergstrom, L. and Lindstrom, U.",
    title = "{Anti-de Sitter space, thermal phase transition, and confinement in gauge theories}",
    eprint = "hep-th/9803131",
    archivePrefix = "arXiv",
    reportNumber = "IASSNS-HEP-98-21",
    doi = "10.4310/ATMP.1998.v2.n3.a3",
    journal = "Adv. Theor. Math. Phys.",
    volume = "2",
    pages = "505--532",
    year = "1998"
}

@article{SUNDBORG2000349,
title = {The Hagedorn transition, deconfinement and N=4 SYM theory},
journal = {Nuclear Physics B},
volume = {573},
number = {1},
pages = {349-363},
year = {2000},
issn = {0550-3213},
doi = {https://doi.org/10.1016/S0550-3213(00)00044-4},
url = {https://www.sciencedirect.com/science/article/pii/S0550321300000444},
author = {Bo Sundborg},
}

@article{Polchinski:2000uf,
    author = "Polchinski, Joseph and Strassler, Matthew J.",
    title = "{The String dual of a confining four-dimensional gauge theory}",
    eprint = "hep-th/0003136",
    archivePrefix = "arXiv",
    reportNumber = "IAS-TH-00-18, NSF-ITP-00-16",
    month = "3",
    year = "2000"
}

@article{Cho:2002hq,
    author = "Cho, Y. M. and Neupane, Ishwaree P.",
    title = "{Anti-de Sitter black holes, thermal phase transition and holography in higher curvature gravity}",
    eprint = "hep-th/0202140",
    archivePrefix = "arXiv",
    reportNumber = "SNUTP-2002-04",
    doi = "10.1103/PhysRevD.66.024044",
    journal = "Phys. Rev. D",
    volume = "66",
    pages = "024044",
    year = "2002"
}

@article{Aharony:2003sx,
    author = "Aharony, Ofer and Marsano, Joseph and Minwalla, Shiraz and Papadodimas, Kyriakos and Van Raamsdonk, Mark",
    editor = "Doebner, H. D. and Dobrev, V. K.",
    title = "{The Hagedorn - deconfinement phase transition in weakly coupled large N gauge theories}",
    eprint = "hep-th/0310285",
    archivePrefix = "arXiv",
    reportNumber = "WIS-29-03-DPP",
    doi = "10.4310/ATMP.2004.v8.n4.a1",
    journal = "Adv. Theor. Math. Phys.",
    volume = "8",
    pages = "603--696",
    year = "2004"
}

@article{SPRADLIN2005199,
title = {The one-loop partition function of N=4 super-Yang–Mills theory on $\mathbb{R} \times S^3$},
journal = {Nuclear Physics B},
volume = {711},
number = {1},
pages = {199-230},
year = {2005},
issn = {0550-3213},
doi = {https://doi.org/10.1016/j.nuclphysb.2005.01.007},
url = {https://www.sciencedirect.com/science/article/pii/S0550321305000088},
author = {Marcus Spradlin and Anastasia Volovich},
}

@article{Karch:2006pv,
    author = "Karch, Andreas and Katz, Emanuel and Son, Dam T. and Stephanov, Mikhail A.",
    title = "{Linear confinement and AdS/QCD}",
    eprint = "hep-ph/0602229",
    archivePrefix = "arXiv",
    reportNumber = "BUHEP-06-02, INT-PUB-06-04",
    doi = "10.1103/PhysRevD.74.015005",
    journal = "Phys. Rev. D",
    volume = "74",
    pages = "015005",
    year = "2006"
}

@article{BallonBayona:2007vp,
    author = "Ballon Bayona, C. A. and Boschi-Filho, Henrique and Braga, Nelson R. F. and Pando Zayas, Leopoldo A.",
    title = "{On a Holographic Model for Confinement/Deconfinement}",
    eprint = "0705.1529",
    archivePrefix = "arXiv",
    primaryClass = "hep-th",
    reportNumber = "MCTP-07-17",
    doi = "10.1103/PhysRevD.77.046002",
    journal = "Phys. Rev. D",
    volume = "77",
    pages = "046002",
    year = "2008"
}

@article{Gursoy:2008za,
    author = "Gursoy, U. and Kiritsis, E. and Mazzanti, L. and Nitti, F.",
    title = "{Holography and Thermodynamics of 5D Dilaton-gravity}",
    eprint = "0812.0792",
    archivePrefix = "arXiv",
    primaryClass = "hep-th",
    reportNumber = "CPHT-RR088-1108, SPIN-08-57, ITP-UU-08-74",
    doi = "10.1088/1126-6708/2009/05/033",
    journal = "JHEP",
    volume = "05",
    pages = "033",
    year = "2009"
}

@article{Marolf:2013ioa,
    author = "Marolf, Donald and Rangamani, Mukund and Wiseman, Toby",
    title = "{Holographic thermal field theory on curved spacetimes}",
    eprint = "1312.0612",
    archivePrefix = "arXiv",
    primaryClass = "hep-th",
    reportNumber = "DCPT-13-51",
    doi = "10.1088/0264-9381/31/6/063001",
    journal = "Class. Quant. Grav.",
    volume = "31",
    pages = "063001",
    year = "2014"
}

@article{Brodsky:2014yha,
    author = "Brodsky, Stanley J. and de Teramond, Guy F. and Dosch, Hans Gunter and Erlich, Joshua",
    title = "{Light-Front Holographic QCD and Emerging Confinement}",
    eprint = "1407.8131",
    archivePrefix = "arXiv",
    primaryClass = "hep-ph",
    reportNumber = "SLAC-PUB-15972",
    doi = "10.1016/j.physrep.2015.05.001",
    journal = "Phys. Rept.",
    volume = "584",
    pages = "1--105",
    year = "2015"
}

@article{PhysRevLett.120.071605,
  title = {Hagedorn Temperature of ${\text{AdS}}_{5}/{\mathrm{CFT}}_{4}$ via Integrability},
  author = {Harmark, Troels and Wilhelm, Matthias},
  journal = {Phys. Rev. Lett.},
  volume = {120},
  issue = {7},
  pages = {071605},
  numpages = {6},
  year = {2018},
  month = {Feb},
  publisher = {American Physical Society},
  doi = {10.1103/PhysRevLett.120.071605},
  url = {https://link.aps.org/doi/10.1103/PhysRevLett.120.071605}
}

@article{Hawking:1999dp,
    author = "Hawking, S. W. and Reall, H. S.",
    title = "{Charged and rotating AdS black holes and their CFT duals}",
    eprint = "hep-th/9908109",
    archivePrefix = "arXiv",
    reportNumber = "DAMTP-R-99-108",
    doi = "10.1103/PhysRevD.61.024014",
    journal = "Phys. Rev. D",
    volume = "61",
    pages = "024014",
    year = "2000"
}

@article{Caceres:2006dj,
    author = "Caceres, Elena and Guijosa, Alberto",
    title = "{Drag force in charged N=4 SYM plasma}",
    eprint = "hep-th/0605235",
    archivePrefix = "arXiv",
    reportNumber = "UTTG-07-06, ICN-UNAM-06-05G",
    doi = "10.1088/1126-6708/2006/11/077",
    journal = "JHEP",
    volume = "11",
    pages = "077",
    year = "2006"
}

@article{DHoker:2009ixq,
    author = "D'Hoker, Eric and Kraus, Per",
    title = "{Charged Magnetic Brane Solutions in AdS (5) and the fate of the third law of thermodynamics}",
    eprint = "0911.4518",
    archivePrefix = "arXiv",
    primaryClass = "hep-th",
    doi = "10.1007/JHEP03(2010)095",
    journal = "JHEP",
    volume = "03",
    pages = "095",
    year = "2010"
}

@article{Ballon-Bayona:2013cta,
    author = "Ballon-Bayona, Alfonso",
    title = "{Holographic deconfinement transition in the presence of a magnetic field}",
    eprint = "1307.6498",
    archivePrefix = "arXiv",
    primaryClass = "hep-th",
    reportNumber = "ICTP-SAIFR-2013-009",
    doi = "10.1007/JHEP11(2013)168",
    journal = "JHEP",
    volume = "11",
    pages = "168",
    year = "2013"
}

@article{Dudal:2014jfa,
    author = "Dudal, David and Mertens, Thomas G.",
    title = "{Melting of charmonium in a magnetic field from an effective AdS/QCD model}",
    eprint = "1410.3297",
    archivePrefix = "arXiv",
    primaryClass = "hep-th",
    doi = "10.1103/PhysRevD.91.086002",
    journal = "Phys. Rev. D",
    volume = "91",
    pages = "086002",
    year = "2015"
}

@article{Evans:2016jzo,
    author = "Evans, Nick and Miller, Carlisson and Scott, Marc",
    title = "{Inverse Magnetic Catalysis in Bottom-Up Holographic QCD}",
    eprint = "1604.06307",
    archivePrefix = "arXiv",
    primaryClass = "hep-ph",
    doi = "10.1103/PhysRevD.94.074034",
    journal = "Phys. Rev. D",
    volume = "94",
    number = "7",
    pages = "074034",
    year = "2016"
}

@article{Li:2016gfn,
    author = "Li, Danning and Huang, Mei and Yang, Yi and Yuan, Pei-Hung",
    title = "{Inverse Magnetic Catalysis in the Soft-Wall Model of AdS/QCD}",
    eprint = "1610.04618",
    archivePrefix = "arXiv",
    primaryClass = "hep-th",
    doi = "10.1007/JHEP02(2017)030",
    journal = "JHEP",
    volume = "02",
    pages = "030",
    year = "2017"
}

@article{Arefeva:2018hyo,
    author = "Aref'eva, Irina and Rannu, Kristina",
    title = "{Holographic Anisotropic Background with Confinement-Deconfinement Phase Transition}",
    eprint = "1802.05652",
    archivePrefix = "arXiv",
    primaryClass = "hep-th",
    doi = "10.1007/JHEP05(2018)206",
    journal = "JHEP",
    volume = "05",
    pages = "206",
    year = "2018"
}

@article{Braga:2018zlu,
    author = "Braga, Nelson R. F. and Ferreira, Luiz F.",
    title = "{Heavy meson dissociation in a plasma with magnetic fields}",
    eprint = "1802.02084",
    archivePrefix = "arXiv",
    primaryClass = "hep-ph",
    doi = "10.1016/j.physletb.2018.06.053",
    journal = "Phys. Lett. B",
    volume = "783",
    pages = "186--192",
    year = "2018"
}

@article{Dudal:2018rki,
    author = "Dudal, David and Mertens, Thomas G.",
    title = "{Holographic estimate of heavy quark diffusion in a magnetic field}",
    eprint = "1802.02805",
    archivePrefix = "arXiv",
    primaryClass = "hep-th",
    doi = "10.1103/PhysRevD.97.054035",
    journal = "Phys. Rev. D",
    volume = "97",
    number = "5",
    pages = "054035",
    year = "2018"
}

@article{Cheng:2025inr,
    author = "Cheng, Wenxing and Zhang, Zi-qiang",
    title = "{Heavy quark potential in STU and AdS$_{5}${\textendash}Reissner{\textendash}Nordstrom backgrounds}",
    doi = "10.1140/epjc/s10052-025-14429-x",
    journal = "Eur. Phys. J. C",
    volume = "85",
    number = "7",
    pages = "774",
    year = "2025"
}

@article{Hawking:1998kw,
    author = "Hawking, S. W. and Hunter, C. J. and Taylor, Marika",
    title = "{Rotation and the AdS / CFT correspondence}",
    eprint = "hep-th/9811056",
    archivePrefix = "arXiv",
    doi = "10.1103/PhysRevD.59.064005",
    journal = "Phys. Rev. D",
    volume = "59",
    pages = "064005",
    year = "1999"
}

@article{Gibbons:2004ai,
    author = "Gibbons, G. W. and Perry, M. J. and Pope, C. N.",
    title = "{The First law of thermodynamics for Kerr-anti-de Sitter black holes}",
    eprint = "hep-th/0408217",
    archivePrefix = "arXiv",
    reportNumber = "DAMTP-2004-87, MIFP-04-17",
    doi = "10.1088/0264-9381/22/9/002",
    journal = "Class. Quant. Grav.",
    volume = "22",
    pages = "1503--1526",
    year = "2005"
}

@article{Bhattacharyya:2007vs,
    author = "Bhattacharyya, Sayantani and Lahiri, Subhaneil and Loganayagam, R. and Minwalla, Shiraz",
    title = "{Large rotating AdS black holes from fluid mechanics}",
    eprint = "0708.1770",
    archivePrefix = "arXiv",
    primaryClass = "hep-th",
    doi = "10.1088/1126-6708/2008/09/054",
    journal = "JHEP",
    volume = "09",
    pages = "054",
    year = "2008"
}

@article{NataAtmaja:2010hd,
    author = "Nata Atmaja, A. and Schalm, K.",
    title = "{Anisotropic Drag Force from 4D Kerr-AdS Black Holes}",
    eprint = "1012.3800",
    archivePrefix = "arXiv",
    primaryClass = "hep-th",
    reportNumber = "ITFA-2010-XXX, ITFA-2010-24",
    doi = "10.1007/JHEP04(2011)070",
    journal = "JHEP",
    volume = "04",
    pages = "070",
    year = "2011"
}

@article{Mcinnes:2018wzw,
    author = "Mcinnes, Brett",
    title = "{Applied holography of the AdS$_5${\textendash}Kerr space{\textendash}time}",
    eprint = "1803.02528",
    archivePrefix = "arXiv",
    primaryClass = "hep-ph",
    doi = "10.1142/S0217751X19501380",
    journal = "Int. J. Mod. Phys. A",
    volume = "34",
    number = "24",
    pages = "1950138",
    year = "2019"
}

@article{Arefeva:2020jvo,
	author = "Aref'eva, Irina Ya. and Golubtsova, Anastasia A. and Gourgoulhon, Eric",
	title = "{Holographic drag force in 5d Kerr-AdS black hole}",
	eprint = "2004.12984",
	archivePrefix = "arXiv",
	primaryClass = "hep-th",
	doi = "10.1007/JHEP04(2021)169",
	journal = "JHEP",
	volume = "04",
	pages = "169",
	year = "2021"
}

@article{Golubtsova:2021agl,
    author = "Golubtsova, Anastasia A. and Gourgoulhon, Eric and Usova, Marina K.",
    title = "{Heavy quarks in rotating plasma via holography}",
    eprint = "2107.11672",
    archivePrefix = "arXiv",
    primaryClass = "hep-th",
    doi = "10.1016/j.nuclphysb.2022.115786",
    journal = "Nucl. Phys. B",
    volume = "979",
    pages = "115786",
    year = "2022"
}

@article{Braga:2022yfe,
	author = "Braga, Nelson R. F. and Faulhaber, Luiz F. and Junqueira, Octavio C.",
	title = "{Confinement-deconfinement temperature for a rotating quark-gluon plasma}",
	eprint = "2201.05581",
	archivePrefix = "arXiv",
	primaryClass = "hep-th",
	doi = "10.1103/PhysRevD.105.106003",
	journal = "Phys. Rev. D",
	volume = "105",
	number = "10",
	pages = "106003",
	year = "2022"
}

@article{Chen:2022mhf,
    author = "Chen, Yidian and Li, Danning and Huang, Mei",
    title = "{Inhomogeneous chiral condensation under rotation in the holographic QCD}",
    eprint = "2208.05668",
    archivePrefix = "arXiv",
    primaryClass = "hep-ph",
    doi = "10.1103/PhysRevD.106.106002",
    journal = "Phys. Rev. D",
    volume = "106",
    number = "10",
    pages = "106002",
    year = "2022"
}

@article{McInnes:2022xxt,
    author = "McInnes, Brett",
    title = "{Why is black hole entropy affected by rotation?}",
    eprint = "2210.11751",
    archivePrefix = "arXiv",
    primaryClass = "gr-qc",
    doi = "10.1007/JHEP02(2023)072",
    journal = "JHEP",
    volume = "02",
    pages = "072",
    year = "2023"
}

@article{Yadav:2022qcl,
    author = "Yadav, Gopal",
    title = "{Deconfinement temperature of rotating QGP at intermediate coupling from M-theory}",
    eprint = "2203.11959",
    archivePrefix = "arXiv",
    primaryClass = "hep-th",
    doi = "10.1016/j.physletb.2023.137925",
    journal = "Phys. Lett. B",
    volume = "841",
    pages = "137925",
    year = "2023"
}

@article{Braga:2023qee,
    author = "Braga, Nelson R. F. and Ferreira, Luiz F. and Junqueira, Octavio C.",
    title = "{Configuration entropy of a rotating quark-gluon plasma from holography}",
    eprint = "2301.01322",
    archivePrefix = "arXiv",
    primaryClass = "hep-th",
    doi = "10.1016/j.physletb.2023.138265",
    journal = "Phys. Lett. B",
    volume = "847",
    pages = "138265",
    year = "2023"
}

@article{Braga:2023fac,
    author = "Braga, Nelson R. F. and Ferreira, Yan F.",
    title = "{Bottomonium dissociation in a rotating plasma}",
    eprint = "2309.11643",
    archivePrefix = "arXiv",
    primaryClass = "hep-ph",
    doi = "10.1103/PhysRevD.108.094017",
    journal = "Phys. Rev. D",
    volume = "108",
    number = "9",
    pages = "094017",
    year = "2023"
}

@article{PhysRevD.107.106017,
	title = {Probing the holographic model of $\mathcal{N}=4$ SYM rotating quark-gluon plasma},
	author = {Golubtsova, Anastasia and Tsegelnik, Nikita},
	journal = {Phys. Rev. D},
	volume = {107},
	issue = {10},
	pages = {106017},
	numpages = {19},
	year = {2023},
	publisher = {American Physical Society},
	doi = {10.1103/PhysRevD.107.106017},
	url = {https://link.aps.org/doi/10.1103/PhysRevD.107.106017}
}

@article{Chen:2022obe,
    author = "Chen, Jun-Xia and Hou, De-Fu",
    title = "{Heavy quark potential and jet quenching parameter in a rotating D-instanton background}",
    eprint = "2202.00888",
    archivePrefix = "arXiv",
    primaryClass = "hep-ph",
    doi = "10.1140/epjc/s10052-024-12708-7",
    journal = "Eur. Phys. J. C",
    volume = "84",
    number = "4",
    pages = "447",
    year = "2024"
}

@article{Chen:2023yug,
    author = "Chen, Jun-Xia and Hou, De-Fu and Ren, Hai-Cang",
    title = "{Drag force and heavy quark potential in a rotating background}",
    eprint = "2308.08126",
    archivePrefix = "arXiv",
    primaryClass = "hep-ph",
    doi = "10.1007/JHEP03(2024)171",
    journal = "JHEP",
    volume = "03",
    pages = "171",
    year = "2024"
}

@article{Braga:2023qej,
    author = "Braga, Nelson R. F. and Junqueira, Octavio C.",
    title = "{Inhomogeneity of a rotating quark-gluon plasma from holography}",
    eprint = "2306.08653",
    archivePrefix = "arXiv",
    primaryClass = "hep-th",
    doi = "10.1016/j.physletb.2023.138330",
    journal = "Phys. Lett. B",
    volume = "848",
    pages = "138330",
    year = "2024"
}

@article{Cai:2023cjl,
    author = "Cai, Yi-ze and Zhang, Zi-qiang",
    title = "{Holographic Schwinger effect in spinning black hole backgrounds}",
    eprint = "2310.13865",
    archivePrefix = "arXiv",
    primaryClass = "hep-ph",
    doi = "10.1088/1674-1137/ad061f",
    journal = "Chin. Phys. C",
    volume = "48",
    number = "1",
    pages = "015102",
    year = "2024"
}

@article{Chen:2024edy,
    author = "Chen, Jun-Xia and Wang, Sheng and Hou, Defu and Ren, Hai-Cang",
    title = "{String tension and Polyakov loop in a rotating background}",
    eprint = "2410.04763",
    archivePrefix = "arXiv",
    primaryClass = "hep-ph",
    doi = "10.1103/PhysRevD.111.026020",
    journal = "Phys. Rev. D",
    volume = "111",
    number = "2",
    pages = "026020",
    year = "2025"
}

@article{Ferreira:2025iqe,
    author = "Ferreira, Luiz F.",
    title = "{Dispersion relations and pole-skipping in a holographic charmonium model with rotating plasma}",
    eprint = "2510.02647",
    archivePrefix = "arXiv",
    primaryClass = "hep-ph",
    doi = "10.1103/32sr-nyrv",
    journal = "Phys. Rev. D",
    volume = "112",
    number = "7",
    pages = "074043",
    year = "2025"
}

@article{Zhu:2025bom,
    author = "Zhu, Zhou-Run and Wang, Sheng and Tian, Man-Li and Hou, Defu",
    title = "{Running coupling constant and jet quenching parameter in the spinning background from holography}",
    eprint = "2512.13076",
    archivePrefix = "arXiv",
    primaryClass = "hep-ph",
    month = "12",
    year = "2025"
}

@article{Braga:2025eiz,
    author = "Braga, Nelson R. F. and Junqueira, Octavio C.",
    title = "{Holographic QCD phase diagram for a rotating plasma in the Hawking-Page approach}",
    eprint = "2501.16446",
    archivePrefix = "arXiv",
    primaryClass = "hep-th",
    doi = "10.1016/j.physletb.2025.139669",
    journal = "Phys. Lett. B",
    volume = "868",
    pages = "139669",
    year = "2025"
}

@article{McInnes:2014haa,
    author = "McInnes, Brett",
    title = "{Angular Momentum in QGP Holography}",
    eprint = "1403.3258",
    archivePrefix = "arXiv",
    primaryClass = "hep-th",
    doi = "10.1016/j.nuclphysb.2014.08.011",
    journal = "Nucl. Phys. B",
    volume = "887",
    pages = "246--264",
    year = "2014"
}

@article{McInnes:2016dwk,
    author = "McInnes, Brett",
    title = "{A rotation/magnetism analogy for the quark{\textendash}gluon plasma}",
    eprint = "1604.03669",
    archivePrefix = "arXiv",
    primaryClass = "hep-th",
    doi = "10.1016/j.nuclphysb.2016.08.001",
    journal = "Nucl. Phys. B",
    volume = "911",
    pages = "173--190",
    year = "2016"
}

@article{Chen:2020ath,
	author = "Chen, Xun and Zhang, Lin and Li, Danning and Hou, Defu and Huang, Mei",
	title = "{Gluodynamics and deconfinement phase transition under rotation from holography}",
	eprint = "2010.14478",
	archivePrefix = "arXiv",
	primaryClass = "hep-ph",
	doi = "10.1007/JHEP07(2021)132",
	journal = "JHEP",
	volume = "07",
	pages = "132",
	year = "2021"
}

@article{Hou:2021own,
    author = "Hou, Defu and Atashi, Mahdi and Bitaghsir Fadafan, Kazem and Zhang, Zi-qiang",
    title = "{Holographic energy loss of a rotating heavy quark at finite chemical potential}",
    doi = "10.1016/j.physletb.2021.136279",
    journal = "Phys. Lett. B",
    volume = "817",
    pages = "136279",
    year = "2021"
}

@article{Mamani:2022qnf,
    author = "Mamani, Luis A. H. and Hou, Defu and Braga, Nelson R. F.",
    title = "{Melting of heavy vector mesons and quasinormal modes in a finite density plasma from holography}",
    eprint = "2204.08068",
    archivePrefix = "arXiv",
    primaryClass = "hep-ph",
    doi = "10.1103/PhysRevD.105.126020",
    journal = "Phys. Rev. D",
    volume = "105",
    number = "12",
    pages = "126020",
    year = "2022"
}

@article{Zhao:2022uxc,
    author = "Zhao, Yan-Qing and He, Song and Hou, Defu and Li, Li and Li, Zhibin",
    title = "{Phase diagram of holographic thermal dense QCD matter with rotation}",
    eprint = "2212.14662",
    archivePrefix = "arXiv",
    primaryClass = "hep-ph",
    doi = "10.1007/JHEP04(2023)115",
    journal = "JHEP",
    volume = "04",
    pages = "115",
    year = "2023"
}

@article{Wang:2024szr,
    author = "Wang, Jia-Hao and Feng, Sheng-Qin",
    title = "{Rotation effect on the deconfinement phase transition in holographic QCD}",
    eprint = "2403.01814",
    archivePrefix = "arXiv",
    primaryClass = "hep-ph",
    doi = "10.1103/PhysRevD.109.066019",
    journal = "Phys. Rev. D",
    volume = "109",
    number = "6",
    pages = "066019",
    year = "2024"
}

@article{Chen:2024jet,
    author = "Chen, Yidian and Chen, Xun and Li, Danning and Huang, Mei",
    title = "{Deconfinement and chiral restoration phase transition under rotation from holography in an anisotropic gravitational background}",
    eprint = "2405.06386",
    archivePrefix = "arXiv",
    primaryClass = "hep-ph",
    doi = "10.1103/PhysRevD.111.046006",
    journal = "Phys. Rev. D",
    volume = "111",
    number = "4",
    pages = "046006",
    year = "2025"
}

@article{Ahmed:2025bwi,
    author = "Ahmed, Hiwa A. and Chen, Yidian and Huang, Mei",
    title = "{Gluon polarization contribution to the spin alignment of vector mesons from holography}",
    eprint = "2501.13401",
    archivePrefix = "arXiv",
    primaryClass = "hep-ph",
    doi = "10.1103/PhysRevD.111.086006",
    journal = "Phys. Rev. D",
    volume = "111",
    number = "8",
    pages = "086006",
    year = "2025"
}

@article{Deser:1997ri,
	author = "Deser, Stanley and Levin, Orit",
	title = "{Accelerated detectors and temperature in (anti)-de Sitter spaces}",
	eprint = "gr-qc/9706018",
	archivePrefix = "arXiv",
	reportNumber = "BRX-TH-415",
	doi = "10.1088/0264-9381/14/9/003",
	journal = "Class. Quant. Grav.",
	volume = "14",
	pages = "L163--L168",
	year = "1997"
}

@article{Hamilton:2006az,
    author = "Hamilton, Alex and Kabat, Daniel N. and Lifschytz, Gilad and Lowe, David A.",
    title = "{Holographic representation of local bulk operators}",
    eprint = "hep-th/0606141",
    archivePrefix = "arXiv",
    reportNumber = "CU-TP-1149",
    doi = "10.1103/PhysRevD.74.066009",
    journal = "Phys. Rev. D",
    volume = "74",
    pages = "066009",
    year = "2006"
}

@article{Podolsky:2006px,
    author = "Podolsky, J. and Griffiths, J. B.",
    title = "{Accelerating Kerr-Newman black holes in (anti-)de Sitter space-time}",
    eprint = "gr-qc/0601130",
    archivePrefix = "arXiv",
    doi = "10.1103/PhysRevD.73.044018",
    journal = "Phys. Rev. D",
    volume = "73",
    pages = "044018",
    year = "2006"
}

@article{Chernicoff:2008sa,
    author = "Chernicoff, Mariano and Guijosa, Alberto",
    title = "{Acceleration, Energy Loss and Screening in Strongly-Coupled Gauge Theories}",
    eprint = "0803.3070",
    archivePrefix = "arXiv",
    primaryClass = "hep-th",
    doi = "10.1088/1126-6708/2008/06/005",
    journal = "JHEP",
    volume = "06",
    pages = "005",
    year = "2008"
}

@article{Russo:2008gb,
    author = "Russo, J. G. and Townsend, P. K.",
    title = "{Accelerating Branes and Brane Temperature}",
    eprint = "0805.3488",
    archivePrefix = "arXiv",
    primaryClass = "hep-th",
    reportNumber = "DAMTP-2008-20, UB-ECM-PF-08-07",
    doi = "10.1088/0264-9381/25/17/175017",
    journal = "Class. Quant. Grav.",
    volume = "25",
    pages = "175017",
    year = "2008"
}

@article{Ghoroku:2010sp,
    author = "Ghoroku, Kazuo and Ishihara, Masafumi and Kubo, Kouki and Taminato, Tomoki",
    title = "{Accelerated Quark and Holography for Confining Gauge theory}",
    eprint = "1010.4396",
    archivePrefix = "arXiv",
    primaryClass = "hep-th",
    reportNumber = "FIT-HE--10-02",
    doi = "10.1103/PhysRevD.83.024020",
    journal = "Phys. Rev. D",
    volume = "83",
    pages = "024020",
    year = "2011"
}

@article{Hirayama:2010xi,
    author = "Hirayama, Takayuki and Kao, Pei-Wen and Kawamoto, Shoichi and Lin, Feng-Li",
    title = "{Unruh effect and Holography}",
    eprint = "1001.1289",
    archivePrefix = "arXiv",
    primaryClass = "hep-th",
    doi = "10.1016/j.nuclphysb.2010.10.018",
    journal = "Nucl. Phys. B",
    volume = "844",
    pages = "1--25",
    year = "2011"
}

@article{Chernicoff:2010yv,
    author = "Chernicoff, Mariano and Paredes, Angel",
    title = "{Accelerated detectors and worldsheet horizons in AdS/CFT}",
    eprint = "1011.4206",
    archivePrefix = "arXiv",
    primaryClass = "hep-th",
    doi = "10.1007/JHEP03(2011)063",
    journal = "JHEP",
    volume = "03",
    pages = "063",
    year = "2011"
}

@article{Czech:2012be,
    author = "Czech, Bartlomiej and Karczmarek, Joanna L. and Nogueira, Fernando and Van Raamsdonk, Mark",
    title = "{Rindler Quantum Gravity}",
    eprint = "1206.1323",
    archivePrefix = "arXiv",
    primaryClass = "hep-th",
    doi = "10.1088/0264-9381/29/23/235025",
    journal = "Class. Quant. Grav.",
    volume = "29",
    pages = "235025",
    year = "2012"
}

@article{Parikh:2011aa,
    author = "Parikh, Maulik and Samantray, Prasant and Verlinde, Erik",
    title = "{Rotating Rindler-AdS Space}",
    eprint = "1112.3433",
    archivePrefix = "arXiv",
    primaryClass = "hep-th",
    doi = "10.1103/PhysRevD.86.024005",
    journal = "Phys. Rev. D",
    volume = "86",
    pages = "024005",
    year = "2012"
}

@article{Fareghbal:2014oba,
    author = "Fareghbal, Reza and Naseh, Ali",
    title = "{Rindler/Contracted-CFT Correspondence}",
    eprint = "1404.3937",
    archivePrefix = "arXiv",
    primaryClass = "hep-th",
    doi = "10.1007/JHEP06(2014)134",
    journal = "JHEP",
    volume = "06",
    pages = "134",
    year = "2014"
}

@article{Almheiri:2014lwa,
	author = "Almheiri, Ahmed and Dong, Xi and Harlow, Daniel",
	title = "{Bulk Locality and Quantum Error Correction in AdS/CFT}",
	eprint = "1411.7041",
	archivePrefix = "arXiv",
	primaryClass = "hep-th",
	reportNumber = "SU-ITP-14-30, SU-ITP-14/30",
	doi = "10.1007/JHEP04(2015)163",
	journal = "JHEP",
	volume = "04",
	pages = "163",
	year = "2015"
}

@article{Astorino:2016xiy,
    author = "Astorino, Marco",
    title = "{CFT Duals for Accelerating Black Holes}",
    eprint = "1605.06131",
    archivePrefix = "arXiv",
    primaryClass = "hep-th",
    reportNumber = "UAI-PHY-16-06",
    doi = "10.1016/j.physletb.2016.07.019",
    journal = "Phys. Lett. B",
    volume = "760",
    pages = "393--405",
    year = "2016"
}

@article{Appels:2016uha,
    author = "Appels, Michael and Gregory, Ruth and Kubiznak, David",
    title = "{Thermodynamics of Accelerating Black Holes}",
    eprint = "1604.08812",
    archivePrefix = "arXiv",
    primaryClass = "hep-th",
    reportNumber = "DCPT-16-15",
    doi = "10.1103/PhysRevLett.117.131303",
    journal = "Phys. Rev. Lett.",
    volume = "117",
    number = "13",
    pages = "131303",
    year = "2016"
}

@Article{Parikh2018,
	author={Parikh, Maulik
	and Samantray, Prasant},
	title={Rindler-AdS/CFT},
	journal={Journal of High Energy Physics},
	year={2018},
	day={19},
	volume={2018},
	number={10},
	pages={129},
	issn={1029-8479},
	doi={10.1007/JHEP10(2018)129},
	url={https://doi.org/10.1007/JHEP10(2018)129}
}

@article{Anabalon:2018qfv,
    author = "Anabal{\'o}n, Andr{\'e}s and Gray, Finnian and Gregory, Ruth and Kubiz{\v{n}}{\'a}k, David and Mann, Robert B.",
    title = "{Thermodynamics of Charged, Rotating, and Accelerating Black Holes}",
    eprint = "1811.04936",
    archivePrefix = "arXiv",
    primaryClass = "hep-th",
    doi = "10.1007/JHEP04(2019)096",
    journal = "JHEP",
    volume = "04",
    pages = "096",
    year = "2019"
}

@article{Anabalon:2018ydc,
    author = {Anabal{\'o}n, Andr{\'e}s and Appels, Michael and Gregory, Ruth and Kubiz{\v{n}}{\'a}k, David and Mann, Robert B. and Ovg{\"u}n, Al{\"\i}},
    title = "{Holographic Thermodynamics of Accelerating Black Holes}",
    eprint = "1805.02687",
    archivePrefix = "arXiv",
    primaryClass = "hep-th",
    doi = "10.1103/PhysRevD.98.104038",
    journal = "Phys. Rev. D",
    volume = "98",
    number = "10",
    pages = "104038",
    year = "2018"
}

@article{Gregory:2019dtq,
    author = "Gregory, Ruth and Scoins, Andrew",
    title = "{Accelerating Black Hole Chemistry}",
    eprint = "1904.09660",
    archivePrefix = "arXiv",
    primaryClass = "hep-th",
    reportNumber = "DCPT-19/09",
    doi = "10.1016/j.physletb.2019.06.071",
    journal = "Phys. Lett. B",
    volume = "796",
    pages = "191--195",
    year = "2019"
}

@article{Ferrero:2020twa,
    author = "Ferrero, Pietro and Gauntlett, Jerome P. and Ipi{\~n}a, Juan Manuel P{\'e}rez and Martelli, Dario and Sparks, James",
    title = "{Accelerating black holes and spinning spindles}",
    eprint = "2012.08530",
    archivePrefix = "arXiv",
    primaryClass = "hep-th",
    doi = "10.1103/PhysRevD.104.046007",
    journal = "Phys. Rev. D",
    volume = "104",
    number = "4",
    pages = "046007",
    year = "2021"
}

@article{Cassani:2021dwa,
    author = "Cassani, Davide and Gauntlett, Jerome P. and Martelli, Dario and Sparks, James",
    title = "{Thermodynamics of accelerating and supersymmetric AdS4 black holes}",
    eprint = "2106.05571",
    archivePrefix = "arXiv",
    primaryClass = "hep-th",
    reportNumber = "Imperial/TP/2021/JG/04",
    doi = "10.1103/PhysRevD.104.086005",
    journal = "Phys. Rev. D",
    volume = "104",
    number = "8",
    pages = "086005",
    year = "2021"
}

@article{Sugishita:2022ldv,
    author = "Sugishita, Sotaro and Terashima, Seiji",
    title = "{Rindler bulk reconstruction and subregion duality in AdS/CFT}",
    eprint = "2207.06455",
    archivePrefix = "arXiv",
    primaryClass = "hep-th",
    reportNumber = "YITP-22-70",
    doi = "10.1007/JHEP11(2022)041",
    journal = "JHEP",
    volume = "11",
    pages = "041",
    year = "2022"
}

@article{Barrientos:2022bzm,
    author = "Barrientos, Jose and Cisterna, Adolfo and Kubiznak, David and Oliva, Julio",
    title = "{Accelerated black holes beyond Maxwell's electrodynamics}",
    eprint = "2205.15777",
    archivePrefix = "arXiv",
    primaryClass = "gr-qc",
    doi = "10.1016/j.physletb.2022.137447",
    journal = "Phys. Lett. B",
    volume = "834",
    pages = "137447",
    year = "2022"
}

@article{Wu:2023meo,
    author = "Wu, Di",
    title = "{Topological classes of thermodynamics of the four-dimensional static accelerating black holes}",
    eprint = "2307.02030",
    archivePrefix = "arXiv",
    primaryClass = "hep-th",
    doi = "10.1103/PhysRevD.108.084041",
    journal = "Phys. Rev. D",
    volume = "108",
    number = "8",
    pages = "084041",
    year = "2023"
}

@article{Ju:2023bjl,
    author = "Ju, Xin-Xiang and Pan, Wen-Bin and Sun, Ya-Wen and Wang, Yuan-Tai",
    title = "{Generalized Rindler Wedge and Holographic Observer Concordance}",
    eprint = "2302.03340",
    archivePrefix = "arXiv",
    primaryClass = "hep-th",
    month = "2",
    year = "2023"
}

@article{Arenas-Henriquez:2023hur,
	author = "Arenas-Henriquez, Gabriel and Cisterna, Adolfo and Diaz, Felipe and Gregory, Ruth",
	title = "{Accelerating Black Holes in $2+1$ dimensions: Holography revisited}",
	eprint = "2308.00613",
	archivePrefix = "arXiv",
	primaryClass = "hep-th",
	doi = "10.1007/JHEP09(2023)122",
	journal = "JHEP",
	volume = "09",
	pages = "122",
	year = "2023"
}

@article{Ju:2023dzo,
    author = "Ju, Xin-Xiang and Liu, Bo-Hao and Pan, Wen-Bin and Sun, Ya-Wen and Wang, Yuan-Tai",
    title = "{Squashed entanglement from generalized Rindler wedge}",
    eprint = "2310.09799",
    archivePrefix = "arXiv",
    primaryClass = "hep-th",
    doi = "10.1007/JHEP09(2025)006",
    journal = "JHEP",
    volume = "09",
    pages = "006",
    year = "2025"
}

@article{Sin:2004yx,
    author = "Sin, Sang-Jin and Zahed, Ismail",
    title = "{Holography of radiation and jet quenching}",
    eprint = "hep-th/0407215",
    archivePrefix = "arXiv",
    reportNumber = "ICTP-IC-2004-58",
    doi = "10.1016/j.physletb.2005.01.020",
    journal = "Phys. Lett. B",
    volume = "608",
    pages = "265--273",
    year = "2005"
}

@article{Herzog:2006gh,
    author = "Herzog, C. P. and Karch, A. and Kovtun, P. and Kozcaz, C. and Yaffe, L. G.",
    title = "{Energy loss of a heavy quark moving through N=4 supersymmetric Yang-Mills plasma}",
    eprint = "hep-th/0605158",
    archivePrefix = "arXiv",
    reportNumber = "NSF-KITP-06-36",
    doi = "10.1088/1126-6708/2006/07/013",
    journal = "JHEP",
    volume = "07",
    pages = "013",
    year = "2006"
}

@article{Herzog:2006se,
    author = "Herzog, Christopher P.",
    title = "{Energy Loss of Heavy Quarks from Asymptotically AdS Geometries}",
    eprint = "hep-th/0605191",
    archivePrefix = "arXiv",
    doi = "10.1088/1126-6708/2006/09/032",
    journal = "JHEP",
    volume = "09",
    pages = "032",
    year = "2006"
}

@article{Gubser:2006bz,
    author = "Gubser, Steven S.",
    title = "{Drag force in AdS/CFT}",
    eprint = "hep-th/0605182",
    archivePrefix = "arXiv",
    reportNumber = "PUPT-2198",
    doi = "10.1103/PhysRevD.74.126005",
    journal = "Phys. Rev. D",
    volume = "74",
    pages = "126005",
    year = "2006"
}

@article{Liu:2006ug,
    author = "Liu, Hong and Rajagopal, Krishna and Wiedemann, Urs Achim",
    title = "{Calculating the jet quenching parameter from AdS/CFT}",
    eprint = "hep-ph/0605178",
    archivePrefix = "arXiv",
    reportNumber = "MIT-CTP-3739, RBRC-601",
    doi = "10.1103/PhysRevLett.97.182301",
    journal = "Phys. Rev. Lett.",
    volume = "97",
    pages = "182301",
    year = "2006"
}

@article{Chernicoff:2012gu,
    author = "Chernicoff, Mariano and Fernandez, Daniel and Mateos, David and Trancanelli, Diego",
    title = "{Jet quenching in a strongly coupled anisotropic plasma}",
    eprint = "1203.0561",
    archivePrefix = "arXiv",
    primaryClass = "hep-th",
    reportNumber = "DAMTP-2012-15, ICCUB-12-098, MAD-TH-12-02",
    doi = "10.1007/JHEP08(2012)041",
    journal = "JHEP",
    volume = "08",
    pages = "041",
    year = "2012"
}

@article{Akhmedov:1998vf,
    author = "Akhmedov, Emil T.",
    title = "{A Remark on the AdS / CFT correspondence and the renormalization group flow}",
    eprint = "hep-th/9806217",
    archivePrefix = "arXiv",
    reportNumber = "ITEP-TH-32-98",
    doi = "10.1016/S0370-2693(98)01270-2",
    journal = "Phys. Lett. B",
    volume = "442",
    pages = "152--158",
    year = "1998"
}

@article{Son:2002sd,
    author = "Son, Dam T. and Starinets, Andrei O.",
    title = "{Minkowski space correlators in AdS / CFT correspondence: Recipe and applications}",
    eprint = "hep-th/0205051",
    archivePrefix = "arXiv",
    reportNumber = "INT-PUB-02-34",
    doi = "10.1088/1126-6708/2002/09/042",
    journal = "JHEP",
    volume = "09",
    pages = "042",
    year = "2002"
}

@article{Polchinski:2001tt,
    author = "Polchinski, Joseph and Strassler, Matthew J.",
    title = "{Hard scattering and gauge / string duality}",
    eprint = "hep-th/0109174",
    archivePrefix = "arXiv",
    reportNumber = "NSF-ITP-01-76, UPR-956-T",
    doi = "10.1103/PhysRevLett.88.031601",
    journal = "Phys. Rev. Lett.",
    volume = "88",
    pages = "031601",
    year = "2002"
}

@article{Boschi-Filho:2002xih,
    author = "Boschi-Filho, Henrique and Braga, Nelson R. F.",
    title = "{Gauge / string duality and scalar glueball mass ratios}",
    eprint = "hep-th/0212207",
    archivePrefix = "arXiv",
    doi = "10.1088/1126-6708/2003/05/009",
    journal = "JHEP",
    volume = "05",
    pages = "009",
    year = "2003"
}

@article{Boschi-Filho:2002wdj,
    author = "Boschi-Filho, Henrique and Braga, Nelson R. F.",
    title = "{QCD / string holographic mapping and glueball mass spectrum}",
    eprint = "hep-th/0209080",
    archivePrefix = "arXiv",
    doi = "10.1140/epjc/s2003-01526-4",
    journal = "Eur. Phys. J. C",
    volume = "32",
    pages = "529--533",
    year = "2004"
}

@article{Kovtun:2004de,
    author = "Kovtun, P. and Son, Dan T. and Starinets, Andrei O.",
    title = "{Viscosity in strongly interacting quantum field theories from black hole physics}",
    eprint = "hep-th/0405231",
    archivePrefix = "arXiv",
    reportNumber = "INT-PUB-04-09, UW-PT-04-04",
    doi = "10.1103/PhysRevLett.94.111601",
    journal = "Phys. Rev. Lett.",
    volume = "94",
    pages = "111601",
    year = "2005"
}

@article{Casalderrey-Solana:2006fio,
    author = "Casalderrey-Solana, Jorge and Teaney, Derek",
    title = "{Heavy quark diffusion in strongly coupled N=4 Yang-Mills}",
    eprint = "hep-ph/0605199",
    archivePrefix = "arXiv",
    doi = "10.1103/PhysRevD.74.085012",
    journal = "Phys. Rev. D",
    volume = "74",
    pages = "085012",
    year = "2006"
}

@article{Liu:2006nn,
    author = "Liu, Hong and Rajagopal, Krishna and Wiedemann, Urs Achim",
    title = "{An AdS/CFT Calculation of Screening in a Hot Wind}",
    eprint = "hep-ph/0607062",
    archivePrefix = "arXiv",
    reportNumber = "MIT-CTP-3757",
    doi = "10.1103/PhysRevLett.98.182301",
    journal = "Phys. Rev. Lett.",
    volume = "98",
    pages = "182301",
    year = "2007"
}

@article{Ryu:2006bv,
    author = "Ryu, Shinsei and Takayanagi, Tadashi",
    title = "{Holographic derivation of entanglement entropy from AdS/CFT}",
    eprint = "hep-th/0603001",
    archivePrefix = "arXiv",
    reportNumber = "NSF-KITP-06-11, NSF-KITP-06-11",
    doi = "10.1103/PhysRevLett.96.181602",
    journal = "Phys. Rev. Lett.",
    volume = "96",
    pages = "181602",
    year = "2006"
}

@article{Shuryak:2005ia,
    author = "Shuryak, Edward and Sin, Sang-Jin and Zahed, Ismail",
    title = "{A Gravity dual of RHIC collisions}",
    eprint = "hep-th/0511199",
    archivePrefix = "arXiv",
    doi = "10.3938/jkps.50.384",
    journal = "J. Korean Phys. Soc.",
    volume = "50",
    pages = "384--397",
    year = "2007"
}

@article{Grumiller:2008va,
    author = "Grumiller, Daniel and Romatschke, Paul",
    title = "{On the collision of two shock waves in AdS(5)}",
    eprint = "0803.3226",
    archivePrefix = "arXiv",
    primaryClass = "hep-th",
    reportNumber = "INT-PUB-08-06, MIT-CTP-3939, YITP-08-18",
    doi = "10.1088/1126-6708/2008/08/027",
    journal = "JHEP",
    volume = "08",
    pages = "027",
    year = "2008"
}

@article{Bobev:2005cz,
    author = "Bobev, N. P. and Dimov, H. and Rashkov, R. C.",
    title = "{Semiclassical strings in Lunin-Maldacena background}",
    eprint = "hep-th/0506063",
    archivePrefix = "arXiv",
    journal = "Bulg. J. Phys.",
    volume = "35",
    pages = "274--285",
    year = "2008"
}

\end{document}